\documentclass[10pt, final, journal, letterpaper, twocolumn]{IEEEtran}

\IEEEoverridecommandlockouts
\usepackage{multirow,array}
\usepackage{booktabs}
\usepackage{mathrsfs}
\usepackage{amsfonts}
\usepackage{array}
\usepackage{algorithmic}
\usepackage{amssymb}
\usepackage{amsmath}
\usepackage{graphicx}
\usepackage{multirow}
\usepackage{amsthm}
\usepackage{color}
\usepackage{mathtools}
\usepackage[table]{xcolor}
\usepackage{empheq}
\usepackage[linesnumbered,lined,vlined,ruled,commentsnumbered]{algorithm2e}
\usepackage{epstopdf}
\usepackage{stmaryrd}
\usepackage{multirow,url}
\usepackage{cite}
\usepackage{bm}
\usepackage{indentfirst}
\usepackage{epsfig}
\usepackage{xcolor}
\usepackage{subfigure}

\newtheorem{theorem}{Theorem}

\newtheorem{proposition}{Proposition}
\newtheorem{lemma}{Lemma}
\newtheorem{game}{Game}
\newtheorem{definition}{Definition}
\newtheorem{assumption}{Assumption}
\newtheorem{observation}{Observation}

\newtheorem{corollary}{Corollary}
\newtheorem{example}{Example}

\ifodd 1

\newcommand{\com}[1]{\textbf{\color{red} (Comment: #1) }}
\newcommand{\comg}[1]{\textbf{\color{blue} (COMMENT: #1)}}
\newcommand{\response}[1]{\textbf{\color{blue} (RESPONSE: #1)}}
\else

\newcommand{\com}[1]{}
\newcommand{\comg}[1]{}
\newcommand{\response}[1]{}
\fi

\newcommand{\bs}{\boldsymbol}

\begin{document}

\title{Hybrid Pricing for Mobile Collaborative Internet Access}
%

\author{\IEEEauthorblockN{Meng Zhang, 
 Lin Gao,~\IEEEmembership{Senior Member,~IEEE},\\
Jianwei Huang,~\IEEEmembership{Fellow,~IEEE}, and Michael L. Honig,~\IEEEmembership{Fellow,~IEEE}}
\thanks{
%
 This work was supported in part by the General Research Fund CUHK 14219016 from Hong Kong UGC, in part by the Presidential Fund from the Chinese University of Hong Kong, Shenzhen, in part by the NSFC under Grant 61771162, and in part by NSF grant AST-134338. 	Part of this work was presented
 at IEEE INFOCOM 2017 \cite{INFOCOM}.

M. Zhang and J. Huang are with the Department of Information
Engineering, The Chinese University of Hong Kong, Hong Kong (e-mail:
zm015@ie.cuhk.edu.hk; jwhuang@ie.cuhk.edu.hk). J. Huang is also with the School of Science and Engineering, The Chinese University of Hong Kong, Shenzhen.

L. Gao is with the School of Electronic and Information Engineering,
Harbin Institute of Technology, Shenzhen 150001, China (e-mail:
gaolin@hitsz.edu.cn).

M. L. Honig is with the Department of Electrical Engineering
and Computer Science, Northwestern University, Evanston, IL 60208 USA (e-mail: mh@eecs.northwestern.edu).

}
%
%
\vspace{-30pt}
}

\maketitle

\begin{abstract}
Mobile Collaborative Internet Access (MCA) enables mobile users to share their Internet through flexible tethering arrangements. This can potentially make better use of network resources.
However, from a mobile network operator's (MNO's) viewpoint, 
it can either reduce revenue or increase congestion, and thus has been blocked by some MNOs in practice.
We propose a hybrid pricing framework for MNOs who charge users separately for access and tethering. 
This scheme serves to coordinate the tethering decisions of mobile users with MNO network management objectives.
We analyze the MNOs' equilibrium pricing strategies in both cooperative and competitive scenarios.
In the cooperative scenario, at the equilibrium, each user's cost is independent of any chosen tethering links. We then characterize the optimal hybrid pricing strategies of MNOs  in this scenario. For the competitive scenario, 
we formulate the MNOs' competitive interactions as a pricing game, and we show that MNO competition leads to equalized prices for users if an equilibrium exists but does not guarantee its existence. Both insights motivate a quantity competition game, which is shown to guarantee an equilibrium.
Simulation results show that in scenarios of interest the proposed hybrid pricing schemes can double both MNOs' profit and users' payoff and such improvements increase with the degree of network heterogeneity.
\end{abstract}
\vspace{-0.1cm}

\section{Introduction}

\subsection{Background and Motivation}

Global mobile data traffic has been experiencing explosive growth and is expected to reach 77.5 exabytes per month by 2022, approximately a sevenfold increase over 2017\cite{Cisco}. However, mobile network capacity is growing relatively slowly \cite{Cisco}, which results in a global mismatch of demand and supply. On the other hand, the heterogeneity of networks and mobile users leads to different types of mismatch even at the same time and location. For example, one user may underutilize network resources in a high-speed network, while another user, connected to a separate overlapping network with less capacity or higher load, may be underserved. This creates opportunities for more effective resource allocation and sharing across networks and users.

One approach to achieving more efficient network resource utilization is the new paradigm of  \textit{user-provided network} (UPN), in which users serve as micro-operators and provide network connections (and resources) for others\cite{UPN2}.
Examples of UPNs include the services enabled by companies such as Karma \cite{Kar} and Open Garden \cite{opengarden}. Specifically,
Open Garden provides a mobile app enabling mobile devices to dynamically form mesh networks (through Bluetooth and Wi-Fi direct links) and flexibly tether data for each other to improve Internet connections across users. The Open Garden enables mobile multi-hop and multi-path connectivity sharing among users and realizes mobile collaborative Internet access (MCA).

\begin{figure}
	\begin{centering}
		\includegraphics[scale=.4]{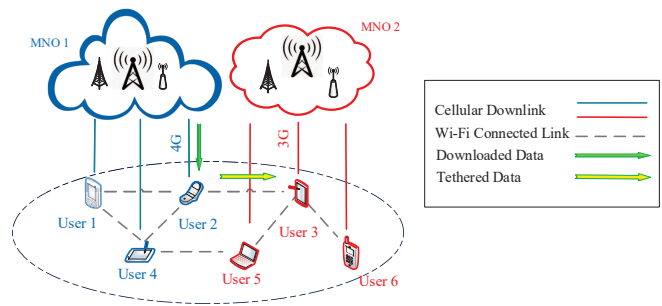}
		\vspace{-0.2cm}
		\caption{Illustration of an Open-Garden-like MCA framework consisting of two MNOs and six users.}
		\label{system}
	\end{centering}
	\vspace{-0.5cm}
\end{figure}

Fig. \ref{system} illustrates an Open-Garden-like MCA service with 2 mobile network operators (MNOs) and 6 users, where MNO 1 and its subscribers are marked in blue and MNO 2 and its subscribers are marked in red. In this example, user 3 has a high data demand that cannot be satisfied by her 3G downlink. In contrast, user 2 is connected via an underutilized 4G link. User 3 therefore requests a tether to user 2.

Although MCA can improve the overall utilization of network resources, MNOs have not been entirely supportive of MCA and the forms of shared mobile Internet access or \emph{tethering}. This is mainly because MCA can effectively exploit the diversity of  users and network environments and bridge heterogeneous networks, which may  reduce revenue to MNOs who charge high access prices and may increase more congestions for MNOs who charge low access prices. In other words,  each user can exploit the MCA to download her dedicated data on a low price downlink, reducing each MNO's profit. 
Consequently, AT\&T in the US requested Google Play to block its subscribers' access to Open Garden \cite{block}.

Even before the emergence of MCA services such as Open Garden, MNOs deployed several methods to restrict tethering (in the form of personal hotspots) among users belonging to different MNOs, mainly by direct prohibition or imposing an additional tethering fee. For example, the MNO au in Japan charges about \$5 per month  for the tethering service\cite{au}. In the US, MNOs such as AT\&T, Verizon Wireless, and Sprint  charge an additional \$10-\$30 per month for user tethering \cite{TetherCharge}. In the UK, some MNOs such as Virgin Mobile and iD Mobile prohibit all forms of tethering \cite{UKTether}, while some other  MNOs (such
as Three) charge users an additional fee for upgrading to data
plans that allow tethering \cite{Three}. 

\begin{table}[]
	\centering
	\caption{Parameters of Example \ref{E1} (All units in \$/GB)}
	\vspace{-0.3cm}
	\label{Ex1}
	\begin{tabular}{c c c c}
		\hline \hline
		MNO                                  & \multicolumn{2}{c}{$1$}           & $2$                    \\ \hline
		Charge for Data Service              & \multicolumn{2}{c}{$10$ } & $15$           \\ \hline
		Operational Cost                     & \multicolumn{2}{c}{$5$ }  & $11$       \\ \hline
		Tethering Charge                     & \multicolumn{2}{c}{$2$ }  & $-$                  \\ \hline
		Profit w/o Tethering                 & \multicolumn{2}{c}{$5$ }  & $4$           \\ \hline Profit w/ Tethering        & \multicolumn{2}{c}{$5\vert7$} & $0$\\ \hline \hline
		User                                 & \multicolumn{2}{c}{$1$}           & $2$                    \\ \hline
		Payment w/o Tethering                & \multicolumn{2}{c}{$10$ } & $15$           \\ \hline
		Payment w/ Tethering                 & \multicolumn{2}{c}{$10$} & $12$          \\ \hline
	\end{tabular}
	\vspace{-0.5cm}
\end{table}

Inspired by existing tethering pricing policies, 
we propose a hybrid pricing framework for MNOs to reconcile the conflicting objectives of MNOs and users of MCA.
The key idea is to separate the pricing for dedicated Internet access and tethering access.  Specifically,
each MNO sets a usage-based \textit{access price} for each subscriber and an additional usage-based \textit{tethering price} associated with MCA. 	
If a pricing scheme accounts for such MCA dynamics and offers proper incentives for users to properly load balance across MNOs, it can lead to a mutual benefit for both the MNOs and users. 


\begin{table*}[t]
	\centering
	\caption{Summary of Key Model Features}
	\vspace{-0.2cm}
	\begin{tabular}{c| c c c c c c}
		\hline
		{Scenario} & {Formulation}& Key Features  & {Existence of Equilibrium} & {Section}\\
		\hline
		\hline
		Cooperative & Optimization & MNOs maximize their total profit & $\surd$ &   \ref{cooperative}  \\
		\hline
		\multirow{ 2}{*}{Competitive} & Price Competition Game (PCG) & MNOs compete on prices &may not exist &  \ref{PC}\\
		\cline{2-5}
		& Quantity Competition Game (QCG)&  MNOs compete on quantity outputs & $\surd$ &  \ref{QC} \\
		\hline
	\end{tabular}
	\label{res}
	\vspace{-0.4cm}
\end{table*}

\subsection{An Illustrative Example}

\begin{figure}
	\begin{centering}
		\includegraphics[scale=.38]{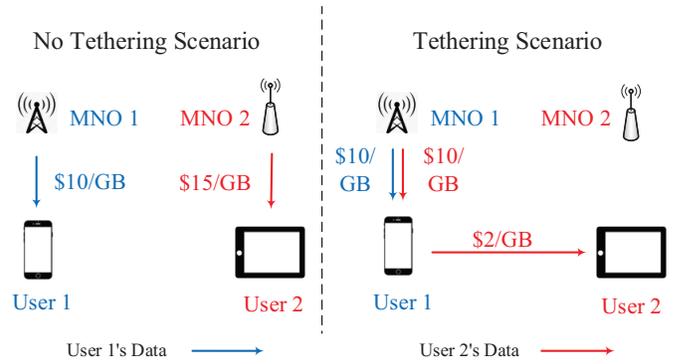}
		\vspace{-0.5cm}
		\caption{Illustration of the pricing scheme where
			no tethering is allowed (left) and the proposed hybrid pricing framework
			(right) for an MCA network with two MNOs and two users. 
		}
		\label{system2}
	\end{centering}
	\vspace{-0.4cm}
\end{figure}

Before presenting our solution and contributions, we first provide an example to illustrate the proposed hybrid pricing framework and how it differs from traditional pricing without tethering. The key parameters are presented in Table \ref{Ex1}.
\begin{example}\label{E1}	
	Consider two MNOs offering Internet access services to two neighboring users, respectively, as shown in Fig. \ref{system2}. Consider the following pricing scenarios:
	\begin{itemize}
		\item \textbf{No Tethering}: MNOs do not allow tethering. Therefore, each user can only download data through her MNO. 
		\item \textbf{Tethering}: 	
		MNO $1$ charges user $1$ a separate fee for  data tethering. Therefore,  user $2$ can access through user $1$'s cellular downlink at a lower cost. 
		For the costs  shown in Table \ref{Ex1}, the tethering pricing scheme offers user $2$ a discount of $\$3$ per GB  ($\$12$ per GB instead of $\$15$ per GB). MNO $1$ receives a profit of $\$5/$GB from  user $1$ and a profit of  $\$7/$GB from  user $2$. Therefore,  the total MNOs' profit from user $2$ with tethering (which is $\$7/$GB)  is higher than that when there is no tethering (which is $\$4/$GB). 
		This illustrates how the pricing scheme with tethering can benefit both MNOs  (as a whole)  and users.
	\end{itemize}
\end{example}

The preceding example also shows that an MNO's profit (such as the one of MNO 2) may decrease by allowing the tethering scheme, as users (e.g., user 2) may suspend their dedicated Internet access.  This incentivizes MNOs either to cooperate and share a higher total profit, or to engage in competitive hybrid pricing strategies to attract more user traffic. 
An example of such an MNO cooperation is the national roaming service, where MNOs (e.g., Reliance and Vodafone in India) provide mobile data service for other MNO's subscribers \cite{roam}. Examples of MNO competition are more common and appear where it is difficult for MNOs to cooperate due to, for example, technical,  policy, and market concerns.

Motivated by the preceding discussion, we ask the following questions  in this paper:
\begin{enumerate}
	\item How would cooperative MNOs set hybrid prices for access and tethering  to maximize their total profit? 
	\item How would competitive MNOs set equilibrium hybrid prices  to maximize their own profits? 
	\item How does the introduction of hybrid pricing impact MNOs' profit and users' surplus in both cooperative and competitive scenarios? 
\end{enumerate}

\subsection{Contributions}


We consider a model with multiple MNOs and users, where users share their Internet access through MCA and MNOs adopt hybrid pricing schemes to charge each user separately for Internet access and tethering. 
We assume MNOs choose pricing policies over a slower time scale relative to the rate at which users choose a dedicated MNO. Hence, we model the interaction between MNOs and users as a two-stage leader-follower (Stackelberg) game\cite{GameTheory}.
In Stage I , the MNOs determine the access and tethering prices. Cooperative MNOs decide their hybrid pricing schemes jointly to maximize their total profit. We also consider a competitive scenario, where each MNO maximizes its own profit. In Stage II, the neighboring users cooperatively decide the amount of traffic to download and to tether to maximize their total payoff. 

The challenges of analyzing equilibria in both cooperative and competitive scenarios mainly lie in the multiple ways in which the users can configure their access connections, either through direct access or tethered to another user.
This combined with coupled constraints make it difficult to characterize users' optimal decisions. The large space of user decisions and interaction among MNOs' pricing decisions further complicate the competitive pricing scheme.

We summarize the key features of the scenarios studied in Table \ref{res}. Our main contributions follow.
\begin{itemize}
	\item \textit{Hybrid Pricing Framework.} 
	We formulate the first model for a hybrid pricing scheme for MCA. It reconciles the conflicting objectives of MNOs and users in MCA.
	
	
	\item \textit{Cooperative Scheme.} We first study hybrid pricing schemes for cooperative MNOs. We show that the optimal (MNO-profit-maximizing) scheme makes each user's cost independent of her selection of tethering links. This allows a transformation of the challenging hybrid pricing problem into a tractable one.
	We show a numerical example in which the cooperative scheme approximately doubles both MNOs' profit and  users' payoff in a practical scenario of interest.

	
	\item \textit{Competitive Scheme.} We then study equilibrium hybrid pricing with competitive MNOs. We formulate the MNOs' interactions as a Price Competition Game (PCG) and show that in equilibrium prices are equalized across different users. However, an equilibrium may
	not exist in some cases. 
	The equalized prices and non-existence equilibrium motivate us to reformulate the price setting of MNOs as a Quantity Competition Game (QCG). We show that the QCG game always has (at least) an equilibrium and approximates the PCG.
	Numerical results show that, compared to the cooperative scheme, the competitive scheme achieves slightly smaller MNO profits but much larger user payoffs , in a  heterogeneous network.
	
	\item \textit{Free-Tethering.} We further study the scheme where tethering is allowed without additional payment. Our analytical and numerical results show that pricing with free tethering is nearly optimal for cooperative MNOs, when users have similar cellular and tethering energy costs.

	%

\end{itemize}

We organize the rest of paper as follows. In Section \ref{Related}, we review related work.
In Section \ref{System}, we introduce the system model and formulate the two-stage Stackelberg game. We study the cooperative hybrid pricing scheme in Section \ref{cooperative}. We study two types of competitive interactions among MNOs, including the PCG and the QCG, in Sections \ref{PC} and \ref{QC}, respectively.
We present the numerical results in Section \ref{Numerical}.
Finally, Section \ref{Conc} presents conclusions.

\section{Related Works}\label{Related}



\subsection{User Provided Networks}
Several incentive mechanisms
for the Open-Garden-like MCA service have been proposed in \cite{Open,bitcoin,Open-2}.
Specifically, in \cite{Open-2}, Iosifidis \textit{et al.} propose a distributed incentive mechanism for encouraging MCA service. In \cite{bitcoin}, Syrivelis \textit{et al.} design a cloud-controlled MCA service and study a coalitional game played among users. 
  In \cite{Open}, Georgiadis \textit{et al.} study incentive mechanisms for services exchange in general networks. However, the preceding work does not consider the impact of MCA on MNOs or the interactions among the MNOs.

There are also two related forms of UPN that have been studied.
The first is \textit{wireless community networks} (e.g. FONs \cite{FON}), where individuals share their private residential Wi-Fi access points. Several existing works have considered pricing and incentive design for UPN \cite{add-1}.
 Afrasiabi \textit{et al.} in \cite{FON1} propose a low
introductory price policy to promote the service adoption.
Ma \textit{et al.} in \cite{MaFON} study the user behavior analysis and the MNOs' pricing design. 
The second form is the Karma-like UPN \cite{Kar}. Specifically, Karma sells mobile devices (that convert 4G cellular signals to Wi-Fi signals) to its subscribers and encourages the subscribers to operate as Wi-Fi hotspots and provide Internet access to non-subscribers.
 In  \cite{Karma},
Gao \textit{et al.} propose a hybrid data pricing scheme motivated by Karma's UPN service, to incentivize users to operate as mobile Wi-Fi hotspots and provide Internet access for other users without direct Internet access.
In \cite{Karma-2}, Khalili \textit{et al.} further study the user behavior dynamics and network evolution under such a hybrid data pricing scheme.
Substantially different from these models, the MCA model allows users to concurrently share flexible direct mobile Internet access.

\subsection{Internet Access Pricing}

Our hybrid pricing scheme is also motivated by related work on  topology-aware pricing schemes for wireless mesh networks \cite{Mesh,Mesh2} and cooperative communication networks \cite{Cooper}.
The focus of those studies is on pricing and incentive issues for user cooperation, without considering tethering pricing schemes.




\section{System Model}\label{System}

\subsection{System Overview}

 \emph{MNOs and Users}: We consider an MCA model with a set $\mathcal{N}=\{1,2,...,N\}$ of MNOs and a set $\mathcal{I}=\{1,2,...,I\}$ of users, as illustrated  in Fig. \ref{system}. 
 We assume each user subscribes to only one MNO, which provides a mobile Internet access service for each of its subscribers.  Let $\mathcal{I}_n$ denote the set of subscribers of MNO $n$, and let $\sigma(i)$ denote the MNO to which user $i$ subscribes, i.e., $i\in\mathcal{I}_{\sigma(i)}$ for all $i\in\mathcal{I}$.\footnote{In Fig. \ref{system}, for example, MNO 2's subscriber set is $\mathcal{I}_2=\{3,5,6\}$, and user
2 subscribes to MNO $\sigma(2)=1$.}

\emph{Cellular Links:} For presentation clarity, we focus on the downlink.\footnote{  The uplink case is similar to the downlink, with minor changes such as smaller uplink capacities. For the simultaneous downlink and uplink case, a user's coupled utility for both downloading and uploading  significantly complicates the modeling. Hence, it is left for future work.} Let $C_{i}$ (MBps) be the maximum rate (capacity) that user $i\in\mathcal{I}$ can achieve from his cellular downlink provided by MNO $\sigma(i)$.

\emph{Wireless Mesh Network:} The users cooperate and provide an Open-Garden-like MCA service to each other. Specifically, users form one wireless mesh network, where all users are connected through Wi-Fi Direct.\footnote{For the case where the users form more than one disjointed wireless mobile network, an MNO can differentially price users in each wireless mesh network. Thus, we can decompose the pricing problem  at each wireless mesh network. Therefore, it is enough to focus on one wireless mobile network.}  Therefore, there is a Wi-Fi (tethering) link between each pair of users, which can be a one-hop or multi-hop Wi-Fi Direct link. 
 We assume the Wi-Fi links have high capacities. Hence, the performance bottleneck in the network is always the cellular links. Note that we can equivalently treat a multi-hop connection as a single-hop one  in our model. This is because (i) 
 a tethering price (to be introduced) is independent of any relay in a multi-hop connection, and (ii)  the capacities of Wi-Fi links are sufficiently high. 
 Hence, we focus on the case where every pair of users are single-hop neighbors for simplicity.

\emph{Time Period:} We consider a time period (e.g. several hours to a day) consisting of several time slots (e.g. several minutes each). We consider a
quasi-static mobility model, where each user moves randomly
across time slots, and remains at the same location within each
time slot. 
The  link capacities are considered constant during each  time slot.


\emph{User Roles:} In each time slot, each user can be a \emph{gateway}, or a \emph{client}, or both. A gateway node downloads data directly from its MNO, and a client node consumes data for some local  mobile applications.



\emph{MNO Cost:} We consider a linear average operational cost $e_i$ of MNO $\sigma(i)$ for transmitting every GB to user $i$.
Sending traffic to different users may incur different operational costs because of the different technologies (3G/LTE) and the different channel conditions.

\emph{User Cost:} We include a user's energy cost $c_{i\leftarrow j}$ per GB for the data transfer over cellular downlink $j$ and the Wi-Fi Direct to user $i$, given by\cite{energy1}
\begin{equation}
c_{i\leftarrow j}=c^\text{Down}_{ j}+c^\text{Wi-Fi}_{i\leftarrow j},
\end{equation}
where $c^\text{Down}_{j}$ denotes the energy cost per GB for user $i$ to download, and $c^\text{Wi-Fi}_{i\leftarrow j}$ includes the energy cost on Wi-Fi Direct for each tethered GB. It captures the aggregate cost experienced  by tethering initiator (user $j$) and the tethering recipient (user $i$).

%

\subsection{Mobile Network Operators}

\emph{Access Prices:} 
Denote $a_i\geq0$ as the \textit{access price} (\$/GB) that MNO $\sigma(i)$ charges user $i$. Here we allow the MNOs to charge different prices to different users, based on users' QoS requirements and network service types. Perfect price
differentiation among users leads to the
maximum design flexibility for the MNOs.\footnote{In
practice, the MNOs may partially
differentiate among users with a limited
number of prices \cite{diff}. An example of partial price differentiation is the student discounts for mobile data plans offered by many MNOs \cite{Inforules}. }

\emph{Tethering Prices:} We further consider a linear tethering price, i.e., the MNOs can charge for each tethered GB in addition to the basic data payment. Let $t_{i\leftarrow j}$ denote the \textit{tethering price} (\$/GB) that MNO $\sigma(j)$ charges user $j$ for each GB that user $j$ tethers to user $i$. Note that the MNO can set a negative tethering price, in which case MNO $\sigma(j)$ will give user $j$ a discount of $|t_{i\leftarrow j}|$ for the data tethered to user $i$. 
 Here we define $t_{i\leftarrow i}\triangleq0$ for each $i$. We further allow the MNO to differentiate not only the gateway users but also the clients,\footnote{In practice, to charge/block tethering data, MNOs may adopt different techniques to detect whether and for whom a user is tethering, such as inspecting MAC addresses and network packets for TTLs \cite{tether}.} \emph{i.e.,} $t_{i\leftarrow j}$ can be different for every $i$ and $j$.

\emph{Hybrid Prices:} Define $\bs{h}\triangleq\{h_{i\leftarrow j}\}_{i,j\in\mathcal{I}}$ as the hybrid price matrix, where $h_{i\leftarrow j}$ denotes the hybrid price that user $j$ needs to pay to her MNO, for each GB she tethers to user $i$. This includes both the access price and the tethering price,
\begin{align}
h_{i\leftarrow j}\triangleq a_j+t_{i\leftarrow j}\geq 0.
\end{align}


\emph{Profit:} With the preceding notation, MNO $n$'s profit is
\begin{equation}
V_n=\sum_{j\in\mathcal{I}_n}\sum_{i\in\mathcal{I}}(h_{i\leftarrow j}-e_j)x_{i\leftarrow j}\label{MNOprofit2},
\end{equation}
where $x_{i\leftarrow j}$ is the user traffic (to be defined next).

\subsection{Mobile Users}

\emph{Traffic Matrix:} Let  $x_{i\leftarrow j}\geq0$ denote the data downloaded by user $j$ and tethered to user $i$ (tethered data) if $j\neq i$, and the data user $i$ downloads for herself (directly downloaded data) if $j=i$. We define $\bs{x}\triangleq\{x_{i\leftarrow j}\}_{i,j\in\mathcal{I}}$ as the traffic matrix.


\emph{Payoff:} Let $J(\cdot)$ denote the users' total payoff, given by
\begin{equation}
J\left(\bs{x};\bs{h}\right)=\sum_{i\in\mathcal{I}}U_i\left(\sum_{j\in\mathcal{I}}x_{i\leftarrow j}\right)-\sum_{i\in\mathcal{I}}\sum_{j\in\mathcal{I}}(h_{i\leftarrow j}+c_{i\leftarrow j})x_{i\leftarrow j},
\end{equation}
where $U_i(\cdot)$ is a positive, increasing, twice-continuously differentiable, and strictly concave function of user $i$. The strict concavity indicates that users have the diminishing marginal satisfaction of additional data consumption. 

We summarize notations in Table \ref{notation}.

\begin{table}[]
	\centering
	\caption{Notation}
		\vspace{-0.3cm}
	\label{notation}
	\begin{tabular}{|c|c|}
		\hline
		\hline
		Symbol              & Physical Meaning                                             \\ \hline
		$\mathcal{I}_n$         & Set of MNO $i$'s subscribers                                \\ \hline
		$\sigma(i)$         & MNO subscribed by user $i$                                   \\ \hline
		$C_i$               & Downlink capacity of user $i$                                \\ \hline
		$e_i$               & Operational cost over downlink $i$, borne by MNO $\sigma(i)$ \\ \hline
		$c_{i\leftarrow j}$ & Energy cost over link $i\leftarrow j$, borne by users        \\ \hline
		$x_{i\leftarrow j}$ & Data downloaded by user $j$ and tethered to user $i$         \\ \hline
		$U_i(\cdot)$        & Utility function for user $i$                                \\ \hline
		$V_n(\cdot)$        & Profit function for MNO $n$                                  \\ \hline
		$a_i$               & Access price over downlink $i$                               \\ \hline
		$t_{i\leftarrow j}$ & Tethering price over link $i\leftarrow j$                    \\ \hline
		$h_{i\leftarrow j}$ & Hybrid price over link $i\leftarrow j$                       \\ \hline
		$p_i$               & Delivered price for user $i$                                           \\ \hline
		$J(\cdot)$          & Users' total payoff                                          \\ \hline
				$d_i(\cdot)$          & User $i$'s demand function                                          \\ \hline
	\end{tabular}
		\vspace{-0.5cm}
\end{table}

\subsection{Two-Stage Stackelberg Game}

MNOs decide the hybrid prices for each time period at the beginning of the entire time period, and users decide the data traffic in each time slot. Since pricing is independent across different time slots, we focus on one time slot.  

\begin{figure}[t]
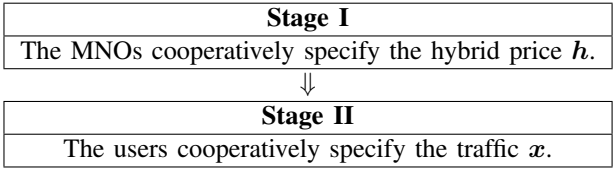

	\centering
	\label{my-label}
	\begin{tabular}{c}
		\hline
		\multicolumn{1}{|c|}{ \textbf{Stage I}}        \\ \hline
		\multicolumn{1}{|c|}{The MNOs cooperatively specify the hybrid price $\bs{h}$.} \\ \hline
		$\Downarrow   $                                  \\ \hline
		\multicolumn{1}{|c|}{\textbf{Stage II}}        \\ \hline
		\multicolumn{1}{|c|}{The users cooperatively specify the traffic $\bs{x}$.}         \\ \hline
	\end{tabular}
	\caption{Two-Stage Stackelberg Game with cooperative MNOs.}\label{Stackelberg}
	\vspace{-0.5cm}
\end{figure}

  We model the interaction between MNOs and users as a two-stage Stackelberg game, as illustrated in Fig. \ref{Stackelberg}. Specifically, in Stage I, the MNOs simultaneously decide the hybrid price matrix $\bs{h}$. In Stage II, the users simultaneously  decide the traffic matrix $\bs{x}$. 
  
  Note that, in Stage II, users only participate in the MCA if they receive higher payoffs than not participating.
  Such user cooperation can be achieved with a self-enforcing bargaining mechanism as in \cite{Open-2}. Such a mechanism  can maximize the aggregate user payoff. The aggregate payoff can be fairly shared among users through proper money transfer by solving the corresponding Nash bargaining problem \cite{Open-2}. For the purpose of designing an MNO's hybrid pricing scheme, it is enough to focus on optimizing the users' traffic $\bs{x}$ through solving the users' payoff maximization problem (without considering how users decide their money transfers).
  
In Stage I, we consider two sets of models where the MNOs are cooperative and competitive, respectively. We assume that the cooperative MNOs jointly choose the hybrid pricing matrix to maximize their total profit. Each competitive MNO sets the data and tethering prices to maximize its own profit.

\section{Cooperative Hybrid Pricing}\label{cooperative}

In this section, motivated by existing MNO cooperation (e.g. national roaming \cite{roam}), we consider the scenario in which MNOs cooperatively determine prices.\footnote{Similar to the users cooperation, the MNO cooperation is achievable by a bargaining mechanism, since each MNO can be better off by participating in such a mechanism.} Specifically, MNOs
cooperatively decide the hybrid price matrix $\bs{h}$ to maximize their total profit in Stage I, and users cooperatively determine the traffic in Stage II.


We will derive the \textit{(subgame perfect) equilibrium} by backward induction, i.e., given the hybrid pricing matrix $\bs{h}$, we characterize the users' traffic decision $\bs{x}^*(\bs{h})$ that maximizes the users' total payoff in Stage II, and then we characterize the MNOs' optimal hybrid pricing matrix $\bs{h}^*$ that maximizes the MNOs' profit.

\subsection{Users' Consumption Decisions in Stage II}\label{MUCom}

We first formulate the Users' Payoff Maximization (UPM) problem and then characterize the optimal traffic under the equilibrium hybrid price matrix $\bs{h}^*$ along with several  structural results.

\subsubsection{UPM}
Under an  arbitrary hybrid price matrix $\bs{h}$, we formulate (UPM) as
\begin{subequations}\label{UPM}
\begin{align}
{\rm (UPM)}~&\max_{\bs{x}}~\sum_{i\in\mathcal{I}} \left[U_i\left(\sum_{l\in\mathcal{I}}x_{i\leftarrow l}\right)-\sum_{j\in\mathcal{I}}(h_{i\leftarrow j}+c_{i\leftarrow j})x_{i\leftarrow j}\right]\label{UPM-Obj}\\
&{\rm s.t.}~~~~~\sum_{l\in\mathcal{I}}x_{l\leftarrow j}\leq C_j, ~~~~~~~~~~~~\forall j\in\mathcal{I},\label{Con1}\\
&~~~~~~~~~~~~~x_{i\leftarrow j}\geq 0, ~~~~~~~~~~~~~~~~\forall i,j\in\mathcal{I}.\label{Con2}
\end{align}
\end{subequations}
Constraint \eqref{Con1} indicates that the sum of user $j$'s direct rate and the rate tethered to other users cannot exceed the capacity of her downlink $C_j$. Let $\bs{x}^*(\bs{h})$ denote the solution to the (UPM) Problem in \eqref{UPM}, given hybrid price matrix $\bs{h}$.

%





Here we user  $\bs{\lambda}\triangleq\{\lambda_i\}_{i\in\mathcal{I}}$ and  $\bs{\mu}\triangleq\{\mu_{i\leftarrow j}\}_{i,j\in\mathcal{I}}$ to denote the dual variables corresponding to constraints \eqref{Con1} and \eqref{Con2}, respectively. The dual variable $\lambda_i$ is also known as the \textit{shadow price} for downlink $i$.  We will explain the properties of these dual variables next.

\subsubsection{Analysis under Equilibrium Hybrid Prices}
 In the following, we assume the MNOs' set the equilibrium price matrix $\bs{h}^*$ (of the entire two-stage game)  and derive structural properties concerning both users' decision $\bs{x}^*(\bs{h}^*)$ and the MNOs' equilibrium price matrix $\bs{h}^*$. These significantly simplify the solution in both Stage I and Stage II. The proofs of all theorems, corollaries, propositions, and lemmata can be found in the Appendix.

\begin{theorem}\label{T1}
	Under MNOs' equilibrium hybrid price matrix $\bs{h}^*$, the optimal shadow price satisfies $\lambda_j^*=0,~j\in\mathcal{I}$.
\end{theorem}

The intuition behind Theorem \ref{T1} is that if $\lambda_j^*>0$ for some downlink $j$, then the MNOs can increase the hybrid price $h_{i\leftarrow j}$ on downlink $j$ for all $i$ to increase their profit.

To uniquely specify the MNOs' equilibrium hybrid pricing scheme, we make an additional assumption  in Assumption \ref{Assump1}. The reason is that if there exist downlinks $j\neq k$ such that $h_{i\leftarrow j}+c_{i\leftarrow j}=h_{i \leftarrow k}+c_{i\leftarrow k}$, (UPM) is not strictly convex  and hence there may be more than one globally optimal solution. This implies that the MNOs' total profit may not be unique for a given $\bs{h}$.
Hence, we adopt the following assumption in the rest of the paper:
\begin{assumption}\label{Assump1}
The users select their traffic decision $\bs{x}^*$ such that\footnote{Note that we will show that through solving their optimal cooperative pricing, the MNOs can compute and send the users' optimal traffic decision that satisfies \eqref{A1} to the users, without the need for the users to know $\{e_j\}_{j\in\mathcal{I}}$.}
\vspace{-0.2cm}
\begin{equation}
\bs{x}^*=\arg\min_{\bs{x}\in\mathcal{X}^o} \sum_{i\in\mathcal{I}}\sum_{j\in\mathcal{I}}(e_j+c_{i\leftarrow j})x_{i\leftarrow j}, \label{A1}
\end{equation}
where $\mathcal{X}^o$ denotes the set of all $\bs{x}$ that solves (UPM) in \eqref{UPM}.
\end{assumption}
We will later show in Section \ref{P-MNO} that Assumption \ref{Assump1} ensures that the $\bs{x}^*$ satisfying \eqref{A1} corresponds to a solution that MNOs want users select among $\mathcal{X}^o$, hence induces a unique MNO profit. Moreover,
Assumption \ref{Assump1} incurs no loss of generality as MNOs can always slightly adjust $\bs{h}$ to ensure that users select the $\bs{x}^*$ satisfying \eqref{A1}.




From Theorem \ref{T1} and Assumption \ref{Assump1}, we can derive the equilibrium price matrix $\bs{h}^*$.
\begin{corollary}\label{P11}
 There exists an equilibrium  price matrix $\bs{h}^*$ such that for each user $i$, we have
\begin{align}
h_{i\leftarrow j}^*+c_{i\leftarrow j}=h_{i \leftarrow k}^*+c_{i\leftarrow k}, ~~~\forall j,k\in\mathcal{I}.
\end{align}
\end{corollary}

Corollary \ref{P11} suggests that one of the equilibrium price solutions for the MNOs is \emph{gateway-independent}, i.e., each user faces the same sum cost across different links (the traffic cost $h_{i\leftarrow j}$ plus the energy cost $c_{i\leftarrow j}$).\footnote{Different users may still face different sum costs.}
This motivates us to combine the hybrid price $h_{i\leftarrow j}$ and the energy cost $c_{i\leftarrow j}$ into a single \textit{delivered price} $p_i$.
\begin{definition}
The delivered price $p_i$ for each user $i$ is
\begin{align}
p_i\triangleq h_{i\leftarrow j}+c_{i\leftarrow j},~~\forall j\in\mathcal{I}.\label{deliver}
\end{align}
\end{definition}
Therefore, to optimize over the price matrix $\bs{h}$,
it is enough to optimize the delivered price vector ${\bs{p}}\triangleq\{{p}_i\}_{i\in\mathcal{I}}$ and then recover the hybrid price matrix by $h_{i\leftarrow j}={p}_i-c_{i\leftarrow j}$  for all $i,j\in\mathcal{I}$.

%

\begin{figure}
\begin{centering}
  \includegraphics[scale=.33]{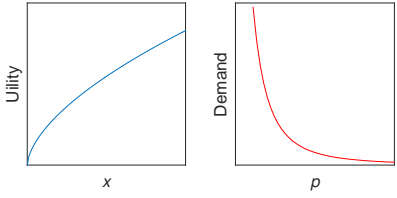}
  \vspace{-0.2cm}
  \caption{An $\alpha$-fair utility function $U(x)=x^{1-\alpha}/(1-\alpha)$ and the corresponding demand function $d(p)=(1/p)^{1/\alpha}$.}
  \label{demand}
  \end{centering}
  	\vspace{-0.5cm}
\end{figure}

To summarize, the equilibrium shadow price is zero (Theorem \ref{T1}) and the equilibrium delivered price ${p}_i$ is gateway-independent (Corollary \ref{P11}). Therefore, we can define a demand function for the user traffic.
\begin{definition}
Given the delivered price ${p}_i$, user $i$'s one-dimensional \textit{demand function} is
\begin{align}
 d_i({p}_i)&\triangleq\sum_{j\in\mathcal{I}}x_{i\leftarrow j}^*(\bs{h}^*)=\arg\max_{x\geq0}~[U_i(x)-x{p}_i].\label{d}
\end{align}
\end{definition}
Since $U_i (x)$ is strictly concave, $U_i'(x)$ is decreasing, and $d_i(\cdot)$ is unique-valued and non-increasing.  We present an illustrative example of a utility function and the corresponding demand function in Fig. \ref{demand}.
By the first-order condition in \eqref{d}, the demand function satisfies
\begin{equation}
U_i'(d_i({p}_i^*))\geq {p}_i^*~\text{ with equality if}~d_i({p}_i^*)>0. \label{decon}
\end{equation}
\begin{proposition}\label{Prop1}
	At an equilibrium, user $i$ has a positive demand if and only if $U_i'(0)>\min_{j}[c_{i\leftarrow j}+e_{j}]$.
\end{proposition}

The proof shows that if user $i$'s marginal utility is too small, then MNOs cannot profit from serving user $i$.

  By Proposition \ref{Prop1}, we can remove all users $i$ with $U_i'(0)\leq\min_{j}[c_{i\leftarrow j}+e_{j}]$ from the set $\mathcal{I}$. In other words, without loss of generality, we can assume that $U_i'(0)>\min_{j}[c_{i\leftarrow j}+e_{j}]$ for all users in the set $\mathcal{I}$. 
Therefore, equality in \eqref{decon} always holds. We thus have the following definition:
  \begin{definition}[Inverse Demand Function]
  User $i$'s inverse demand function is her marginal utility function $U_i'(\cdot)$.
  \end{definition}


%
%
%

We next show that it is enough for MNOs to optimize their total profit based on the delivered price and the users' inverse demand functions.

%

\subsection{Cooperative MNOs}\label{P-MNO}

In this subsection, we derive  the cooperative pricing scheme to maximize MNOs' total profit by  problem transformations.

\subsubsection{Operator Profit Maximization (OPM)} Given $\bs{x}^*(\bs{h})$  derived in Stage II, we can derive the optimal hybrid pricing matrix $\bs{h}^*$ by solving (OPM) in Stage I,
\begin{subequations}\label{OPM}
\begin{align}
{\rm (OPM)}~~&\max_{\bs{h}}~~ \sum_{i\in\mathcal{I}}\sum_{j\in\mathcal{I}}(h_{i\leftarrow j}-e_j) x^*_{i\leftarrow j}(\bs{h})\\
{\rm s.t.}~~&h_{i\leftarrow j}\geq 0,~\forall i,j\in\mathcal{I}
\end{align}
\end{subequations}

\subsubsection{Revised OPM (R-OPM)}
Define MNO $\sigma(j)$'s \textit{delivered (operational) cost} as $\tilde{e}_{i\leftarrow j}=e_j+c_{i\leftarrow j}$ for each GB user $j$ tethers to user $i$. We can transform (OPM) into a new problem where the traffic ${\bs{x}}$ are the optimization variables.
The key idea is to exploit the gateway-independent pricing suggested by Corollary \ref{P11}. It allows us to further adopt the inverse demand function $U_i'(\cdot)$ to determine the delivered price. We have
\vspace{-0.2cm}
\begin{subequations}\label{R-OPM}
	\begin{align}
	{\rm (R-OPM)}~
	&\max_{\bs{x}}~~\sum_{j\in\mathcal{I}}\sum_{i\in\mathcal{I}}\left(U_i' \left(\sum_{l\in\mathcal{I}}x_{i\leftarrow l}\right)-\tilde{e}_{i\leftarrow j}\right)x_{i\leftarrow j}\label{R-OPM1}\\
	&~{\rm s.t.}~~\sum_{l\in\mathcal{I}}x_{l\leftarrow j}\leq C_j,~x_{i\leftarrow j}\geq 0, \forall i,j.\label{R-OPM3}
	\end{align}
\end{subequations}

%


%


The objective in (R-OPM) in \eqref{R-OPM} may not be concave in general, which may make it difficult to find a globally optimal solution to (R-OPM).\footnote{Some of the results in this section only apply to the case where MNOs set optimal prices. Since (R-OPM) is not convex, we may obtain a locally optimal solution. In this case, those results do not hold.} We denote
\begin{align}
P_i(x)=-x\frac{U_i'''(x)}{U_i''(x)},~\forall i\in\mathcal{I}.
\end{align}
In the economics literature, $U_i'''(x)$ stands for user $i$'s \emph{prudence} and $P_i(x)$ is the \emph{coefficient of relative prudence} \cite{prudence}. We use it to measure the concavity of user $i$'s marginal utility (or the inverse demand function).
We introduce the following assumption to ensure the convexity of (R-OPM):
\begin{assumption}\label{A2}
Each user $i$'s coefficient of relative prudence satisfies
$P_i(x)\leq 2,$ for any $x\geq0$  and any $i\in\mathcal{I}.$
\end{assumption}
Assumption \ref{A2} is satisfied by  a wide range of utility functions, some of which are listed in Table \ref{utility}.
\begin{table}[t]
	\centering
	\caption{Examples of Utility Functions Satisfying Assumption \ref{A2}}
	\vspace{-0.15cm}
    \label{utility}
	\begin{tabular}{|c|c|c|}
		\hline
		Name                  & Utility Function                                                  & Ref.                \\ \hline\hline
		$\alpha$-fair  & $x^{1-\alpha}/(1-\alpha),~\alpha\in[0,1)$                         & \cite{mo2000fair,Fair} \\ \hline
		logarithmic    & $\theta\log(a+x),~\theta>0, a\geq 0$                              & \cite{Open-2}          \\ \hline
		exponential   & $1-\exp(-\theta x),~0\leq \theta\leq 2/\sum_{i\in\mathcal{I}}C_i$ & \cite{Jianwei}         \\ \hline
		quadractic    & $ax^2+bx$, $0\leq -2a\sum_{i\in\mathcal{I}}C_i\leq b$       & \cite{Jianwei}         \\ \hline
	\end{tabular}
	\vspace{-0.15cm}
\end{table}
\setlength{\textfloatsep}{5pt}
\begin{algorithm}[tb]
	\caption{Cooperative Pricing Scheme}\label{Algorithm1}
		  \SetKwInOut{Input}{Input}\SetKwInOut{Output}{Output}
		  \Input{$\{U_i(\cdot)\}_{i\in\mathcal{I}}$, $\{\tilde{e}_{i\leftarrow j}\}_{i,j\in\mathcal{I}}$, $\{C_i\}_{i\in\mathcal{I}}$}
		  \Output{~Equilibrium hybrid price matrix $\bs{h}^*$ and equilibrium traffic matrix $\bs{x}^*(\bs{h}(\bs{p}^*))$}
	The MNOs obtain $\bs{x}^R$ via solving (R-OPM)\;
	The MNOs obtain the optimal delivered price vector ${\bs{p}}^*$ from \eqref{T2E} and then the optimal hybrid price matrix $\bs{h}^*$ from \eqref{deliver}\;
	The MNOs send users the optimal traffic decision $\bs{x}^*(\bs{h}(\bs{p}^*))=\bs{x}^R$\;
\end{algorithm}
We can readily verify that $U_i'(x)x$ is concave if and only if Assumption \ref{A2} holds.
Therefore, (R-OPM) is convex, which can be easily solved by convex optimization solvers such as CVX \cite{CVX}. 
\begin{theorem}[Equivalence of (R-OPM) and (OPM)]\label{T2}
For any optimal solution $\bs{x}^{R}$ to (R-OPM),
\begin{enumerate}
  \item the optimal delivered price vector $\bs{p}^*=\{p_i^*\}_{i\in\mathcal{I}}$ in \eqref{deliver} is, for each user $i$,
\begin{equation}
{p}_i^*=U'_i \left(\sum_{j\in\mathcal{I}}x_{i\leftarrow j}^{R}\right); \label{T2E}
\end{equation}
  \item given $\bs{p}^*$ in \eqref{T2E}, $\bs{x}^{R}$ is the optimized user traffic, i.e, $\bs{x}^*(\bs{h}(\bs{p}^*))=\bs{x}^{R}$.
\end{enumerate}

\end{theorem}
The significance of Theorem \ref{T2} is two-fold. First, 
we can solve (R-OPM) efficiently and then use \eqref{T2E} to obtain the optimal delivered price vector $\bs{p}^*$ and then the
optimal hybrid price matrix $\bs{h}^*$ by \eqref{deliver}.
Second, it implies that MNOs can directly send the solution to (R-OPM) $\bs{x}^R$ to users, without the need to let users independently solve (UPM).\footnote{This is possible as we have considered the complete information setting in this paper, i.e., MNOs know the users' utility functions. This is reasonable as the users' mobile application information (i.e., whether the user is watching YouTube or downloading emails) is often communicated to the MNOs by the devices.}   We summarize this procedure in Algorithm \ref{Algorithm1}.


%

\subsection{Cooperative MNOs with Free Tethering}\label{FTPCoop}

  In this subsection, we further consider a free-tethering  (FT) cooperative pricing  scheme when there is a strict regulation on the tethering pricing \cite{fcc}. Specifically,  the tethering price satisfies $t_{i\leftarrow j}=0$ for all users $i$ and $j$, and the MNOs optimize the access prices. Due to the space limit, we present the detailed descriptions in Appendix \ref{FTPscheme}.

We show that such an FT scheme can lead to the optimal prices under some conditions on users' utility and energy cost.

\begin{corollary}\label{coro1}
The optimized FT scheme corresponds to the optimal hybrid pricing to the OPM Problem in \eqref{OPM} when both of the following conditions are satisfied:
\begin{enumerate}
  \item Users' energy costs for Wi-Fi links are zero and those for cellular downlinks are the same, i.e., $c_{i\leftarrow j}^\text{Wi-Fi}=0$ and $c_{i}^\text{Down}=c_{j}^\text{Down}$ for all $i,j\in\mathcal{I}$.
  \item Each user $i$ has an isoelastic utility, i.e., $U_i(x)=\frac{\theta_ix^{1-\alpha}+\Xi_i}{1-\alpha}$ (with $\theta_i$ and $\Xi_i$ being non-negative and $\alpha\in[0,1)$).\footnote{The widely used $\alpha$-fair utility function in Table \ref{utility} is a special subclass of this class of functions.}
\end{enumerate}
\end{corollary}
Condition 1) is justified by measurements showing that the average energy cost on Wi-Fi Direct $c_{i\leftarrow j}^\text{Wi-Fi}$ is much less than that for cellular connections $c_j^\text{Down}$ \cite{energy1}. Moreover, the traffic cost in general is much larger than energy cost for users.

We explain Corollary \ref{coro1} as follows.
When condition 1) is satisfied, Proposition 1 implies that  gateway-independent pricing requires $a_i^*=a_j^*+t_{i\leftarrow j}^*$. In addition, the weighted $\alpha$-fair utility function exhibits a constant price elasticity of demand, i.e., heterogeneous users' (in terms of $\theta_i$ and $b_i$) responses to the delivered price are similar. This equalizes the optimal delivered prices across heterogeneous users. Hence, we have $p_i^*=p_j^*$ and thus $t_{i\leftarrow j}^*=0$.

\section{Competitive MNOs: Price Competition}\label{PC}

\begin{figure}[t]
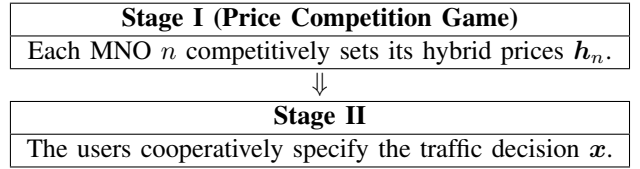

\centering
\begin{tabular}{c}
\hline
\multicolumn{1}{|c|}{ \textbf{Stage I (Price Competition Game)}}        \\ \hline
\multicolumn{1}{|c|}{Each MNO $n$ competitively sets its hybrid prices $\bs{h}_n$.} \\ \hline
$\Downarrow   $                                  \\ \hline
\multicolumn{1}{|c|}{\textbf{Stage II}}        \\ \hline
\multicolumn{1}{|c|}{The users cooperatively specify the traffic decision $\bs{x}$.}         \\ \hline
\end{tabular}
\caption{Two-Stage Stackelberg Game for the PCG.}\label{Stackelberg-2}
	\vspace{-0.1cm}
\end{figure}

We now analyze the MNOs and users' decisions in the competitive MNOs models. In Stage I, each MNO sets hybrid prices independently to maximize its own profit. Hence,
 MNOs participate in a \textit{price competition game} (PCG). In Stage II, the  users \textit{cooperatively} determine traffic. Fig. \ref{Stackelberg-2} summarizes the  interactions among competitive MNOs and users.

Introducing competition among MNOs increases the difficulty of analysis. This is because, each MNO needs to consider users' multi-dimensional decisions and also the hybrid pricing strategies of other MNOs. For analytical tractability, we adopt the following assumption in the rest of this paper:
\begin{assumption}\label{Asump3}
  User energy costs for the Wi-Fi links are zero, i.e., $c_{i\leftarrow j}^\text{Wi-Fi}=0$ for all $i,j\in\mathcal{I}$.
\end{assumption}
As previously discussed, the energy cost on a Wi-Fi link $c_{i\leftarrow j}^\text{Wi-Fi}$ is typically much less than that for cellular connections $c_j^\text{Down}$ and the traffic cost.
By Assumption \ref{Asump3}, we have $c_{i\leftarrow j}=c_{k\leftarrow j}$ and $\tilde{e}_{i\leftarrow j}=\tilde{e}_{k\leftarrow j}$ for all $i,k\in\mathcal{I}$. Thus, we drop the index $i$ in $c_{i\leftarrow j}$ and $\tilde{e}_{i\leftarrow j}$ and have $c_{j}$ and $\tilde{e}_j$.

%



Similar to the cooperative-MNO scenario, we will derive the equilibrium by backward induction, i.e., given the hybrid pricing matrix $\bs{h}$, we characterize the users' traffic decision $\bs{x}^{\star}(\bs{h})$ in Stage II. We then characterize each MNO's equilibrium prices. We next formally define the PCG and its corresponding equilibrium:
\begin{game}[PCG] Price Competition Game consists of
\begin{itemize}
  \item \textit{Players}:\textit{ the set $\mathcal{N}$ of MNOs.}
  \item \textit{Strategies}: $\bs{h}_{n}\triangleq\{h_{i\leftarrow j}\}_{i\in\mathcal{I}, j\in\mathcal{I}_n}$.
  \item \textit{Payoffs}: for each MNO $n$:
  \begin{align}
  V_n(\bs{h})=\sum_{j\in\mathcal{I}_n}\sum_{i\in\mathcal{I}}(h_{i\leftarrow j}-\tilde{e}_j)x_{i\leftarrow j}^\star(\bs{h}),
  \end{align}
\end{itemize}
where $x_{i\leftarrow j}^\star(\bs{h})$ is the traffic optimized in Stage II as derived in Section \ref{cooperative}.
\end{game}

Let $\star$ denote the competitive equilibrium values, to differentiate from 
 the cooperative equilibrium values denoted by $*$.

\begin{definition}\label{D-PCE}
A price competitive equilibrium (PCE) for the PCG is a hybrid price matrix $\bs{h}^\star$ such that, for every MNO $n$,
\begin{equation}
V_n(\bs{h}^\star)\geq V_n(\bs{h}_n,\bs{h}_{-n}^\star),~\forall \bs{h}_n.
\end{equation}
\end{definition}

\subsection{Users' Consumption Decisions in Stage II}

Similar to the interaction in Stage II for the cooperative-MNO model, the users cooperatively optimize $\bs{x}$ by solving (UPM) in \eqref{UPM} for a given $\bs{h}$. 

The PCG shares part of the equilibrium results with the cooperative hybrid pricing scheme. Specifically,  Theorem \ref{T1} also holds in the PCG. That is, for each downlink $j$, the equilibrium shadow price is zero, i.e., $\lambda_j^\star=0$ for all $j\in\mathcal{I}$. Therefore,
  similar to Corollary \ref{P11}, we can prove the following  gateway independence result in a similar way:
\begin{proposition}[Gateway Independence]\label{Prop3}
If a PCE $\tilde{\bs{h}}^\star$ exists, there exists a PCE $\bs{h}^\star$ such that 
\begin{itemize}
  \item  $\bs{h}^\star$ is \textit{gateway-independent}, i.e., for each user $i$,
\begin{align}
h_{i\leftarrow j}^\star+c_{j}=h_{i\leftarrow k}^\star+c_{k},~\forall j,k\in\mathcal{I}; \label{GWI}
\end{align}
  \item  every MNO $n$ receives the same profit with $\bs{h}^\star$ and $\tilde{\bs{h}}^\star$, i.e., $V_n(\tilde{\bs{h}}^\star)=V_n(\bs{h}^\star)$.
\end{itemize}
\end{proposition}



Proposition \ref{Prop3} implies that among all possible PCEs, it is enough to focus on the \emph{gateway-independent} ones. Let $\bs{p}^\star=\{p_{i}^\star\}_{i\in\mathcal{I}}$ denote the equilibrium delivered price vector, i.e.,
 \begin{equation}
 p_{i}^\star\triangleq h_{i\leftarrow j}^\star+c_j,~\forall i\in\mathcal{I}, ~\forall j\in\mathcal{I}.\label{PricingINDE}
 \end{equation}
The gateway-independent PCE indicates that we can adopt the one-dimensional demand function $d_i(p_i^\star)$ defined in \eqref{d} to characterize each user's traffic decision.




\subsection{MNOs' Price Competition in Stage I}

In this subsection, we study the interactions among MNOs through price competition. 
We sort the delivered costs $\{\tilde{e}_j\}$ in ascending order without loss of generality, i.e., $\tilde{e}_1<\tilde{e}_2<...<\tilde{e}_j$.\footnote{Here, we assume that all downlinks have different delivered costs, since the probability of two parameters being exactly the same is zero.}
The indices will denote the corresponding channels.
\begin{definition}[Market Clearing Price]\label{D5}
We define $\zeta_s$ as the \textbf{market clearing (delivered) price} for the first $s$ channels, which satisfies
$\sum_{i\in\mathcal{I}}d_i(\zeta_s)=\sum_{\ell=1}^sC_\ell$ for all $s\in\mathcal{I}.$
\end{definition}

Given a fixed $s$, the market clearing price $\zeta_s$ denotes the delivered price under which the users' aggregate demand of the first $s$ downlinks is equal to the total capacity of these downlinks.  Since $d_i(\cdot)$ is decreasing, $\zeta_s$ also decreases in $s$  for a proper value of $s$. We will show that the delivered prices at a PCE are given by a market clearing price.
By exploiting the continuity and strict concavity of $U_i(\cdot)$, we can prove:
\begin{lemma}\label{L11}
There exists a unique $\zeta_s$ for every $s\in\mathcal{I}$.
\end{lemma}

To characterize PCEs, we further define the following
\begin{definition}[Threshold Downlink]
The \textbf{threshold downlink} ${g^{\rm thr}}$ is the  downlink with the smallest delivered cost among all MNOs other than MNO $\sigma(1)$, i.e., ${g^{\rm thr}}\triangleq \arg\min_{i\in\mathcal{I}/\mathcal{I}_{\sigma(1)}}\tilde{e}_i$.
\end{definition}
Note that MNO $\sigma(1)$ is the most competitive MNO, in the sense that it possesses the downlink with the least delivered cost. In this sense, $\sigma(g^{\rm thr})$ can be considered the second most competitive  MNO.\footnote{MNO $\sigma(g^{\rm thr})$ may not be MNO $\sigma(2)$, since the user $2$ may be MNO $\sigma(1)$'s subscriber, i.e., $\sigma(1)=\sigma(2)\neq \sigma(g^{\rm thr})$.} Thus,
the delivered cost of the threshold downlink $\tilde{e}_{{g^{\rm thr}}}$ measures how competitive MNO $\sigma({g^{\rm thr}})$ is. As we will show, a large $\tilde{e}_{{g^{\rm thr}}}$ implies MNO $\sigma({g^{\rm thr}})$ offers little competition, in which case MNO $\sigma(1)$ effectively monopolizes the market. Otherwise, MNO $\sigma(1)$ shares the market with at least  MNO $\sigma({g^{\rm thr}})$.


The following definition characterizes an MNO which captures a positive market share at a PCE.

\begin{definition}[Traffic-Supporting MNO]
An MNO $n$ is \textbf{traffic-supporting} at a PCE if some downlinks of users subscribing to MNO $n$ are active, i.e., $\sum_{j\in\mathcal{I}_n}\sum_{i\in\mathcal{I}}x_{i\leftarrow j}(\bs{h}^\star)>0$.
\end{definition}

We  characterize the PCEs as follows:



\begin{proposition}\label{LVI-1}
If a PCE exists in a PCG, then it belongs to one of the following two types:
\begin{itemize} 
  \item \textit{Single-Operator PCE}: when the threshold downlink's delivered cost is high, i.e., $ \zeta_{{g^{\rm thr}}-1}\leq \tilde{e}_{g^{\rm thr}}$, there exists a unique gateway-independent PCE. Such a PCE admits only one traffic-supporting MNO  and satisfies $p_{i}^\star\leq \tilde{e}_{g^{\rm thr}}$ for all $i\in\mathcal{I}$;
  \item \textit{Multi-Operator PCE}: when the threshold downlink's delivered cost is low, i.e., $ \zeta_{{g^{\rm thr}}-1}> \tilde{e}_{g^{\rm thr}}$, a PCE (if any) admits more than one traffic-supporting MNOs and satisfies $p_{i}^\star \geq \tilde{e}_{g^{\rm thr}}$ for all $i\in\mathcal{I}$.
\end{itemize}

\end{proposition}

The intuition is that when the delivered cost of the threshold downlink is so high that no other MNO than MNO $\sigma(1)$ is competitive, then MNO $\sigma(1)$ monopolizes the market.
On the other hand, if the delivered cost of the threshold downlink is low enough, then at least MNO $\sigma({g^{\rm thr}})$ competes with MNO $\sigma(1)$.
  Through this observation, we further characterize the two types of PCEs according to the relation between $\zeta_{{g^{\rm thr}}-1}$ and $\tilde{e}_{g^{\rm thr}}$.


\subsubsection{Single-Operator Equilibrium ($\zeta_{g^{\rm thr}-1}\leq \tilde{e}_{g^{\rm thr}}$)}

In this case, MNO $\sigma(1)$ acts as a monopolist and is the only traffic-supporting MNO. To characterize the equilibrium price, let us first consider a monopolist delivered price vector $\bs{\tilde{p}}^\star=\{\tilde{p}_i^\star\}_{i\in\mathcal{I}}$ which is equivalent to the cooperative pricing scheme and thus can be obtained by
 solving (OPM) via Algorithm \ref{Algorithm1}.
The relation between the monopolist delivered price $\tilde{p}_i^\star$ and the delivered costs are given in the following proposition.



\begin{proposition}\label{T3}
When $\zeta_{{g^{\rm thr}}-1}\leq \tilde{e}_{g^{\rm thr}}$, there exists a unique gateway-independent equilibrium delivered price $p_i^\star$ such that, for each user $i$,
\begin{align}
 \hspace{-0.2cm}{p}_{i}^\star=\begin{cases}\tilde{p}_i^{\star},~~&{\rm if}~~\tilde{p}_{i}^\star<\tilde{e}_{g^{\rm thr}}~~\text{(perfect monopoly)},\\
 \tilde{e}_{g^{\rm thr}},~~&{\rm if}~~\tilde{p}_{i}^\star\geq\tilde{e}_{g^{\rm thr}}~~\text{(depressed monopoly)}.\end{cases}\label{T3-Eqqq}
  \end{align}
\end{proposition}

 That is, MNO $\sigma(1)$ adopts the monopoly delivered price $\tilde{p}_i^\star$ for user $i$ when $\tilde{p}_i^\star$ is sufficiently low, since no other MNOs has a downlink cost low enough to compete with MNO $\sigma(1)$  for this user. However, MNO $\sigma({g^{\rm thr}})$  competes with  MNO $1$ when the monopoly delivered price $\tilde{p}_i^\star$ exceeds $\tilde{e}_{g^{\rm thr}}$. Therefore, although MNO $\sigma(1)$ is still the only traffic-supporting MNO, the equilibrium delivered price is $\tilde{e}_{g^{\rm thr}}$  for user $i$ due to competition from MNO $\sigma({g^{\rm thr}})$.

\subsubsection{Multi-Operator Equilibrium ($\zeta_{{g^{\rm thr}}-1}> \tilde{e}_{g^{\rm thr}}$)}
In contrast with the single-operator case, a PCE may not exist when $\zeta_{{g^{\rm thr}}-1}> \tilde{e}_{g^{\rm thr}}$.  We will characterize the necessary condition for the multi-operator PCE, based on which we discuss the reason why a  multi-operator PCE may not exist.
 \begin{definition}[Critical Downlink] The critical downlink $\hat{s}$ is the downlink such that
  $\zeta_{\hat{s}}$ satisfies:   
\begin{align}
\begin{cases}
\tilde{e}_{\hat{s}}\leq \zeta_{\hat{s}}\leq \tilde{e}_{{\hat{s}}+1},~ &{\rm if} ~  \hat{s}<|\mathcal{I}|\\
\tilde{e}_{\hat{s}}\leq \zeta_{\hat{s}}, ~ &{\rm if} ~ \hat{s}=|\mathcal{I}|\end{cases}, \label{s}
\end{align}
where $\zeta_{s}$ is the market clearing price in Definition \ref{D5}.
 \end{definition}
Since $\zeta_{s}$ decreases in $s$, and $\tilde{e}_s$ increases in $s$, there is at most one $\zeta_{\hat{s}}$ satisfying \eqref{s}.\footnote{
	For example, when $\tilde{e}_1=0.1$, $\tilde{e}_2=0.6$, and $\zeta_1=0.5$, we have $\hat{s}=1$.
	Note that it is possible that $\hat{s}$ does not exist. An example is an MCA with two users with $\tilde{e}_1=0.1$, $\tilde{e}_2=0.6$, $\zeta_1=1$, and $\zeta_2=0.5$. We see that $\hat{s}$ cannot be $1$ or $2$. In such a case, no pricing profile can satisfy the necessary condition in Proposition \ref{T5}, implying that there is no PCE.} The following theorem states that the equilibrium delivered prices clear the first $\hat{s}$ downlinks.

\begin{theorem} [User Independence and Uniqueness]\label{T5}  When $\zeta_{{g^{\rm thr}}-1}> \tilde{e}_{g^{\rm thr}}$, if a PCE exists, there exists a unique equilibrium delivered price vector $\bs{p}^\star$ such that 
	$p_{i}^{\star}= \zeta_{\hat{s}}$  for all $i\in\mathcal{I}.$

\end{theorem}

%

By Theorem \ref{T5}, the gateway-independent delivered prices are also \textit{user-independent}, i.e., the delivered prices are equal across different users. Specifically, suppose the delivered price for one user $i$ is higher than that of another user $k$. Some traffic-supporting MNO $n$ can always slightly decrease its hybrid price for user $i$ but slightly increase  the price for user $k$. This can attract traffic demanded by  users who will pay a higher price, and hence increases its profit.
Uniqueness is mainly because there is at most one $\hat{s}$ as we have mentioned.

Theorem \ref{T5} only states the necessary condition of a multi-operator PCE. That is, a PCE may not exist (with an example presented in Appendix \ref{AL}). 
As an illustration, next we characterize the PCE existence conditions for a simple example.

\begin{figure}
	\begin{centering}
		\includegraphics[scale=.38]{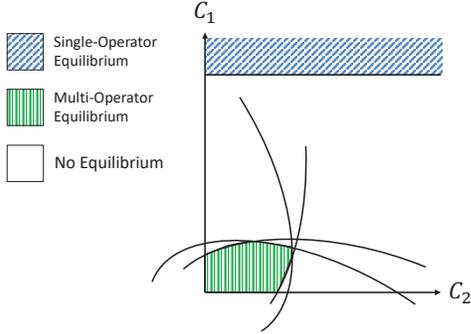}
		\vspace{-0.3cm}
		\caption{Characterizations (in terms of different capacity combinations) of the PCE existence conditions for an example with two MNOs and two users. }
		\label{Chara11}
	\end{centering}
\end{figure}


\subsubsection{Simple Example}
It is difficult to completely characterize a multi-operator PCE, due to the complexity of the multi-dimensional strategy that each MNO has to determine. We hence characterize the existence conditions (in terms of $C_1$ and $C_2$) for a system with two MNOs and two users, as shown in Fig. \ref{Chara11}. We set $\tilde{e}_1<\tilde{e}_2$ (so that $\sigma(1)=1$ and $\sigma(g^{\rm thr})=2$). It follows
\begin{itemize}
	\item One gateway-independent single-operator PCE lies in the region with a large enough $C_1$. This corresponds to a sufficient small $\zeta_1$ such that $\zeta_1\leq \tilde{e}_2$.
	\item One gateway-independent multi-operator PCE lies in the region with small enough $C_1$ and $C_2$. This leads to a sufficiently large $\zeta_2$ such that $\zeta_2\geq \tilde{e}_2$.
	\item There is a large region for no PCE. This is consistent with the well-known Edgeworth paradox in economics \cite{edgeworth}. That is, the situation where MNO $1$ monopolizes the market is not an equilibrium as $\zeta_1$  is too high. Nor is
	the situation where both MNOs charge the delivered prices that fully utilize their capacities (satisfying Theorem \ref{T5}). This is because $\zeta_2$ is so low that at least one MNO is willing to increase its prices to generate more profits.
\end{itemize}
Due to space limitations, we present the detailed analysis 
 in Appendix \ref{DANE}.

Similar to Section \ref{FTPCoop}, we also derive conditions under which the gateway-independent PCE $\bs{h}^\star$ leads to zero tethering prices, as shown in Appendix \ref{FTCOMP}.


%

Regarding the multi-operator PCE, the induced user-independent delivered prices and the possible non-existence of PCE motivate us to reformulate the MNO competition as a quantity game.
\section{Competitive MNOs: Quantity Competition}\label{QC}

In this section, we consider an alternative way for modeling a multi-operator PCE. Specifically, we introduce the Quantity Competition Game (QCG). 
We show that there exists at least an equilibrium for the QCG. Moreover, a multi-operator PCE, if exists, is exactly an equilibrium outcome of the QCG. We then propose an algorithm to compute the equilibrium. 

\subsection{Quantity Competition Game}\label{QC-A}

\textit{Uniform Delivered Price:} A multi-operator PCE equalizes the delivered prices across users and downlinks but may not exist. This fact motivates us to reformulate MNOs' interactions into quantity competition based on the uniform delivered pricing. Specifically, each MNO $n$ decides its quantity (total output traffic) $q_n$ of all its downlinks and compete with each other.\footnote{To realize the output control in practice, each MNO can manipulate the maximum data access speed of each downlink by setting a data speed cap.} Given these quantity choices, they follow a uniform delivered price $\pi(\cdot)$ which adjusts to the level that users' aggregate demands $D(p)=\sum_{i\in\mathcal{I}}d_i(p)$ is equal to the MNOs' aggregate output, i.e., 
$\pi(\cdot)\triangleq D^{-1}(\cdot)$,
where $f^{-1}(\cdot)$ denotes the inverse function of  $f(\cdot)$.
%

\textit{Aggregate Delivered Cost:} MNO $n$ can choose the output $z_{i}$ for each downlink $i\in\mathcal{I}_n$. However, due to the uniform delivered price, we can see that MNO $n$ should always fully utilizes its low delivered cost downlinks first before providing data at high delivered cost downlinks.
Hence, we can derive the aggregate delivered (operational) cost as
\begin{align}
E_n(q_n)\triangleq&\min_{\{z_{i}\}_{i\in\mathcal{I}_n}}\sum_{i\in\mathcal{I}_n}z_{i}\tilde{e}_{i},~~{\rm s.t.}~~\sum_{i\in\mathcal{I}_n}z_{i}=q_n\label{EEE}.
\end{align}


\textit{Smoothed Approximation:} The above $E_n(q_n)$ contains several non-differentiable points. For analytical tractability, we replace it with an approximation $\tilde{E}_n(q_n)$, which smooths $E_n(q_n)$ in the neighborhood of every non-differentiable point $\bar{q}\in\mathcal{D}_n$, where $\mathcal{D}_n$ denotes the set of non-differentiable points. An example of such a  smoothed function $\tilde{E}_n(q_n)$ is
\begin{align}
\hspace{-0.2cm}\tilde{E}_n(q_n)=\begin{cases} E_n(q_n),~~~{\rm if}~\forall\bar{q}\in \mathcal{D}_n,~ q_n\notin(\bar{q}-\varepsilon,\bar{q}+\varepsilon),\\
E_n(\bar{q}-\varepsilon)+\frac{E_n(\bar{q}+\varepsilon)-E_n(\bar{q}-\varepsilon)}{8\varepsilon^3}(q_n-\bar{q}-\varepsilon)^3,\\~~~~~~~~~~~~~{\rm if}~\exists\bar{q}\in \mathcal{D}_n,~ q_n\in(\bar{q}-\varepsilon,\bar{q}+\varepsilon).~\end{cases}\label{smooth}
\end{align}
We can see $\lim_{\epsilon \rightarrow 0}\tilde{E}_n(q_n)\rightarrow E_n(q_n)$. Different from $E_n(q_n)$, the new function $\tilde{E}_n(q_n)$ admits a continuous second order derivative.  We formally define
\begin{game}[QCG] The Quantity Competition Game consists of 
\begin{itemize}
  \item Players. MNO $n\in\mathcal{N}$.
  \item Strategy space. MNO $n$ chooses its quantity $q_n$ (its total traffic to download) from  $\mathcal{Q}_n\triangleq\left[0,\sum_{i\in\mathcal{I}_n}C_i\right]$.
  \item Payoff function. Each MNO $n$ has 
\begin{equation}
V_n\left(q_n,\sum_{\ell \neq n} q_\ell\right)=q_n \pi\left(\sum_{\ell\in\mathcal{N}}q_\ell\right)-\tilde{E}_n(q_n).\label{profit4}
\end{equation}
\end{itemize}
\end{game}

Accordingly, we introduce the definition of the Quantity Competition Equilibrium (QCE) for the QCG in the following.
\begin{definition}[QCE]
 A QCE for the QCG is a quantity profile $\bs{q}^\star\triangleq\{q_n^\star\}_{n\in\mathcal{N}}$ such that, for each MNO $n$,
\begin{equation}
V_n\left(q_n^\star,\sum_{\ell \neq n} q_\ell^\star\right)\geq V_n\left(q_n,\sum_{\ell \neq n} q_\ell^\star\right), \forall q_n\in\mathcal{Q}_n.
\end{equation}
\end{definition}

We show that
\begin{proposition}\label{PP5}
	The limit of a QCE (as $\epsilon$ approaches $0$) with a smoothed approximation (in \eqref{smooth}) is a QCE without one (i.e., the QCE with $E_n(q_n)$ in \eqref{profit4}).
\end{proposition}

We focus on the smoothed approximation (with $\epsilon$ approaching $0$) of the aggregate cost functions in the following.

\subsection{Equilibrium Properties}\label{QC-B}

In this subsection, we introduce a simple one-dimensional mapping to facilitate  characterizing and computing the QCE. We then analyze
the properties of the QCE, including existence, a sufficient condition for uniqueness, and equivalence to a multi-operator PCE. 

Notice that each MNO $n$'s profit is simply a function of two valuables, i.e., its output $q_n$ and the sum of the total output $\sum_{n\in\mathcal{N}}q_n$. Since all MNOs' profit functions share the same latter valuable, we adopt the following functions as a simpler equivalent characterization of the QCE:\footnote{The mapping $\varphi_n(b)$ is a (one-to-one) function because the objective in \eqref{varphi} is strictly quasi-concave for any given $b$.}
\begin{align}
\varphi_n(b)&\triangleq\arg\max_{q_n\in\mathcal{Q}_n}\left[q_n\pi(b)+\frac{q_n^2}{2}\pi'(b)-\tilde{E}_n(q_n)\right],\label{varphi}\\
\text{and}~
\Phi(b)&\triangleq\sum_{n\in\mathcal{N}}\varphi_n(b).\label{Varphi}
\end{align}
Function  $\varphi_n(b)$ corresponds to MNO $n$'s optimal output given the total output $b$. Variable  $b$ can represent the QCE total output if it is a fixed point of $\Phi(b)$ as we will show.

We adopt the following  assumption:
\begin{assumption}\label{Assump4}
	MNOs' aggregate revenue $\pi(x)x$ is strictly quasi-concave in $x$.
\end{assumption}
Assumption \ref{Assump4} is widely adopted in the literature (see \cite{IterCour} and the reference  therein) and it
holds, for example, when all users have the same type of utility functions chosen from Table \ref{utility} (but can still have different parameters).
We are ready to introduce the following lemma: 
\begin{lemma}\label{L3}
	A strategy profile $\bs{q}^\star$ is a QCE if and only if  $\sum_{\ell \in\mathcal{N}}q_\ell^\star=\Phi\left(\sum_{\ell \in\mathcal{N}} q_\ell^\star\right)$.
\end{lemma}
It is straightforward to prove Lemma \ref{L3} by comparing the optimality conditions of \eqref{varphi} and that of each MNO's best strategy.
Lemma \ref{L3} suggests that we can simply characterize a QCE by a fixed point $b^\star$ of a one-dimensional mapping $\Phi(b)$.\footnote{This  differs substantially from the general $I$-player game, in which case we need to characterize an equilibrium by a fixed point of a mapping with at least $I$ dimensions in general. } Meanwhile, MNO $n$'s QCE output is $q_n^\star=\varphi_n(b^\star)$. The continuity of $\Phi(b)$ further gives

\begin{proposition}[Existence]\label{P6}
There always exists a QCE $\bs{q}^\star$.
\end{proposition}


Such a property is not true for the PCG due to the discontinuity of  each MNO's profit in its strategy, i.e.,
a slight change in its pricing may dramatically change an MNO's profit.



%

The following proposition characterizes a sufficient condition where such a QCE is unique:
\begin{proposition}[Conditions for Uniqueness]\label{P7}
The QCG admits a unique QCE if there are only  two MNOs, i.e., $|\mathcal{N}|=2$.
\end{proposition}
We prove Proposition \ref{P7} by showing that a two-MNO system satisfies the diagonal-dominance condition in \cite{moulin1984dominance}, under which it guarantees the uniqueness.





The following theorem shows the equivalent relation between the multi-operator PCE for the PCG and the QCE.

\begin{theorem}[Equivalence]\label{T11}
A multi-operator PCE is a QCE (as $\epsilon$ approaches $0$). Mathematically, if there exists a
multi-operator PCE $\bs{h}^\star$ of the PCG, then there exists a QCE $\bs{q}^\star$ such that $q_n^\star=\sum_{i\in\mathcal{I}_n}\sum_{j\in\mathcal{I}}x_{i\leftarrow j}(\bs{h}^\star)$ for all $n\in\mathcal{N}$.
\end{theorem}
Theorem \ref{T11} indicates that the QCG is a good approximation to the PCG.\footnote{The  equivalence result in Theorem \ref{T11} is interesting since it differs from the existing result for the classical models and those for many variations. That is, the pricing competition (or the Bertrand competition) and quantity competition (or the Cournot competition) typically lead to different outcomes in general (e.g., \cite{comp1,comp2}).}
Intuitively, an MNO's strategy space  in a PCG possesses more dimensions than that in a QCG. Hence, if MNO $n$ cannot achieve a higher profit by changing its pricing strategy in a PCG, nor can it find a more profitable quantity strategy in a QCG. Note that A QCE may not be a multi-operator PCE, because a QCE always exists (Proposition \ref{P6}) but a multi-operator PCE may not.

\subsection{Algorithm}\label{QC-C}

\begin{algorithm}[tb]\label{Algo2}
\caption{Computation of the QCE}\label{Algorithm2}
 \SetAlgoLined
 \SetKwInOut{Input}{Input}\SetKwInOut{Output}{Output}
 \Input{$\Phi(\cdot)$ in \eqref{Varphi}} \Output{QCE quantity $\bs{q}^\star$} 
 Initialize the iteration index $t\leftarrow 0$, a stopping criteria $\epsilon>0$, and a convergence flag ${\rm Conv\_flag}\leftarrow 0$, and randomly picks a $b(0)\in[0,\sum_{\ell\in\mathcal{I}_n}C_\ell]$\;
 \While{${\rm Conv\_flag}=0$}{
 	$t\leftarrow t+1 $\;
 	MNO $k$ computes
 	\begin{align}
 	b(t+1)=(1/t)\Phi(b(t))+(1-1/t)b(t); \label{dynam}
 	\end{align}
 	\If{$|b(t)-\Phi(b(t))|\leq b(t)\epsilon$}{
 		${\rm Conv\_flag}\leftarrow 1$;
 	}
}
 Compute the QCE output level $q_n^\star=\varphi_n(b^\star)$ using \eqref{varphi}\;
%
\end{algorithm}

Due to the difficulty of expressing the QCE in a closed form, we further design a simple and convergent algorithm to compute the QCE.

We  compute the QCE based on the Mean-Value Iterative Dynamics \cite{meanvalue}, which essentially computes a fixed point of $\Phi(b)$, as summarized in Algorithm \ref{Algo2}. The key step of this algorithm is in \eqref{dynam}, which is to average $b(t)$ of all previous $t$ iterations and let mean value be the input at iteration $t+1$. The above algorithm leads to the following convergence property.
\begin{proposition}\label{P8}
Algorithm \ref{Algo2} converges to a QCE $\bs{q}^\star$.
\end{proposition}
The proof mainly relies on \cite{meanvalue}, which proved that the  Mean-Value Iterative Dynamics is convergent for computing a fixed point of a continuous and compact self-mapping. 
We note that the convergence result does not require
the uniqueness of the QCE, as shown in \cite{meanvalue}.\footnote{Such property
differs from the widely considered best response dynamics
analyzed in \cite{moulin1984dominance}, of which the convergence requires the
uniqueness of the equilibrium in general.}



\subsection{Summary of the Competitive Scheme}

We summarize the competitive pricing scheme in
Algorithm \ref{Algo3}. Specifically, we first check whether there exists
a single-operator PCE for the PCG by Proposition \ref{LVI-1}. If so,
Algorithm \ref{Algo3} returns the single-operator
PCE through the PCG. If not, we adopt Algorithm
2 to compute the QCE as an approximation to a multi-operator PCE.

\begin{algorithm}[tb]\label{Algo3}
	\caption{The Competitive Hybrid Pricing}\label{Algorithm3}
	\SetAlgoLined
	\SetKwInOut{Input}{Input}\SetKwInOut{Output}{Output}
	\Input{$\{\varphi_n(\cdot)\}_{n\in\mathcal{N}}$, $\{\tilde{e}_{j}\}_{j\in\mathcal{I}}$, $\{d_{i}(\cdot)\}_{i\in\mathcal{I}}$, $\{C_j\}_{j\in\mathcal{I}}$}
	\Output{Equilibrium hybrid price matrix $\bs{h}^\star$}
	
	\eIf{There is a single-operator PCE ($\zeta_{g^{\rm thr}-1}\leq \tilde{e}_{g^{\rm thr}}$) according to Proposition \ref{LVI-1}}{
		Compute PCE delivered prices $\bs{p}^\star$  by \eqref{T3-Eqqq}\;
	}{Compute QCE $\bs{q}^\star$ by Algorithm 2\;
	Compute the delivered pricing by
	$p_i^\star=\pi\left(\sum_{n\in\mathcal{N}}{q}_n^\star\right)$ for all $i\in\mathcal{I}\;$}

Compute the hybrid pricing $\bs{h}^\star$ from \eqref{PricingINDE}\;

\end{algorithm}

\section{Numerical Results}\label{Numerical}

In this section, we perform numerical studies to evaluate the performances of the proposed schemes (the cooperative pricing scheme, the competitive pricing scheme, and the FT scheme) compared with two benchmarks, and study the impact of network heterogeneity on their performances.

\subsection{Benchmarking Schemes}

We consider two benchmarking schemes, namely the \textit{no-tethering pricing}  scheme and the \textit{social welfare maximization} scheme for comparison.

\subsubsection{No-Tethering Pricing  (NTP)} For the NTP scheme, MNOs do not allow tethering (which is equivalent to set $t_{i\leftarrow j}$ to be infinity for all $i\neq j, i,j\in\mathcal{I}$) and then optimize the access prices by maximizing their own profits.

 \subsubsection{Social Welfare Maximization (SWM)} Before we introduce the SWM scheme, we first define
 the social welfare as the aggregate payoff of all MNOs and all users, denoted by
 \begin{align}
 \Psi(\bs{x})&=\sum_{i\in\mathcal{I}}U_i\left(\bs{x}\right)-\sum_{i\in\mathcal{I}}\sum_{j\in\mathcal{I}}\tilde{e}_{i\leftarrow j}x_{i\leftarrow j}.\label{SW}
 \end{align}

The SWM scheme is to maximize the social welfare $\Psi(\bs{x})$ subject to the capacity constraints $\sum_{i\in\mathcal{I}}x_{i\leftarrow j}\leq C_j,~\forall j\in\mathcal{I}$ and non-negative constraints $x_{i\leftarrow j}\geq 0,~\forall i,j\in\mathcal{I}$.

\subsection{Simulation Setup}
We perform numerical studies for an MCA involving $|\mathcal{N}|=2$ MNOs\footnote{The two-MNO setting is a reasonable consideration and has been widely adopted in the literature (e.g. \cite{Duo}).} and $|\mathcal{I}|=10$ users and study their interactions for $T=10$ minutes.  We randomly select 5 users as the subscribers of MNO 1 and the remaining 5 users as the those of MNO 2. 

\textit{Truncated Normal Distributions}:  We assume that users' utility parameters, downlink capacities $C_i$, and operational costs $e_i$ follow the respective i.i.d. truncated normal distributions. Let $TN(\mu,\kappa^2)$ denote a truncated normal distribution over the interval $[\mu-2\kappa^2,\mu+2\kappa^2]$, with $\mu$ being the mean and $\kappa^2$ being the variance.

\textit{Parameter Settings:} We consider the following parameter settings unless stated otherwise.  
Each user $i$ has the weighted $\alpha$-fair utility function $U_i(x)=\theta_i x^{1-\alpha}/(1-\alpha)$, where $\alpha=0.4$.
We set $\theta_i\in TN(550,200)$. We consider two types of users including 3G users and LTE users. We select a user as a LTE user with a probability of $\rho_{\rm LTE}$ and a 3G user with a probability of $1-\rho_{\rm LTE}$. We set $\rho_{\rm LTE}=40\%$.
LTE downlinks have relatively higher capacities and incur lower operational costs than the 3G downlinks \cite{4G,3G,SenzaFili}. Specifically,
the downlink capacity (in terms of Mbps) distributions for LTE downlinks and 3G downlinks are $C_{\rm LTE}\sim TN(14,3)$ and $C_{\rm 3G}\sim TN(1,0.3)$, respectively. We choose the average speed according to field experiments \cite{3G,4G}. 
The operational cost distributions for LTE downlinks and 3G downlinks are $e_{\rm LTE}\sim TN(80,10)$ and $e_{\rm 3G}\sim TN(350,40)$, respectively, suggested in \cite{SenzaFili}. Each user experiences the same cellular energy cost and zero Wi-Fi energy cost, and we set $c_i=7.5$ \cite{energy1}.
We run the experiment 10,000 times for each $\left(\rho_{\rm LTE}, \{\theta_i\}_{i\in\mathcal{I}},\{C_i\}_{i\in\mathcal{I}},\{e_i\}_{i\in\mathcal{I}},\{\mathcal{I}_n\}_{n\in\mathcal{N}}\right)$ realization  and show the average together with the error bars. 

\subsection{Results}

\begin{figure}[!t]
	\begin{centering}
			\subfigure []{\includegraphics[scale=.29]{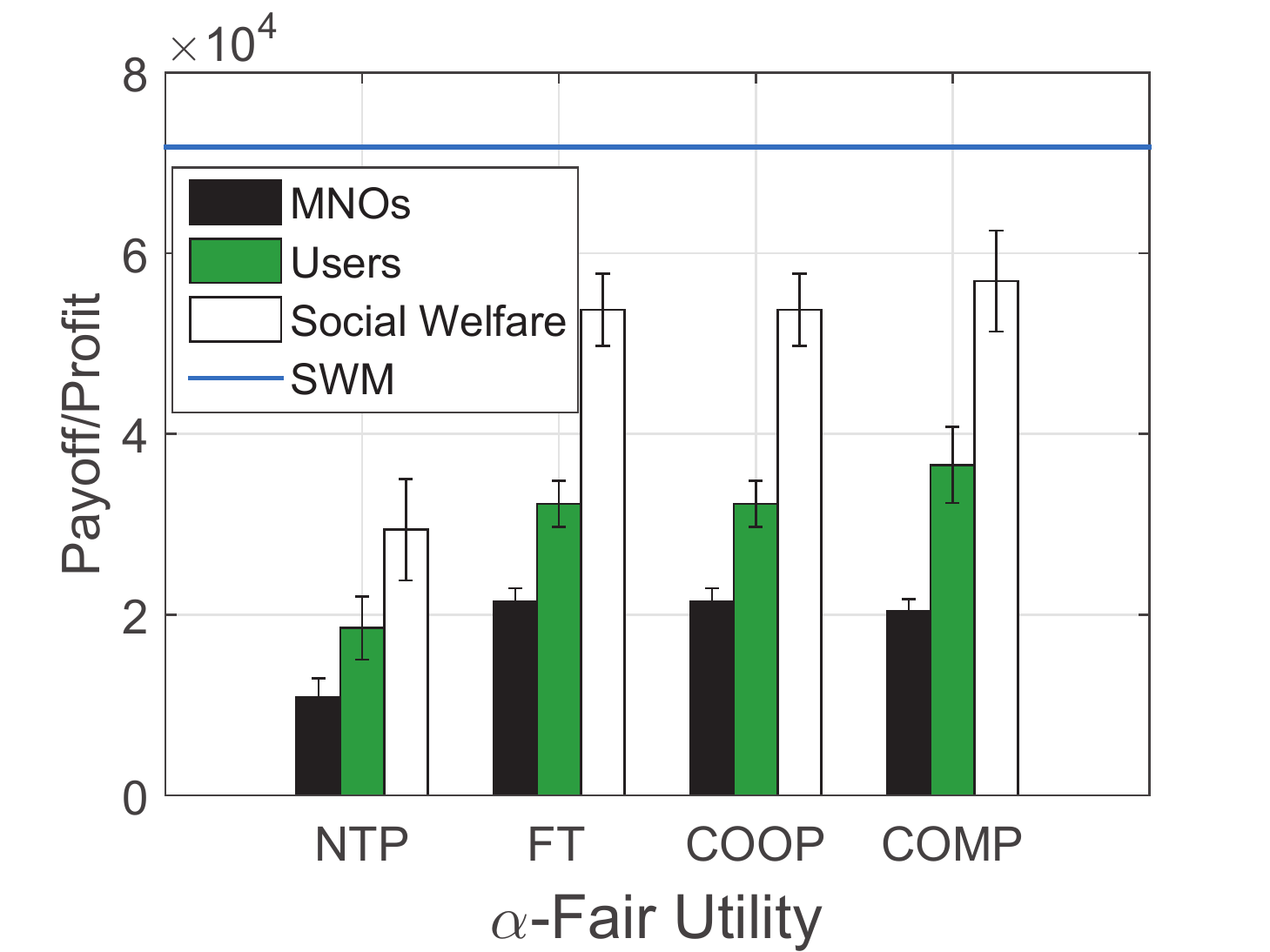}}
				\subfigure []{\includegraphics[scale=.29]{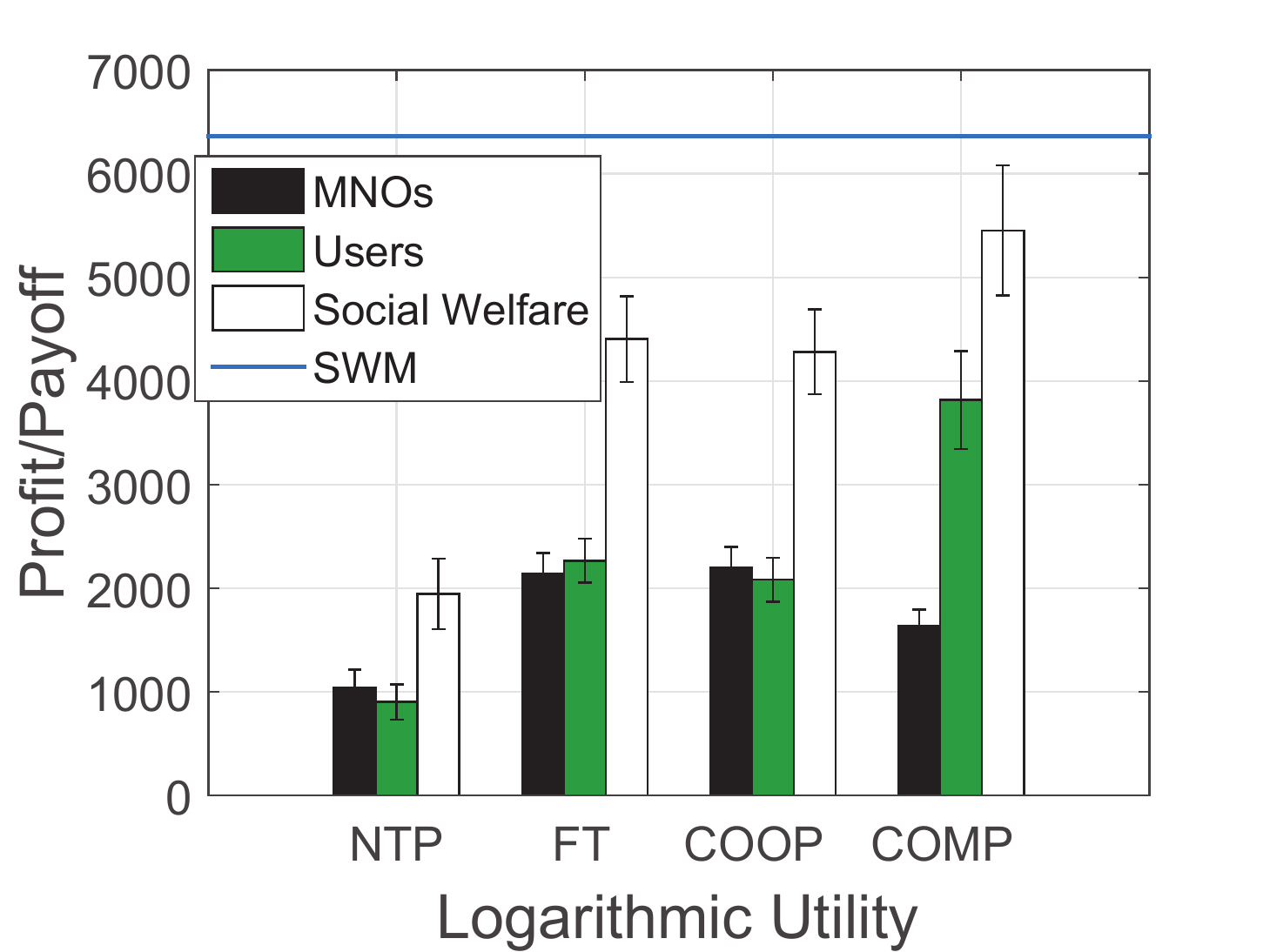}}
		\vspace{-0.25cm}		
		\caption{Performance comparison of the proposed schemes against two benchmarks for (a) $\alpha$-fair utility and (b) logarithmic utility.}
		\label{f3}
		\vspace{-0.1cm}
	\end{centering}
\end{figure}


\subsubsection{Performance Comparison}

In this part, we compare the performances of the cooperative scheme (denoted by COOP), competitive (denoted by COMP), and the FT scheme with the two benchmarks. We study both an $\alpha$-fair utility scenario and a logarithmic utility scenario.


We first consider the $\alpha$-fair utility scenario.
We plot the users' payoff, MNOs' profit, and the social welfare
for users in Fig. \ref{f3}(a).
  First, we observe that the cooperative scheme approximately doubles both MNOs' profit, compared with the NTP scheme. Comparing with the cooperative scheme, the competitive scheme only incurs a slight MNOs' profit loss. For the FT scheme, we see that it achieves exactly the same performance  as the cooperative scheme. This agrees with Corollary \ref{coro1}, i.e., the iso-elasticity of the $\alpha$-fair utility functions, zero Wi-Fi links energy costs $e_{i\leftarrow j}^\text{Wi-Fi}$, and homogeneous cellular downlink energy costs $e_{j}^\text{Down}$ lead to the optimum.
  
  \begin{figure}[!t]
  	\begin{centering}
  		\subfigure [] {\includegraphics[scale=.29]{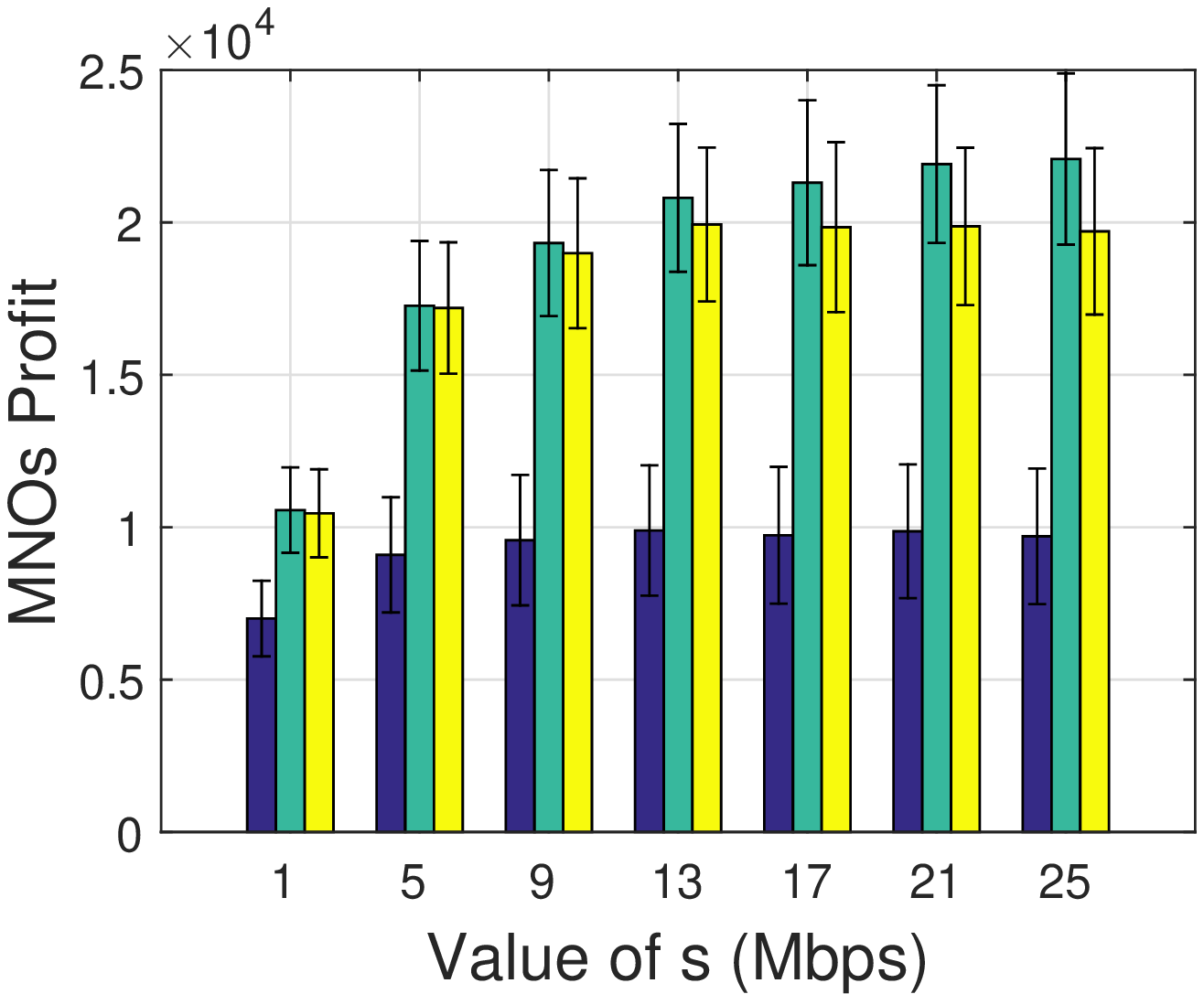}}
  		\subfigure []{\includegraphics[scale=.29]{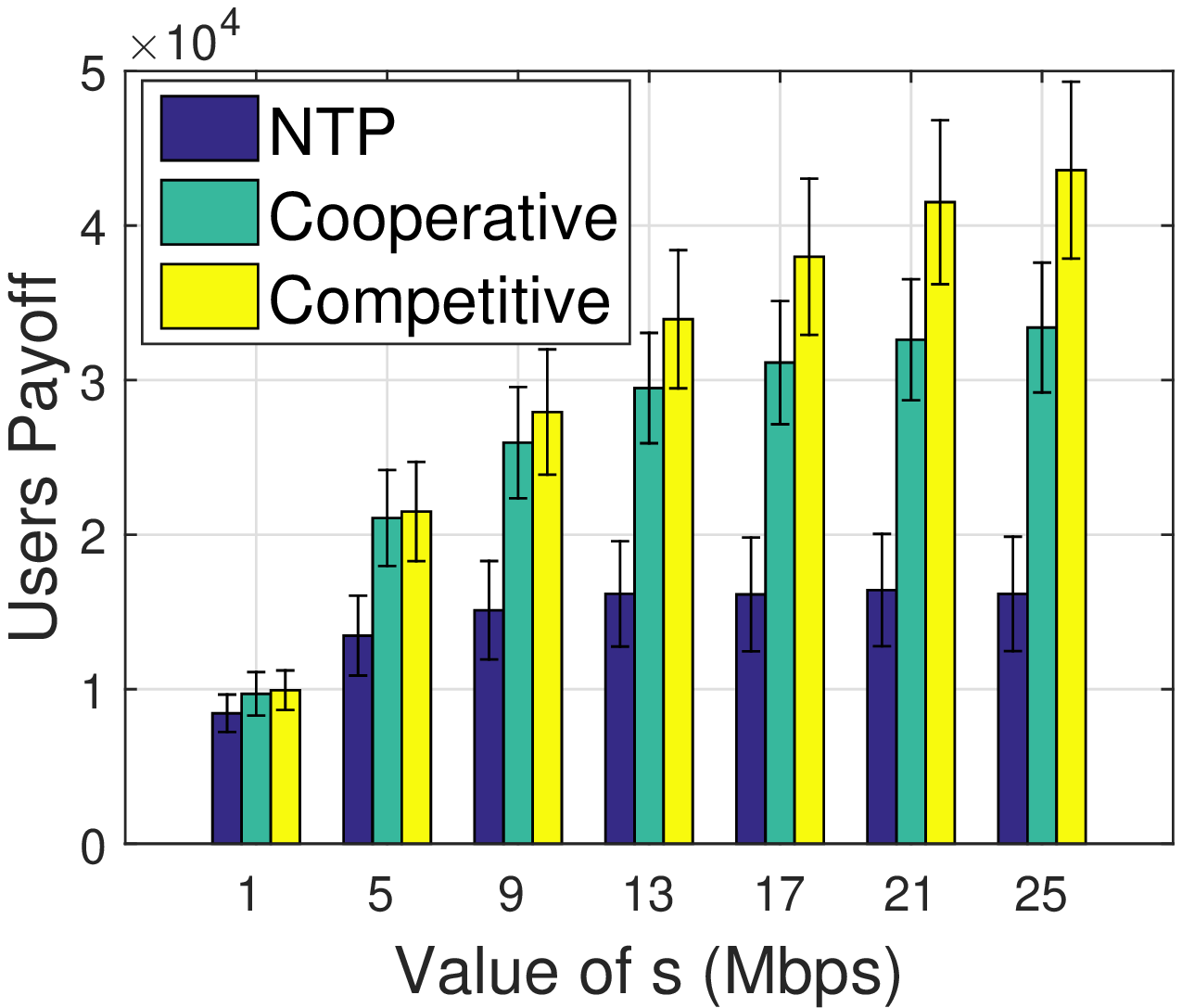}}
  		\vspace{-0.25cm}
  		\caption{MNOs' profit (a) and users' payoff (b) versus the value of $s$, where $s$ is the mean of downlink capacity for LTE users.  }
  		\label{f5}
  		\vspace{-0.1cm}
  	\end{centering}
  \end{figure}

  We observe a similar trend for the users' payoff. Namely, both cooperative and competitive schemes approximately double both users' payoff, compared with  the NTP scheme. The competitive scheme achieves a relatively higher  users' payoff compared with  the cooperative scheme. 
  
  In terms of the social welfare, we observe that the cooperative and competitive schemes perform better than the NTP scheme. Moreover, the competitive scheme achieves $28\%$ more social welfare than the cooperative scheme. Intuitively, the competitive MNOs compete down the hybrid prices so that the achieved social welfare is closer to the SWM scheme. 
  
  To conclude,
  \begin{observation}
  	Compared with the conventional NTP scheme, in the $\alpha$-fair utility setting, both competitive and cooperative schemes can approximately double both MNOs' profit and users' payoff. 
  \end{observation}
  
We then study the logarithmic utility scenario. Each user has a utility given by $U_i(x)=\theta_i \log(1+x)$ with $\theta_i\sim TN(550,200)$. The cooperative scheme approximately doubles the MNOs' profit compared with the FT scheme, while the competitive scheme decreases the MNOs' profit compared with the cooperative scheme. 
A key observation is that the FT scheme perform close to the optimal and can achieve $97.5\%$ of the maximal MNOs' profit. 
Therefore, we highlight
  \begin{observation}
  	The FT scheme performs close to the optimal cooperative scheme with low tethering energy cost and homogeneous cellular energy cost, even without isoelastic utility.
  \end{observation}
  
  In terms of users' payoff, the cooperative scheme approximately doubles the NTP scheme's performance. The competitive scheme further improves the performance of  the cooperative scheme by $88\%$.



\begin{figure}[!t]
	\begin{centering}
		\subfigure [] {\includegraphics[scale=.29]{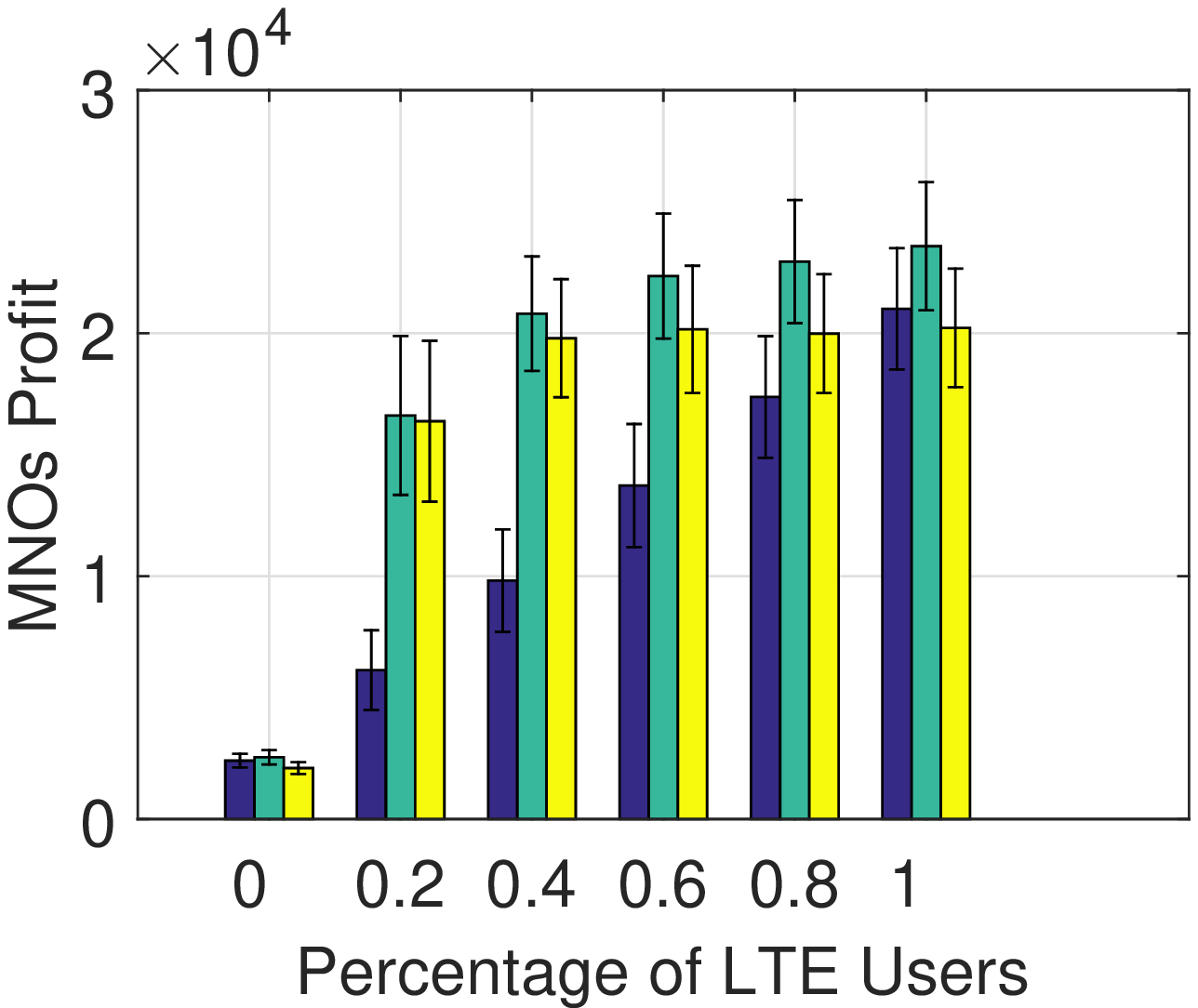}}
		\subfigure []{\includegraphics[scale=.29]{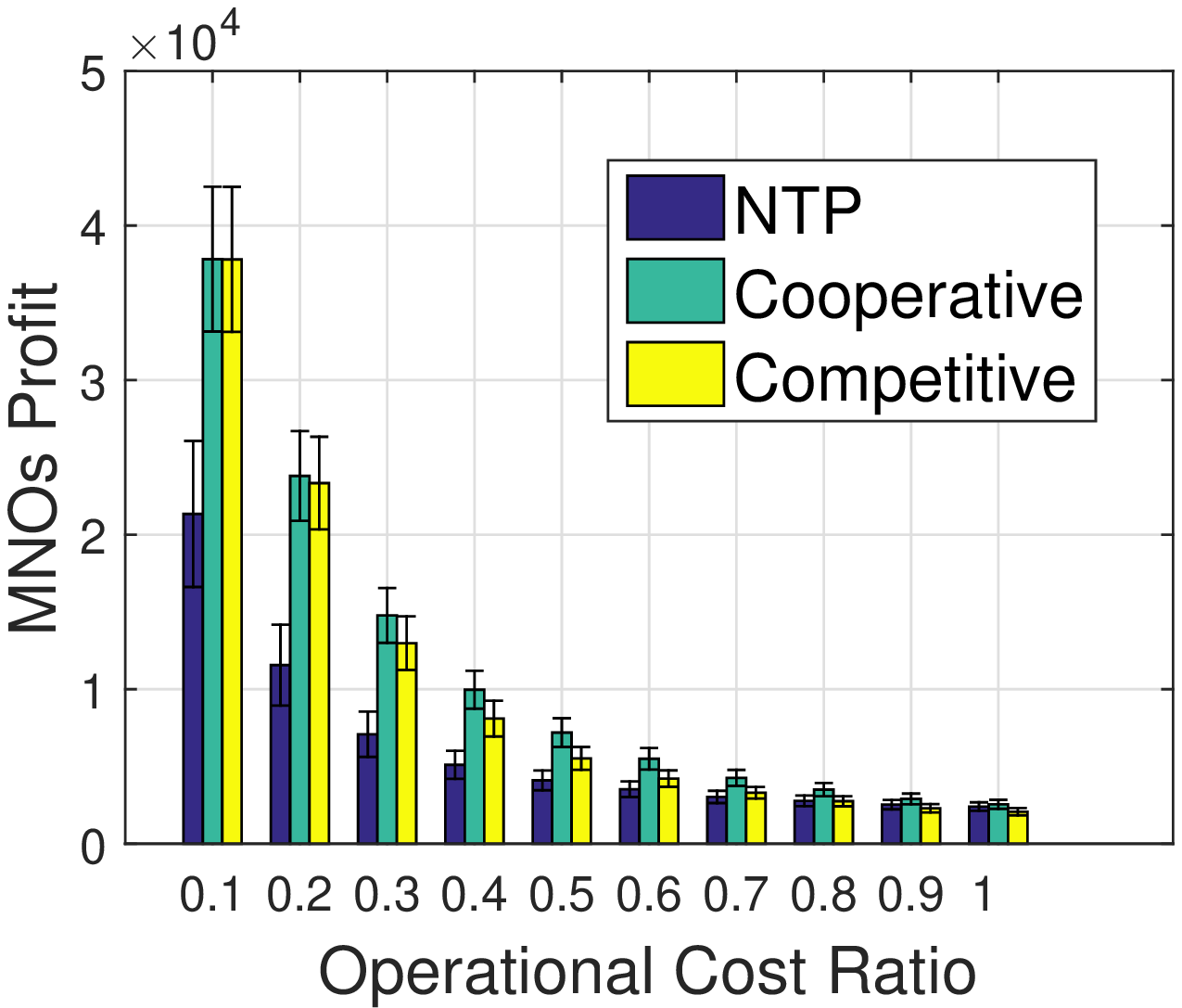}}
		\vspace{-0.25cm}		
		\caption{MNOs' profit versus (a) the percentage of the LTE users and (b) the ratio of the means of LTE operational cost and 3G operational cost (the LTE operational cost distribution $e_{\rm LTE}\sim TN(\eta\times 350,\eta\times 40)$).}
		\label{f6}
	\end{centering}
\end{figure}

\subsubsection{Impact of Network Heterogeneity}

We next perform three sets of experiments to investigate how the performance benefits of the MCA
service depend on the network heterogeneity. Specifically, 
heterogeneities are in terms of the downlink capacity, the portion of LTE users, and the MNOs' operational cost. 


\textbf{Downlink capacity}: For the first experiment, we fix the mean of LTE downlink to be $s$ Mbps and the 3G downlinks  to be $1$ Mbps. Fig. \ref{f5}(a) shows that,  as the capacities of the low-cost LTE channels increase, the MNOs' profit benefits of the cooperative and competitive schemes over the NTP scheme become more significant.
Precisely, when $s=25$ Mbps,
 compared with the conventional NTP scheme, the cooperative and the competitive schemes improve the MNOs' profit by $125\%$ and $101\%$, respectively.  This is because, in the proposed schemes, users prioritize the channels of low operational costs. Hence, the capacity increase of these channels will reduce the MNOs' operational costs and thus the delivered prices to users.
 
 Fig. \ref{f5}(b) shows that compared with the conventional NTP scheme, the cooperative and the competitive schemes' improvements increase in $s$ and can improve
 the users' payoff by $95\%$ and $148\%$, respectively, when $s=25$ Mbps. On the other hand, comparing with the cooperative scheme, the competitive scheme also achieves a substantial gain up to $27\%$ in term of users' payoff when $s$ is large whereas the loss in MNOs' profit is relatively small. This mainly results from the larger social welfare achieved by the competitive scheme.

 In the following, we will study the performances in terms of MNOs' profit.
 The results in terms of users' payoff  for the next two sets of experiments
 are similar to the first set.
 We thus present the numerical results for users' payoff in Appendix \ref{SNR}.


\textbf{Percentage of LTE users}: Fig. \ref{f6}(a) shows  a similar trend regarding the network heterogeneity in terms of the percentage of LTE users. As shown in Fig. \ref{f6}(a), when the percentage of LTE users is between $20\%$-$60\%$, the cooperative and competitive schemes' benefits in terms of MNOs' profits  increase are large, compared with the NTP scheme. When users are all LTE users or all 3G users,  however,  the NTP scheme achieves higher MNOs' profits than the competitive scheme. This implies that when users are homogeneous, each MNO's profit gained from the MCA due to resource pooling is limited while the competitive MNOs also suffer from the additional  competition  enabled by the MCA. 


\textbf{LTE operational cost}: Finally, we study the LTE operational cost's impact on performance in Fig. \ref{f6}(b). Specifically, we consider an operational cost ratio $\eta\in[0,1]$ and set $e_{\rm LTE}\sim TN(\eta\times 350,\eta\times 40)$. Therefore, the a larger $\eta$ leads to a more similar LTE operational cost to the 3G operational cost. We show that as $\eta$ increases, MNOs' profit improvements for both proposed schemes decrease compared with the NTP scheme in Fig. \ref{f6}(b). To summarize, we have

\begin{observation}
The degree of network heterogeneity increases the performance improvement of both cooperative and competitive schemes, comparing with the NTP scheme.
\end{observation}

\begin{observation}
	Compared to the cooperative scheme, the competitive scheme achieves a slightly smaller MNOs' profit but a significantly larger users' payoff, in heterogeneous networks.
\end{observation}

As today's wireless networks are becoming increasingly heterogeneous due to the coexistence of legacy and new technologies, we believe that the MCA with the proposed pricing schemes will be beneficial to both MNOs and users.

%

%

\section{Conclusion}\label{Conc}

In this paper, we proposed a hybrid pricing framework for the mobile collaborative Internet access and studied both cooperative and competitive interactions among MNOs. We showed that there exists an equilibrium hybrid pricing scheme making each user's cost independent of the selection of tethering links, for both cooperative and competitive scenarios. 
We verified that the proposed cooperative and competitive pricing schemes improve MNOs' profit and users' payoff as compared with the no-tethering scheme  in practically interesting scenarios. Moreover, the performance gain increases when the network becomes more diverse. 
Our results also showed that, under some mild conditions, it is possible  to achieve most of the benefits of MCA without imposing additional fees on user tethering. This result is encouraging, as it allows the MNOs to improve their profits (compared to no tethering case) even if they are prohibited from charging for tethering \cite{fcc}.

For the future work, we are particularly interested in studying the impact of incomplete information, where the users' utility functions and topology are private information.
 Another promising direction is to implement the proposed schemes in systems. Inspired by TUBE in \cite{Tube}, one approach is to design an architecture solution that  contains not only the MCA function but also a feedback loop between the MNOs' hybrid price computation and users, which will facilitate MNOs to implement the adaptive and topology-based pricing.


\vspace{-1cm}
\begin{IEEEbiography}[{\includegraphics[width=1in,height=1.25in,clip,keepaspectratio]{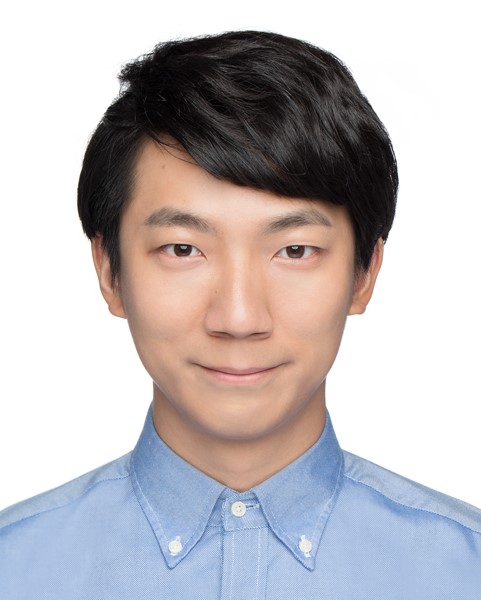}}]{Meng Zhang} (S'15) is a PhD
	student in the Department of Information Engineering  at the Chinese
	University of Hong Kong since 2015. He was a visiting student research collaborator with the Department of Electrical Engineering at Princeton University, from 2018 to 2019.
	His research interests include network
	economics, with emphasis on mechanism design for wireless networks.
\end{IEEEbiography}

\vspace{-1cm}
\begin{IEEEbiography}[{\includegraphics[width=1in,height=1.25in,clip,keepaspectratio]{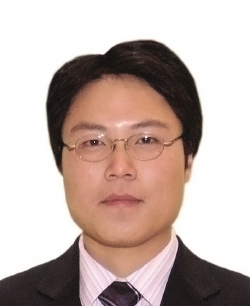}}]{Lin Gao}  (S'08-M'10-SM'16) received the PhD
	degree in electronic engineering from Shanghai
	Jiao Tong University in 2010 and worked as a
	postdoctoral research fellow with the Chinese
	University of Hong Kong from 2010 to 2015.
	He is an associate professor in the School of
	Electronic and Information Engineering, Harbin
	Institute of Technology, Shenzhen, China. He received
	the IEEE ComSoc Asia-Pacific Outstanding
	Young Researcher Award in 2016. His main
	research interests are in the area of network
	economics and games, with applications in wireless communications
	and networking. He has co-authored 3 books including the textbook
	Wireless Network Pricing by Morgan \& Claypool (2013). 
	He is the corecipient
	of three Best (Student) Paper Awards from WiOpt 2013, 2014,
	2015, and is a Best Paper Award Finalist from IEEE INFOCOM 2016.
\end{IEEEbiography}

\vspace{-1cm}
\begin{IEEEbiography}[{\includegraphics[width=1in,height=1.25in,clip,keepaspectratio]{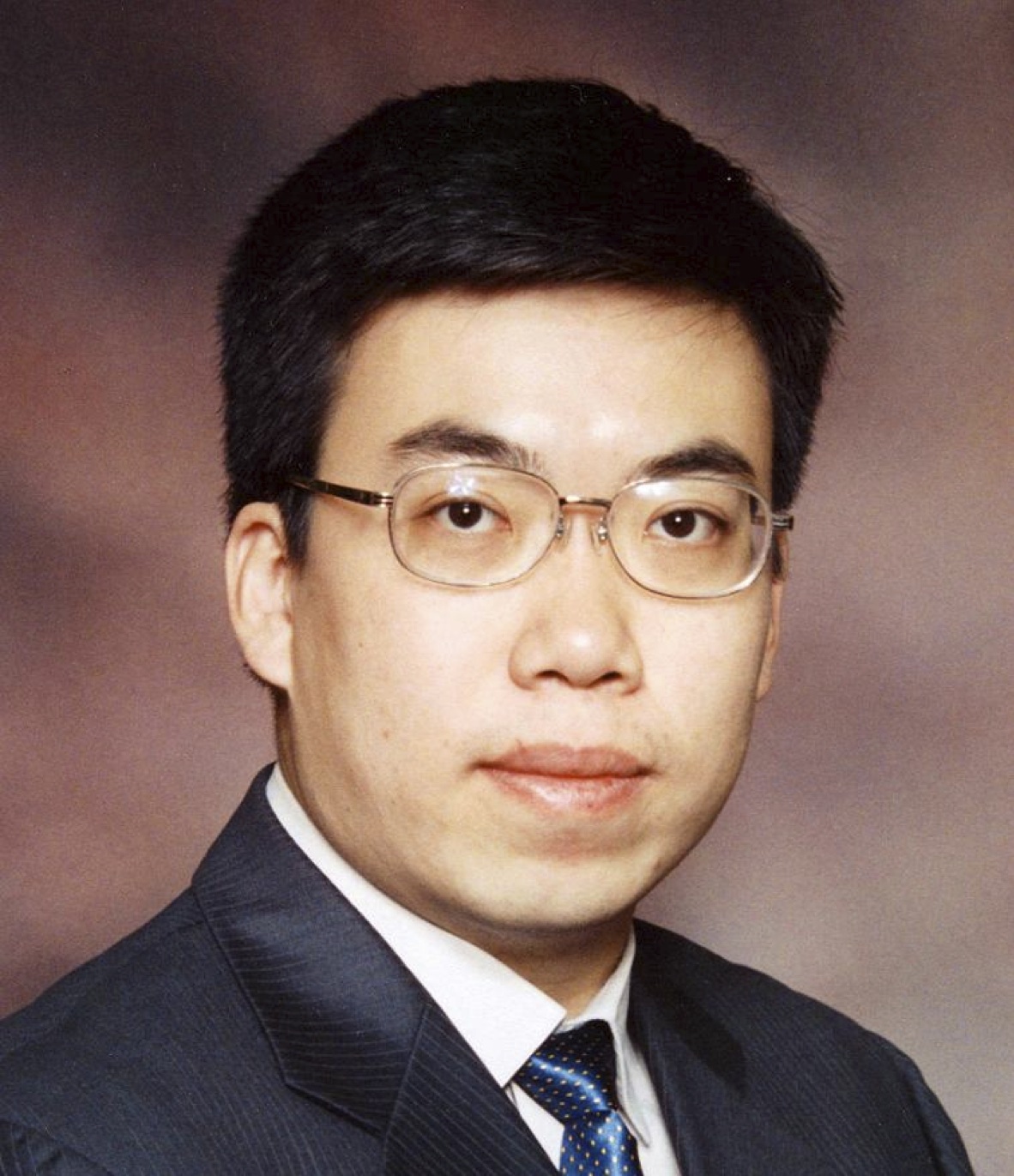}}]{Jianwei Huang} (F'16) is a Presidential Chair Professor and the Associate Dean of the School of Science and Engineering, The Chinese University of Hong Kong, Shenzhen. He is also a Professor in the Department of Information Engineering, The Chinese University of Hong Kong. He received the Ph.D. degree from Northwestern University in 2005, and worked as a Postdoc Research Associate at Princeton University during 2005-2007. He has been an IEEE Fellow, a Distinguished Lecturer of IEEE Communications Society, and a Clarivate Analytics Highly Cited Researcher in Computer Science. He is the co-author of 9 Best Paper Awards, including IEEE Marconi Prize Paper Award in Wireless Communications in 2011. He has co-authored six books, including the textbook on "Wireless Network Pricing." He received the CUHK Young Researcher Award in 2014 and IEEE ComSoc Asia-Pacific Outstanding Young Researcher Award in 2009. He has served as an Associate Editor of IEEE Transactions on Mobile Computing, IEEE/ACM Transactions on Networking, IEEE Transactions on Network Science and Engineering, IEEE Transactions on Wireless Communications, IEEE Journal on Selected Areas in Communications - Cognitive Radio Series, and IEEE Transactions on Cognitive Communications and Networking. 
	He has served as an Editor of Wiley Information and Communication Technology Series, Springer Encyclopedia of Wireless Networks, and Springer Handbook of Cognitive Radio. He has served as the Chair of IEEE ComSoc Cognitive Network Technical Committee and Multimedia Communications Technical Committee. He is the recipient of IEEE ComSoc Multimedia Communications Technical Committee Distinguished Service Award in 2015 and IEEE GLOBECOM Outstanding Service Award in 2010.  
\end{IEEEbiography}

\vspace{-1cm}

\begin{IEEEbiography}[{\includegraphics[width=1in,height=1.25in,clip,keepaspectratio]{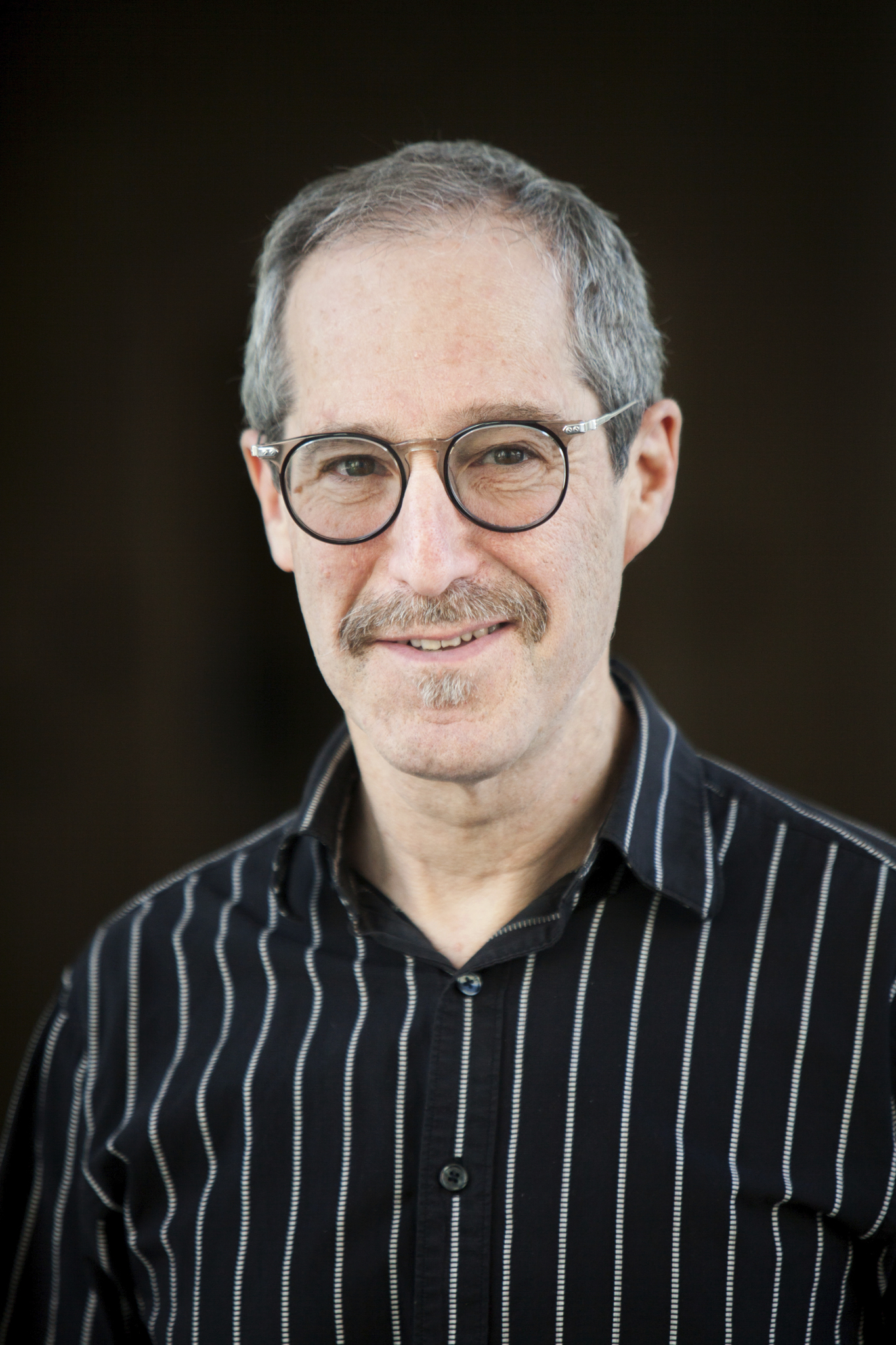}}]{Michael L. Honig} (F'97)  is a Professor in the Department of Electrical Engineering and Computer Science 
	at Northwestern University. Prior to joining Northwestern he worked in the Systems Principles 
	Research Division at Bellcore in Morristown, NJ, and at Bell Laboratories in Holmdel, NJ. 
	His recent research has focused on wireless networks, including interference mitigation 
	and resource allocation, and market mechanisms for dynamic spectrum allocation. 
	He received the B.S. degree in electrical engineering from Stanford University in 1977, 
	and the Ph.D. degree in electrical engineering from the University of California, Berkeley, in 1981. 
	He has held visiting scholar positions at the Naval Research Laboratory (San Diego), 
	the University of California, Berkeley, the University of Sydney, Princeton University, 
	the Technical University of Munich, and the Chinese University of Hong Kong. He has also worked 
	as a free-lance trombonist. Dr. Honig has served as editor and guest editor for several journals, 
	and as a member of the Board of Governors for the IEEE Information Theory Society.
	He is a Fellow of IEEE, the recipient of a Humboldt Research Award for Senior U.S. Scientists, 
	and a co-recipient of the 2002 IEEE Communications Society and Information Theory Society 
	Joint Paper Award and the 2010 IEEE Marconi Prize Paper Award.
\end{IEEEbiography}

\appendix
%

\subsection{KKT conditions and Lemma \ref{L1}}\label{P1}

We characterize the optimal solution $\bs{x}^*(\bs{h})$ to (UPM) in \eqref{UPM}  for any given $\bs{h}$ in the following.
Due to the concavity of the objective and the convexity of the constraint set,  (UPM) is a convex problem. It is easy to verify that it also satisfies the Slater's condition, and hence the Karush-Kuhn-Tucker (KKT) conditions are sufficient and necessary for global optimality \cite{Boyd}: for every user $i$,
\begin{subequations}\label{KKT}
	\begin{align}
\hspace{-0.5cm}	U_i'\left(\sum_{l\in\mathcal{I}}x_{i\leftarrow l}\right)-h_{i\leftarrow j}-c_{i\leftarrow j}-\lambda_j+\mu_{i\leftarrow j}&=0,~\forall j,\label{KKTf}\\
	\lambda_j\left(\sum_{i\in\mathcal{I}}x_{i\leftarrow j}-C_j\right)&=0,~\forall j,\label{KKTs}\\
	x_{i\leftarrow j}\mu_{i\leftarrow j}&=0,~\forall j,\label{KKTm}\\
	\lambda_{i}&\geq 0,~\\
	\mu_{i\leftarrow j}&\geq 0,~\forall j,\label{KKTl}
	\end{align}
\end{subequations}

Based on the KKT conditions in \eqref{KKT}, we have the following lemma for facilitating the following analysis.
\begin{lemma}\label{L1}
	Given any hybrid pricing matrix $\bs{h}$ and $h_{i\leftarrow k}+c_{i\leftarrow k}>h_{i \leftarrow j}+c_{i\leftarrow j}$, the optimal traffic matrix $\bs{x}^*$ satisfies
	\begin{align}
	x_{i\leftarrow k}^*>0~~{\rm~only~if}~~\sum_{l\in\mathcal{I}}x_{l \leftarrow j}^*=C_j.
	\end{align}
\end{lemma}
Intuitively, users purchase the data from the downlink with a lower hybrid price first (and fully utilize that downlink's capacity) before purchasing data from a higher price downlink. 


\begin{IEEEproof}
When $h_{i\leftarrow j}+c_{i\leftarrow j}<h_{i \leftarrow k}+c_{i\leftarrow j}$, by  \eqref{KKTf}, we have:
\begin{equation}
\lambda_j^*-\mu_{i\leftarrow j}^*>\lambda_k^*-\mu_{i\leftarrow k}^*. \label{proof1}
\end{equation}
When $x_{i\leftarrow k}^*>0$, combining  \eqref{KKTm}-\eqref{KKTl} and \eqref{proof1}, we have
\begin{equation}
\lambda_j^*>\lambda_k^*+\mu_{i\leftarrow j}^*\geq0,
\end{equation}
which indicates that $\sum_{i\in\mathcal{I}}x_{i\leftarrow j}^*-C_j=0$ according to  \eqref{KKTs}.
\end{IEEEproof}

\subsection{Proof of Theorem \ref{T1}}\label{P2}

We prove Theorem \ref{T1} for both cooperative and competitive MNOs, respectively.

\subsubsection{Cooperative MNOs}
 Suppose that, given $\bs{h}^*$, there exists an optimal solution $\{\bs{x}^*,\bs{\mu^*},\bs{\lambda}^*\}$ (that satisfies the KKT conditions) with  $\lambda_j^*>0$ for some downlink $j \in \mathcal{I}.$
In this case we can see that if MNOs increase $h_{i\leftarrow j}^*$ to $h_{i\leftarrow j}'\triangleq h_{i\leftarrow j}^*+\lambda_j^*$ for all $i\in\mathcal{I}$, the solution $\{\bs{x}^*,\bs{\mu}^*,\bs{\lambda}'\triangleq\bs{0}\}$ will still satisfy the KKT conditions under  the updated price, implying that it is an optimal solution under the updated price. Furthermore, Assumption 1 implies that the resulted total MNOs' profit is unique in $\bs{h}$. Hence, the  MNOs' profit increases from $$
\sum_{j\in\mathcal{I}}\sum_{i\in\mathcal{I}}(h_{i\leftarrow j}^*-e_j)x^*_{i\leftarrow j}$$ to $$
\sum_{j\in\mathcal{I}}\sum_{i\in\mathcal{I}}(h_{i\leftarrow j}'-e_j)x^*_{i\leftarrow j},$$
which implies that  $\bs{h}^*$ is not  the equilibrium (best) price matrix for the MNOs. Hence, we can see that under the equilibrium price matrix  $\bs{h}^*$, the optimal solution $\{\bs{x}^*,\bs{\mu^*},\bs{\lambda}^*\}$  must satisfy $\lambda_i^*=0$ for all $i\in\mathcal{I}$. 

\subsubsection{Competitive MNOs}
Similar to the cooperative case, let  $\lambda_j^\star>0$ for some downlink $j \in \mathcal{I}.$
In this case we can see that if MNO $\sigma(j)$ increase $h_{i\leftarrow j}^\star$ to $h_{i\leftarrow j}'\triangleq h_{i\leftarrow j}^\star+\lambda_j^\star$ for all $i\in\mathcal{I}$. Furthermore, the resulted MNO $\sigma(j)$'s profit increases from $$
\sum_{k\in\mathcal{I}_{\sigma(j)}}\sum_{i\in\mathcal{I}}(h_{i\leftarrow k}^\star-e_k)x^\star_{i\leftarrow j}$$ to $$
\sum_{k\in\mathcal{I}_{\sigma(j)}}\sum_{i\in\mathcal{I}}(h_{i\leftarrow k}'-e_k)x^\star_{i\leftarrow j},$$
which implies that  $\bs{h}^\star$ is not at the equilibrium price matrix for the MNOs. This completes the proof. 

\subsection{Proof of Corollary \ref{P11}}\label{P3}

 We prove by construction. We define $\rho_{i\leftarrow j}\triangleq h_{i\leftarrow j}+c_{i\leftarrow j}$ for notational simplification.
  According to Theorem \ref{T1}, there exists an  equilibrium such that $\bs{\lambda}^* = 0$. For each user $i$,
  \begin{itemize}
    \item Suppose $x_{i\leftarrow j}^*=0$ for all users $j$. Let $k=\arg \max_{j}\rho_{i\leftarrow j}$. Then we let $h_{i\leftarrow j}'\leftarrow\rho_{i\leftarrow k}-c_{i\leftarrow j}$ for all $j\in\mathcal{I}$ and the $\mu_{i\leftarrow j}'\leftarrow \rho_{i\leftarrow k}-\rho_{i\leftarrow j}$ for all $j\in\mathcal{I}$. We see that the new solution still satisfies the KKT conditions in \eqref{KKTf}-\eqref{KKTl}. After such an operation, $\rho_{i\leftarrow j}=\rho_{i\leftarrow l},$ for all $j,l\in\mathcal{I}$, $\bs{x}^*$ is still the optimal traffic solution, and each MNO's profit maintains the same.
    \item Suppose there exists a user $j$ such that $x_{i\leftarrow j}^*>0$.
    Note that any other $k$ such that $x_{i\leftarrow k}^*>0$, it also satisfies that $\rho_{i\leftarrow j}=\rho_{i\leftarrow k}$ due to \eqref{KKTf}. For the remaining $k$ such that $x_{i\leftarrow k}^*=0$, let  $h_{i\leftarrow k}'\leftarrow h_{i\leftarrow k}-\mu_{i\leftarrow k}$ and the $\mu_{i\leftarrow k}'\leftarrow 0$. Hence, $h_{i\leftarrow k}'+c_{i\leftarrow k} = \rho_{i\leftarrow j}$. We see that the new solution also still satisfies the KKT conditions in \eqref{KKTf}-\eqref{KKTl}. After such an operation, $\rho_{i\leftarrow j}=\rho_{i\leftarrow l}$ for all $j,l\in\mathcal{I}$, $\bs{x}^*$ is still the optimal traffic solution, and each MNO's profit maintains the same.
  \end{itemize}

Therefore, we note that for any equilibrium $(\bs{h}^*,\bs{x}^*)$, we can always construct another $\tilde{\bs{h}}$ such that for each user $i$, $\tilde{h}_{i\leftarrow j}+c_{i\leftarrow j}=\tilde{h}_{i\leftarrow k}+c_{i\leftarrow k},$ for all $j,k\in\mathcal{I}$, $\bs{x}^*$ is still the optimal traffic solution, and each MNO's profit maintains the same. Hence, there always  exists such a pricing $\tilde{\bs{h}}$ that is the optimal cooperative pricing scheme.

\subsection{Proof of Proposition \ref{Prop1}}\label{P-Prop2}

We will prove that $U_i'(0)>\min_{j}[c_{i\leftarrow j}+e_{j}]$ is both sufficient and necessary for user $i$ to have a positive demand at an equilibrium.

\textit{Necessity:} Suppose that, for user $i$, we have  $U_i'(0)\leq\min_{l}[c_{i\leftarrow l}+e_{l}]$.
By the KKT conditions in \eqref{KKTf}, for every downlink $j$, we have
\begin{align}
\mu_{i\leftarrow j}&=-U_i'\left(\sum_{l}x_{i\leftarrow l}^*\right)+\lambda_j+c_{i\leftarrow j}+h_{i\leftarrow j}^*,\nonumber\\
&\overset{(a)}{\geq} -U_i'\left(\sum_{l}x_{i\leftarrow l}^*\right)+c_{i\leftarrow j}+e_j,\nonumber\\
&\overset{(b)}{>}-U_i'(0)+\min_{l}[c_{i\leftarrow l}+e_{l}],\nonumber\\
&\geq0.\label{Equ-P-Prop2}
\end{align}
Inequality $(a)$ is because the equilibrium hybrid price $h_{i\leftarrow j}^*$ should be no less than the operational cost and $\lambda_j\geq 0$. Inequality $(b)$ is because of the strict concavity of $U_i(\cdot)$ from Assumption \ref{Assump1}. We further see $x_{i\leftarrow j}^*=0$ for all  $j\in\mathcal{I}$ (user $i$ has zero demand) due to \eqref{Equ-P-Prop2} and the complementary slackness condition $x_{i\leftarrow j}\mu_{i\leftarrow j}=0$. This proves the  necessity.

\textit{Sufficiency:}
As Proposition \ref{P11} suggests, we focus on the gateway-independent delivered pricing scheme $\{p_i\}_{i\in\mathcal{I}}$. As we previously discussed, each user's traffic at the equilibrium can be fully characterized by a delivered price $p_i$.
When $U_i'(0)>\min_{j}[c_{i\leftarrow j}+e_{j}]$, there exists an arbitrarily small $\epsilon$ such that under an delivered price vector ${\bs{p}}=\{{p}_i=U_i'(0)-\epsilon\}_{i\in\mathcal{I}}$, each user $i$ demands a small amount of data. In this case, user $i$ requests data
from downlink $j^*$ with the lowest delivered cost, i.e., $j^*=\min_{j}\tilde{e}_{i\leftarrow j}$ by Assumption \ref{Assump1}. This leads to
 the positive (though small) total MNOs' profit. Hence, the optimal delivered price $p_i^*$ for each user $i$ should yield the profit no less than that of $\{{p}_i=U_i'(0)-\epsilon\}_{i\in\mathcal{I}}$, in which case each user should also have a positive demand. Hence, under MNOs' optimal pricing scheme, MNOs' profit gained from each user is positive which only happens if each user $i$ has a positive demand.

\subsection{Proof of Theorem \ref{T2}}\label{PT2}

To prove Theorem \ref{T2}, we first introduce an intermediate transformation of (OPM) and then prove the equivalence between the transformation and the (R-OPM).

\subsubsection{Transformed OPM (T-OPM)} 

Let $\bs{h}({\bs{p}})\triangleq\{h_{i\leftarrow j}={p}_i-c_{i\leftarrow j}\}_{i,j\in\mathcal{I}}$ be the hybrid price matrix associated with the delivered price vector $\bs{p}$. 

We reformulate (OPM) as
\begin{subequations}\label{MNOprofit}
\begin{align}
{\rm (T-OPM)}~&\max_{{\boldsymbol{p}}}~\sum_{i\in\mathcal{I}}d_i(p_i){p}_i-\sum_{j\in\mathcal{I}}\sum_{i\in\mathcal{I}}\tilde{e}_{i\leftarrow j}x_{i\leftarrow j}^*(\bs{h}({\bs{p}})),\\
&~{\rm s.t.} ~~p_{i}\geq 0,~\forall i\in\mathcal{I}.
\end{align}
\end{subequations}
Assumption \ref{Assump1} ensures that users select $\bs{x}^*$ from $\mathcal{X}^o$ to minimize the aggregate delivered cost $\sum_{j\in\mathcal{I}}\sum_{i\in\mathcal{I}}\tilde{e}_{i\leftarrow j}x_{i\leftarrow j}$ in \eqref{MNOprofit}. In addition,
since $d_i(p_i)$ is a (one-to-one) function and users select $\bs{x}^*$ according to Assumption \ref{Assump1}, the objective in \eqref{MNOprofit} has a unique maximum.
Based on the gateway-independent pricing structure in Corollary \ref{P11}, we have
\begin{proposition}[Equivalence of (OPM) and (T-OPM)]\label{PROPM}
For any optimal solution $\bs{p}^*$ to (T-OPM), $h_{i\leftarrow j}^*=p_{i}^*+c_{i\leftarrow j}$ for all $i,j\in\mathcal{I}$, is the optimal  solution to (OPM). 
\end{proposition}
From Proposition \ref{PROPM}, we can interpret the delivered cost $\tilde{e}_{i\leftarrow j}$ as the operational cost for each unit $x_{i\leftarrow j}$ in analogy with ${p}_i$ being the unit revenue gained from user $i$.

The physical interpretation is that we now consider an equivalent system where MNOs' operational cost is $\tilde{e}_{i\leftarrow j}$ for each byte downloaded by user $j$ and tethered to user $i$ (and there is no energy cost for user). By considering such an equivalent system, MNOs are able to optimize over the effective  price vector $\tilde{p}$.

\subsubsection{Equivalence between (T-OPM) and (R-OPM) }
Let $\nu_1^*$ be the optimal value of (T-OPM) and $\nu_2^*$ be that of (R-OPM). We will prove $\nu_1^*=\nu_2^*$ by first showing $\nu_1^*\leq \nu_2^*$ and then showing $\nu_1^*\geq \nu_2^*$. And then we show statements i) and ii).

Let $\bs{p}^*$ be the optimal solution to (T-OPM). Note that users' decision $\bs{x}^*(\bs{h}(\bs{p}^*))$ is a feasible solution to (R-OPM), since (R-OPM) and (UPM) share the same constraints. Given $\bs{p}^*$, the objective value of (R-OPM) can be expressed by
\begin{align}
&~~~~\nu_2(\bs{x}^*(\bs{h}(\bs{p}^*)))\nonumber\\
&=\sum_{j\in\mathcal{I}}\sum_{i\in\mathcal{I}}\left(U'_i \left(\sum_{j\in\mathcal{I}}x_{i\leftarrow j}^*(\bs{h}(\bs{p}^*))\right)-\tilde{e}_{i\leftarrow j}\right)x_{i\leftarrow j}^*(\bs{h}(\bs{p}^*))\nonumber\\
&\overset{(a)}{=} \sum_{j\in\mathcal{I}}\sum_{i\in\mathcal{I}}[U'_i\left(d_i(p_i^*)\right)-\tilde{e}_{i\leftarrow j}]x_{i\leftarrow j}^*(\bs{h}(\bs{p}^*))\nonumber\\
&\overset{(b)}{=}\sum_{i\in\mathcal{I}} p_i^*d_{i}(p_i^*)-\sum_{j\in\mathcal{I}}\sum_{i\in\mathcal{I}}\tilde{e}_{i\leftarrow j}x_{i\leftarrow j}^*(\bs{h}(\bs{p}^*))=\nu_1^*,
\end{align}
where $(a)$ is due to \eqref{d} and $(b)$ is because $U'_i(\cdot)$ is the inverse function of $d_i(\cdot)$. We have $\nu_1^*=\nu_2(\bs{x}^*(\bs{h}(\bs{p}^*)))\leq \nu_2^*$.

Next, we prove $\nu_2^*\leq \nu_1^*$.
The (R-OPM) Problem can be equivalently reformulated as
\begin{subequations}
\begin{align}
{\rm (R-OPM-2)}~
~&\max_{\bs{x},\bs{y}}~~\sum_{i\in\mathcal{I}}U'_i(y_i)y_i-\sum_{j\in\mathcal{I}}\sum_{i\in\mathcal{I}}\tilde{e}_{i\leftarrow j}x_{i\leftarrow j}\nonumber\\
&~~~{\rm s.t.}~~y_i=\sum_{j\in\mathcal{I}}x_{i\leftarrow j}, \forall i\in\mathcal{I},\\
&~~~~~~~~~\sum_{j\in\mathcal{I}}x_{i\leftarrow j}\leq C_j, \forall j\in\mathcal{I},\label{P2-C1}\\
&~~~~~~~~~\sum_{j\in\mathcal{I}}x_{i\leftarrow j}\geq 0, \forall i,j\in\mathcal{I}\label{P2-C2}.
\end{align}
\end{subequations}
Denote  $\bs{y}^*=\{y_i^*\}$ as the optimal solution to (R-OPM-2). Then we can express  the optimal solution $\bs{x}^{R}$ to (R-OPM)  as
\begin{align}
\bs{x}^{R}=&\arg \min_{\bs{x}}~\sum_{j\in\mathcal{I}}\sum_{i\in\mathcal{I}}\tilde{e}_{i\leftarrow j}x_{i\leftarrow j}\label{Z1}\\
&~~{\rm s.t.}~~\eqref{P2-C1},~~\eqref{P2-C2} \nonumber\\
&~~~~~~y_i^*=\sum_{j\in\mathcal{I}}x_{i\leftarrow j}, \forall i\in\mathcal{I} \nonumber.
\end{align}
Suppose $\tilde{p}_i=U'_i(\sum_{i\in\mathcal{I}}x_{i\leftarrow j}^{R})=U'_i(y^*)$ for every user $i$. Comparing \eqref{Z1} and \eqref{A1}, we can see $\bs{x}^*=\bs{x}^{R}$ under the price $\tilde{\bs{p}}=\{\tilde{p}_i\}_{i\in\mathcal{I}}$. This indicates that
\begin{align}
\nu_2^*&=\sum_{j\in\mathcal{I}}\sum_{i\in\mathcal{I}}\left(U'_i \left(\sum_{j\in\mathcal{I}}x_{i\leftarrow j}^{R}\right)-\tilde{e}_{i\leftarrow j}\right)x_{i\leftarrow j}^{R}\nonumber\\
&{=} \sum_{j\in\mathcal{I}}\sum_{i\in\mathcal{I}}[U'_i\left(\bs{x}^{R}\right)-\tilde{e}_{i\leftarrow j}]x_{i\leftarrow j}^*(U'_i(\bs{x}^{R}))\nonumber\\
&=\nu_1(\tilde{\bs{p}})\leq \nu_1^*.
\end{align}
That is, $\tilde{\bs{p}}$ corresponds to a feasible solution to (T-OPM).

Combining the above discussion, we have that $\nu_1^*=\nu_2^*$. Hence, $\tilde{\bs{p}}$ is exactly the optimal solution to (T-OPM), i.e., the optimal delivered price for user $i$ is
\begin{equation}
p_i^*=U'_i \left(\sum_{j\in\mathcal{I}}x_{i\leftarrow j}^{R}\right),
\end{equation}
and $\bs{x}^{R}$ is the optimal solution to (UPM) under $\bs{p}^*$. This completes the proof of statement i). Moreover, as previously discussed, the fact that $\bs{x}^*=\bs{x}^{R}$ under the price $\tilde{\bs{p}}=\bs{p}^*$ completes the proof of statement ii).

\subsection{Free-Tethering Pricing Scheme and Proof of Corollary \ref{coro1}}\label{FTPscheme}

In this subsection, we first elaborate the FT scheme and then prove Corollary \ref{coro1}.
\subsubsection{Free-Tethering Pricing Scheme}
Define the uniform delivered price function as the inverse function of the users' aggregate demand function, i.e.,
\begin{align}
\pi\left(\cdot\right)=\left(\sum_{i\in\mathcal{I}}d_i(\cdot)\right)^{-1}.
\end{align}
Such a uniform delivered price $\pi\left(\cdot\right)$ corresponds to a delivered price that equalizes users' aggregate demand function and the total downloads, i.e., $\sum_{i\in\mathcal{I}}d_i\left(\pi\left(\sum_{j\in\mathcal{I}}\sum_{i\in\mathcal{I}}x_{i\leftarrow j}\right)\right)=\sum_{j\in\mathcal{I}}\sum_{i\in\mathcal{I}}x_{i\leftarrow j}$. Similar to (R-OPM) in \eqref{R-OPM1}-\eqref{R-OPM3}, we formulate the  (FT) Problem as follows
\begin{subequations}
	\begin{align}
	{\rm (FT)}~
	~&\max_{\bs{x}}~~\sum_{j\in\mathcal{I}}\sum_{i\in\mathcal{I}}\left(\pi \left(\sum_{k\in\mathcal{I}}\sum_{l\in\mathcal{I}}x_{k\leftarrow l}\right)-\tilde{e}_{i\leftarrow j}\right)x_{i\leftarrow j}\\
	&~~~{\rm s.t.}~~\sum_{l\in\mathcal{I}}x_{l\leftarrow j}\leq C_j, \forall j\in\mathcal{I},\\
	&~~~~~~~~~~~~~~x_{i\leftarrow j}\geq 0, \forall i,j\in\mathcal{I}.
	\end{align}
	\end{subequations}
	Let $\bs{x}^F$ denote the optimal solution to the above problem. Then we can write the corresponding FT hybrid price as
	\begin{equation}
	h_{i\leftarrow j}^F=\pi \left(\sum_{k\in\mathcal{I}}\sum_{l\in\mathcal{I}}x_{k\leftarrow l}^F\right)-c^\text{Down}_j,~\forall i\in\mathcal{I}. 
	\end{equation}
	Since $h_{i\leftarrow j}^F=a_{j}^F+t_{i\leftarrow j}^F$, we further have $h_{i\leftarrow j}^F=a_{j}^F$ and thus $t_{i\leftarrow j}^F=0$.

\subsubsection{Proof of Corollary \ref{coro1}}
			
				We first show that the isoelastic utility functions lead to the same delivered prices across users.
				For an ioselastic utility function, the corresponding demand function is
				\begin{align}
				d_i(p_i)=\left(\frac{\theta_i}{p_i}\right)^{1/\alpha},~\forall i\in\mathcal{I}.
				\end{align}
We define user $i$'s price elasticity of demand as
\begin{align}
\mathcal{E}_i\triangleq\frac{\partial d_i(p_i)}{\partial p_i}\cdot \frac{p_i}{d_i}=-\frac{1}{\alpha},~\forall i\in\mathcal{I}.
\end{align}
 We can derive the optimal solution $p_i^*$ of (T-OPM) as
\begin{align}
p^*_i=\frac{\tilde{e}_{k} \mathcal{E}_i}{1+\mathcal{E}_i}, ~\forall i\in\mathcal{I},
\end{align}
for some downlink $k$.
Since $\mathcal{E}_i$ is a constant for every user $i$, the optimal delivered prices are the same across users. 

				Proposition \ref{Prop1} suggests that the optimal hybrid pricing satisfies
				\begin{align}
				p_i^*=a_i^*+c_{i}^\text{Down}=a_j^*+t_{i\leftarrow j}^*+c_{j}^\text{Down}+c_{i\leftarrow j}^\text{Wi-Fi}, \forall i,j\in\mathcal{I}.\label{free-the11}
				\end{align}
Hence, when $c_{i\leftarrow j}^{\rm Wi-Fi}=0$ and $c_{i}^\text{Down}=c_{j}^\text{Down}$ for any $i\in\mathcal{I}$, \eqref{free-the11} leads to
 $a_i^*=a_j^*$ and $t_{i\leftarrow j}^*=0$ for all users $i$ and $j$.

\subsection{Proof of Proposition \ref{Prop3}} \label{P-Prop3}

Let $\tilde{\bs{h}}^\star$ be an equilibrium price matrix and $\bs{x}^\star$ be the equilibrium traffic matrix under $\tilde{\bs{h}}^\star$. We first prove that, for any $\tilde{\bs{h}}^\star$, there always exists a gateway-independent hybrid price matrix ${\bs{h}}^\star$ that maintains both users' payoff and MNOs' profit. We then show that such a gateway-independent hybrid price matrix ${\bs{h}}^\star$ is a PCE.

\subsubsection{Constructing the  Gateway-Independent Pricing}
 By Theorem \ref{T1}, we have:
\begin{align}
\tilde{h}_{i\leftarrow j}^\star+c_{j}\geq U_i'(\bs{x}^\star)+\tilde{\mu}_{i\leftarrow j}^\star, \forall i\in\mathcal{I}, \nonumber
\end{align}
with $\tilde{\bs\mu}^\star$ being the equilibrium dual variables satisfying
$\tilde\mu_{i\leftarrow j}^\star=0$ if ${x}_{i\leftarrow j}^\star>0$.
 

From Assumption 2, at an equilibrium, each user $i$ must have a positive demand, i.e., $x_{i\leftarrow m(i)}^\star>0$ must hold for some $m(i)\in\mathcal{I}$. Consider a gateway-independent hybrid price matrix $\bs{h}^\star$ such that, for each user $i$,
\begin{align}
h_{i\leftarrow j}^\star+c_j=\tilde{h}_{i\leftarrow m(i)}+c_{m(i)}\triangleq p_i^\star,~~\forall j\in\mathcal{I}.
\end{align}
To show that such a gateway-independent pricing scheme maintains the same users' payoff and MNOs' profits, we exploit the KKT conditions in \eqref{KKTf}-\eqref{KKTl}, which leads to 
\begin{align}
h_{i\leftarrow j}^\star=\tilde{\mu}_{i\leftarrow j}^\star+\tilde{h}_{i\leftarrow j}^\star\label{P-Prop37}.
\end{align}

We can see $V_n(\bs{h}^\star)=V_n(\tilde{\bs{h}}^\star)$ for every MNO $n$. This is because, given the price $\bs{h}^\star$,
($\bs{x}^\star$, $\{\mu_{i\leftarrow j}'=0\}_{i,j}$, $\{\lambda_{j}^\star=0\}_{j}$) satisfies the KKT conditions in \eqref{KKTf}-\eqref{KKTl}, similar  as ($\bs{x}^\star$, $\bs{\mu}^\star$, $\{\lambda_j^\star=0\}_{j}$) satisfies the same set of KKT conditions  under the price $\tilde{\bs{h}}^\star$.

\subsubsection{Gateway-Independent Hybrid Price Matrix is a PCE}
Next, we show that the gateway-independent hybrid price matrix $\bs{h}^\star$ is an equilibrium by contradiction. Suppose that for some MNO $n$, there exists a strategy $\bs{h}'_n\triangleq\{{h}_{i\leftarrow j}'\}_{i,j}$ such that
\begin{align}
V_n(\bs{h}'_n;\bs{h}_{-n}^\star)>V_n(\bs{h}_{n}^\star;\bs{h}_{-n}^\star).
\end{align} 
Let $(\bs{x}',\bs{\lambda}',\bs{\mu}')$ be users' optimal (primal-dual) decision under the hybrid price $(\bs{h}'_n;\bs{h}_{-n}^\star)$, satisfying the KKT conditions:
\begin{subequations}
\begin{align}
U_i'(\bs{x}')-h_{i\leftarrow j}'-c_{j}+\lambda_j'+\mu_{i\leftarrow j}'=0,~\text{if}~ j\in\mathcal{I}_n\label{P-Prop3-2},\\
U_i'(\bs{x}')-h_{i\leftarrow j}^\star-c_{j}+\lambda_j'+\mu_{i\leftarrow j}'=0,~\text{if}~ j\notin\mathcal{I}_n\label{P-Prop3-3}.
\end{align}
\end{subequations}
The MNO's corresponding profit is $V_n(\bs{h}'_n;\bs{h}_{-n}^\star)=\sum_{i\in\mathcal{I}}\sum_{j\in\mathcal{I}_n}  h_{i\leftarrow j}'x_{i\leftarrow j}'$.

If MNO $n$ plays $\bs{h}'_n$ and the remaining MNOs' strategies are $\tilde{\bs{h}}_{-n}^\star$, users' (primal-dual) decision $(\bs{x}'',\bs{\lambda}'',\bs{\mu}'')$ satisfies:
\begin{subequations}
\begin{align}
U_i'(\bs{x}'')-h_{i\leftarrow j}'-c_{j}+\lambda_j''+\mu_{i\leftarrow j}''=0,~\text{if}~ j\in\mathcal{I}_n\label{P-Prop3-4},\\
U_i'(\bs{x}'')-\tilde{h}_{i\leftarrow j}^\star-c_{j}+\lambda_j''+\mu_{i\leftarrow j}''=0,~\text{if}~j\notin\mathcal{I}_n\label{P-Prop3-5}.
\end{align}
\end{subequations}
Consider  \eqref{P-Prop3-2}-\eqref{P-Prop3-3} and \eqref{P-Prop37}, we see that $(\bs{x}'',\bs{\lambda}'',\bs{\mu}'')$ satisfies $\bs{x}''=\bs{x}', \bs{\lambda}''=\bs{\lambda}'$, and $\mu_{i\leftarrow j}''=\mu_{i\leftarrow j}'$ if $j\in\mathcal{I}_n$ and $\mu_{i\leftarrow j}''=\mu_{i\leftarrow j}'+\mu_{i\leftarrow j}^\star$ if $j\notin\mathcal{I}_n$. Hence, $\bs{x}'$ is also the users' best decision under hybrid price $\{\bs{h}'_n;\tilde{\bs{h}}_{-n}^\star\}$. Hence, it leads to $V_n(\bs{h}'_n;\tilde{\bs{h}}_{-n}^\star)=\sum_{i\in\mathcal{I}}\sum_{j\in\mathcal{I}_n}  h_{i\leftarrow j}'x_{i\leftarrow j}'=V_n(\bs{h}'_n;\bs{h}_{-n}^\star)>V_n(\tilde{\bs{h}}^\star)=V_n(\tilde{\bs{h}}^\star)$, which contradicts with the fact that $\tilde{\bs{h}}^\star$ is at the equilibrium. 

Therefore, we proved that the gateway-independent hybrid price matrix $\bs{h}^\star$ in \eqref{P-Prop37} is also at an equilibrium.

\subsection{Proof of Lemma \ref{L11}}

 Due to the continuity of $U_i(\cdot)$, there exists a Lipschitz constant $L_c$ such that, 
\begin{align}
|U_i(x_1)-U_i(x_2)|\leq L_c |x_1-x_2|,~\forall x_1>0, x_2>0.
\end{align}
Therefore, we have
\begin{align}
\lim_{\Delta x\rightarrow 0} \frac{U_i(x+\Delta x)-U_i(x)}{\Delta x}=U_i'(x)\leq L_c,
\end{align}
 i.e., $U_i'(x)$ is bounded for all $x>0$. Due to the strict concavity of $U_i(\cdot)$, $U_i'(x)$ is continuous and strictly decreasing for all $x>0$. The strictly decreasing and bounded $U_i'(x)$ implies that, for any $x>0$, there exists a variable $y$ such that $U_i'(x)=y$. Since $U_i'(\cdot)$ is the inverse function of $d_i(\cdot)$, we have that there exists a unique $y$ such that $d_i(y)=x$, for any $x\in(0,+\infty)$. Similarly, there exists a unique $\zeta_s$ such that $\sum_{i\in\mathcal{I}}d_i(\zeta_s)=\sum_{\ell=1}^sC_\ell$, for any $\sum_{\ell=1}^sC_\ell\in(0,+\infty)$.
 Therefore, we have shown the existence and uniqueness of $\zeta_s$ for any $\sum_{\ell=1}^sC_\ell$.

\subsection{Proof of Proposition \ref{LVI-1}}\label{P-LVI-1}

%

We first prove for the case $\zeta_{g^{\rm thr}-1}\leq \tilde{e}_{g^{\rm thr}}$, it admits only one traffic-supporting MNO  and satisfies $p_{i}^\star\leq \tilde{e}_{g^{\rm thr}},$ for all $i\in\mathcal{I}$. Consider the following two subcases:

\begin{itemize}
  \item Suppose that there exists an gateway-independent equilibrium delivered price $\bs{p}^{\star}$ with $p_{i}^{\star}=p_{j}^{\star}>\tilde{e}_{g^{\rm thr}},$ for all $i,j\in\mathcal{I}$. The total demand of all users satisfies
  \begin{equation}
  \sum_{i\in\mathcal{I}}d_{i}\left(p_{i}^{\star}\right)\overset{(b)}{<}\sum_{i\in\mathcal{I}}d_{i}(\tilde{e}_{g^{\rm thr}})\overset{(c)}{\leq}\sum_{l=1}^{g^{\rm thr}-1}C_l,\label{L2-E1}
  \end{equation}
  where $(b)$ is because $d_i(\cdot)$ is decreasing and (c) holds due to the case condition $\zeta_{{g^{\rm thr}}-1}\leq \tilde{e}_{g^{\rm thr}}$.
   Inequality \eqref{L2-E1} implies that users do not fully utilize the first ${g^{\rm thr}}-1$ downlinks.
  In this case, MNO $\sigma({g^{\rm thr}})$ can always reduce its hybrid prices $h_{i\leftarrow {g^{\rm thr}}}^{\star}\leftarrow p_{i}^{\star}-\tilde{e}_{g^{\rm thr}}-\epsilon$ for all users $i$, where $\epsilon>0$ is an arbitrarily small value. The new hybrid prices can attract downloads on downlink ${g^{\rm thr}}$. Since $ p_{i}^{\star}-\epsilon>\tilde{e}_{g^{\rm thr}}$, this improves MNO ${g^{\rm thr}}$'s profit and
 contradicts with the fact that $\bs{p}^{\star}$ is the equilibrium.
  \item Suppose that there exists $i$ such that $p_{i}^{\star}>\tilde{e}_{g^{\rm thr}}$ and $p_{i}^{\star}\neq p_{j}^{\star}$ for some $j$, then:
  \begin{itemize}
    \item if $\sum_{i\in\mathcal{I}}x_{i\leftarrow {g^{\rm thr}}}^\star<C_{g^{\rm thr}}$, then MNO $\sigma({g^{\rm thr}})$ can always set the price $h_{i\leftarrow {g^{\rm thr}}}^{\star}\leftarrow p_{i}^{\star}-\epsilon$ to attract more traffic demand according to Lemma 1, where $\epsilon>0$ is arbitrarily small. This increases its profit and contradicts with the fact that $\{\bs{p}^{\star}\}$ is the equilibrium.
    \item if $\sum_{i\in\mathcal{I}}x_{i\leftarrow {g^{\rm thr}}}^\star=C_{g^{\rm thr}}$ and $\sum_{j=1}^{{g^{\rm thr}}-1}\sum_{i\in\mathcal{I}}x_{i\leftarrow j}^\star\leq \sum_{j=1}^{{g^{\rm thr}}-1}C_j$, then MNO $\sigma(1)$ can always set  the delivered price
 $p_{i\leftarrow j}\leftarrow p_{i}^\star-\tilde{e}_{i\leftarrow j}-\epsilon$ to attract more traffic demand, which increases its profit and contradicts with the fact that $\bs{p}^{\star}$ is at the equilibrium.
  \end{itemize}
\end{itemize}

Thus, when $\zeta_{{g^{\rm thr}}-1}\leq \tilde{e}_{g^{\rm thr}}$, any equilibrium must satisfy $p_{i,\sigma(1)}^{\star}\leq \tilde{e}_{g^{\rm thr}},$ for all $i$. In addition, the equilibrium delivered price is lower than $\tilde{e}_{g^{\rm thr}}$, so that no other MNOs have the incentive to set an even lower delivered price to compete with MNO $\sigma(1)$. Hence, MNO $\sigma(1)$ is the only traffic-supporting MNO at any equilibrium.

Next, we prove that when $\zeta_{{g^{\rm thr}}-1}> \tilde{e}_{g^{\rm thr}}$, any equilibrium must satisfy $p_{i}^{\star}> \tilde{e}_{g^{\rm thr}},$ for all $i\in\mathcal{I}$ by contradiction. Suppose when $\zeta_{{g^{\rm thr}}-1}> \tilde{e}_{g^{\rm thr}}$, there exists an equilibrium with $p_{i}^{\star}\leq \tilde{e}_{g^{\rm thr}}$ for some $i$ then the total demand of all users satisfies
  \begin{equation}
 \sum_{i\in\mathcal{I}}d_{i}\left(p_{i}^{\star}\right)\geq \sum_{i\in\mathcal{I}}d_{i}(\tilde{e}_{g^{\rm thr}}) {>}\sum_{\ell=1}^{{g^{\rm thr}}-1}C_\ell,
  \end{equation}
which indicates that the users' total demand exceeds MNO $\sigma(1)$'s first ${g^{\rm thr}}-1$ downlinks and there must be some MNO $n\neq \sigma(1)$ serving some users. However, such an MNO $n$ incurs profit loss, since the delivered price is lower than the delivered cost for MNO $n$. Hence, MNO $n$ can always increase the hybrid prices to avoid such a profit loss, which contradicts with the fact that $\bs{p}^\star$ is an equilibrium.


Therefore, when $\zeta_{{g^{\rm thr}}-1}> \tilde{e}_{g^{\rm thr}}$, any equilibrium must satisfy $p_{i}^{\star}> \tilde{e}_{g^{\rm thr}}$ for each user $i$.

\subsection{Proof of Proposition \ref{T3}} \label{P-P3}

%

As Lemma \ref{LVI-1} asserts, when $\zeta_{{g^{\rm thr}}-1}\leq \tilde{e}_{g^{\rm thr}}$, any equilibrium pricing profile should $\bs{p}^{\star}$ satisfy $p_{i}^\star\leq \tilde{e}_{g^{\rm thr}}$ for all $i\in\mathcal{I}$ and MNO $\sigma(1)$ is the only traffic-supporting MNO in this case.

Hence, if the monopolist delivered price $\tilde{p}_i^*<\tilde{e}_{g^{\rm thr}}$ for each user $i$, then MNO $\sigma(1)$'s best strategy is to set $p_{i}^\star=\tilde{p}_{i}^\star$, since the remaining MNOs do not have any incentive to set lower prices due to their high operational cost.

 If MNO $\sigma(1)$ sets the price $p_{i}$ below $\tilde{e}_{g^{\rm thr}}$ for some $i$, then it has the incentive to increase $p_{i}$, since there is no competition from the remaining MNOs and the monopoly delivered price $\tilde{p}_i^\star$ is higher than $\tilde{e}_g$. Hence, at any equilibrium, the only possible equilibrium price for MNO $\sigma(1)$ is $p^{\star}_{i}=\tilde{e}_{g^{\rm thr}}$ for every user $i\in\mathcal{I}$.

\subsection{Proof of Theorem \ref{T5}} \label{P-P4}

Let $\mathcal{N}_e=\{\sigma(i):1\leq i\leq \hat{s}\}$ denote the set of all traffic-supporting MNOs. To prove Proposition \ref{T5}, we need the following lemmata.

\subsubsection{User Independence}
\begin{lemma}\label{L5}
Any gateway-independent multi-operator equilibrium prices $\bs{p}^\star$ are user-independent, i.e., $p_{i}^{\star}=p_{j}^{\star}$ for any users $i$ and $j$.
\end{lemma}


\begin{IEEEproof}
 We show this by contradiction.
Suppose that the equilibrium price matrix $p_{i}^{\star}<p_{j}^{\star}$ for some $i$ and $j$.
 Suppose that MNO $n$ is not the only MNO that supports user $j$'s traffic, i.e., $\sum_{j}\sum_{l\in\mathcal{I}}x_{j\leftarrow l}^\star<\sum_{j}\sum_{l\in\mathcal{I}_{n}}x_{j\leftarrow l}^\star$.
 It is easy to see, by setting $h_{j\leftarrow k}^{\star}\leftarrow p_{j}^{\star}-\tilde{c}_j-\epsilon$ for some $k\in\mathcal{I}_n$, user $j$ will prioritize the utilization of channel $k$ according to Lemma \ref{L5}. It can increase MNO $n$'s profit and
 contradicts with the fact that $\bs{p}^\star$ is the equilibrium delivered price. In other words, if the total received data of the user of higher price is not completely supported by MNO $n$, then MNO $n$ can slightly reduce the price to increase his profit. In this case, to maximize the total users' payoff, the user receiving a higher price will decide to receive all traffic via MNO $n$'s subscribers. Meanwhile, some of the remaining users will receive traffic via other MNOs so that the total downloaded data remains the same. Hence, MNO $n$ can increase her profit by doing so. This completes the proof of  Lemma \ref{L5}. 
\end{IEEEproof}

\subsubsection{Market Clearing Price}
We then introduce the following lemma.
\begin{lemma}\label{L6}
Any multi-operator equilibrium prices $\bs{p}^{\star}$ clear the market of first several downlinks, i.e., $\sum_{i\in\mathcal{I}}d_{i}(p_i^\star)=\sum_{\ell=1}^sC_\ell$ for some $s$.
\end{lemma}
\begin{IEEEproof}
	We prove Lemma \ref{L6} by contradiction. More specifically, consider that at an equilibrium  where  $\sum_{\ell=1}^{s-1}C_\ell<\sum_{i\in\mathcal{I}}d_{i}(p_i^\star)<\sum_{\ell=1}^{s}C_\ell$ for some $s$. We will show that  MNO $\sigma(s)$ can always increase its profit by setting a different pricing scheme.
 We must have that MNO $\sigma(s)$'s channel $s$ is not fully utilized, because users will utilize the low delivered cost downlinks first according to Assumption  1. Since MNO $\sigma(s)$ is not the only traffic-supporting MNO, there exists an user $i$ who does not receive the all her data from MNO $\sigma(s)$, i.e., $\sum_{i}\sum_{j\in\mathcal{I}}x_{i\leftarrow j}^*<\sum_{i}\sum_{j\in\mathcal{I}_{\sigma(s)}}x_{i\leftarrow j}^\star$. In this case, MNO $\sigma(s)$ can set $h_{i\leftarrow s}\leftarrow p_{i}^{\star}-c_j-\varepsilon$ for some traffic-supporting MNO $n\in\mathcal{N}_e$. This increases MNO $\sigma(s)$'s profit because
 \begin{itemize}
 	\item according to Lemma 1, user $i$ will first utilize channel $s$, and thus MNO $\sigma(s)$ can sell additional downloading data;
 	\item  MNO $\sigma(s)$'s delivered price is slightly decreased but still larger than $\tilde{e}_g$.
 \end{itemize}
  Hence, the fact that  MNO $\sigma(s)$'s profit increases contradicts with the fact that $\bs{p}^\star$ is at an equilibrium.
  Hence, any multi-operator equilibrium satisfies $\sum_id_{i}(p_i^\star)=\sum_{i=\ell}^sC_\ell$, i.e.,
$$p_{i}^\star=p_{j}^{\star}=\zeta_s,~\forall i,j\in\mathcal{I},$$
where $\zeta_s$ is the market-clearing delivered price defined in Definition \ref{D5}.
\end{IEEEproof}

\subsubsection{Critical Downlink}
We are ready to prove the remaining part of Theorem \ref{T3}.
According to Lemmata  \ref{L5} and \ref{L6}, 
 if $s<|\mathcal{I}|$ and $\zeta_{s}> e_{s+1}$, given the equilibrium $\bs{p}^\star$, MNO $\sigma(s+1)$ can set the hybrid price ${h}_{i\leftarrow s+1}\leftarrow \zeta_{s}-c_{s+1}-\epsilon$ for all $i\in\mathcal{I}$ to attract more traffic demand $\sum_{i\in\mathcal{I}}x_{i\leftarrow s+1}^\star>0$, resulting in a profit increase.
 Hence, we have: $\zeta_{s}\leq \tilde{e}_{s+1}$ if $s<|\mathcal{I}|$.  Note that $\zeta_{s}\geq \tilde{e}_{s}$, otherwise MNO $\sigma(s)$ can properly increase $h_{i\leftarrow s}$ for all $i\in\mathcal{I}$ to discourage user $s$ to download and tether, which increases MNO $\sigma(s)$'s profit.
Thus, we can see that $s$ must satisfy the conditions in \eqref{s}. In other words, a multi-operator equilibrium must satisfy:
\begin{equation}
p_{i}^\star=p_{j}^{\star}=\zeta_{\hat{s}},~\forall i,j\in\mathcal{I}.
\end{equation}

\subsection{Non-Existence of a Multi-Operator PCE}\label{AL}

The following example illustrates that  it is possible that a multi-operator PCE may not exist:

\begin{example}\label{E2}
	Consider an MCA of two users with utility functions $U_i(x)=4\log(1+x)$ for both $i\in\mathcal{I}$. User $1$ subscribes to MNO $1$ and user $2$ subscribes to MNO $2$. Each user $i$ has the downlink capacity of $C_i=1$ and zero cellular downlink energy cost. MNOs' operational costs are $e_1=1$ and $e_2=2$, respectively. 
\end{example}

Note that $\hat{s}=2$ satisfying that $2=\tilde{e}_{\hat{s}}\leq\zeta_{\hat{s}}=3$. By Proposition \ref{T5}, the only possible gateway-independent equilibrium hybrid prices are $h_{i\leftarrow j}=3=\zeta_{\hat{s}}$ for both $i,j\in\{1,2\}$. This leads to users' traffic decision such that
$\sum_{j\in\{1,2\}}\sum_{i\in\{1,2\}}x_{i\leftarrow 1}=2$ and MNO 2's profit $V_2=1$.

Now, consider another hybrid prices $h_{i\leftarrow 1}=3$ for both $i\in\{1,2\}$ and $h_{i\leftarrow 2}=4$ for both $i\in\{1,2\}$. This leads to users' traffic decision such that
$\sum_{i\in\{1,2\}}x_{i\leftarrow 1}=1$ and $\sum_{i\in\{1,2\}}x_{i\leftarrow 2}=3/5$ and MNO $2$'s profit $V_2=6/5$. We can verify that the first hybrid pricing scheme satisfies \eqref{s} and Proposition \ref{T5}, which indicates that it is the only possible gateway-independent pricing profile satisfying the necessary conditions in Proposition \ref{T5}. However, by increasing hybrid prices from $3$ to $4$, MNO $2$ can strictly increase its profit. Hence, there is no gateway-independent PCE or any arbitrary PCE according to Proposition \ref{Prop3}.

\subsection{Analysis of the Non-Equilibrium Simple Example} \label{DANE}

\begin{figure}
	\begin{centering}
		\includegraphics[scale=.38]{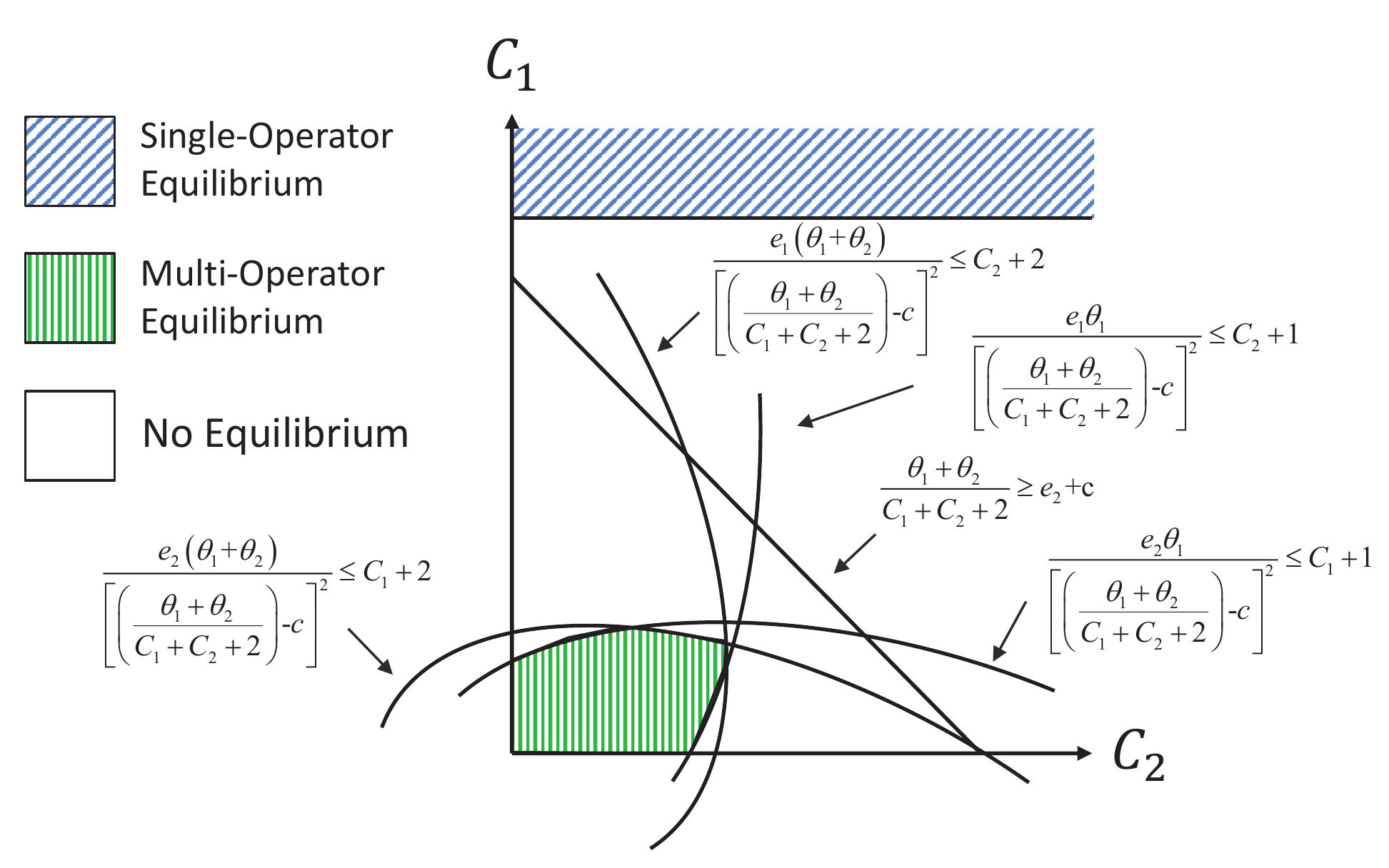}
		\vspace{-0.15cm}
		\caption{Full characterizations (in terms of different capacity combinations) of the PCE existence conditions for a simple example. There are two MNOs and two users. }
		\label{Chara}
	\end{centering}
\end{figure}

In this subsection, we present an illustrative example with two MNOs and two users. User $1$ subscribes to MNO $1$ and user $2$ subscribes to MNO $2$. We assume that user $i$'s utility function is 
\begin{align}
U_i(\cdot)=\theta_{i}\log\left(1+\sum_{j\in\{1,2\}}x_{i\leftarrow j}\right),~\forall i\in\{1,2\}.\label{loguu}
\end{align}
Without loss of generality, we assume 
 and $\tilde{e}_1<\tilde{e}_2$. We first characterize the PCE pricing profile  (if it exists),  and then present the sufficient and necessary conditions of the existence.
 
 \subsubsection{PCE Pricing Profile}It follows
\begin{itemize}
  \item when  $\zeta_{1}=(\theta_1+\theta_2)/(C_1+2)\leq \tilde{e}_2$, Proposition \ref{LVI-1} suggests that the single-operator PCE leads to the gateway-independent pricing strategy, given by 
\begin{align}
p_i^{\star}&=\min\{\sqrt{(e_1+c)\theta_i},\tilde{e}_2\}, &\forall i\in\{1,2\},
\end{align}
according to Proposition \ref{T3};

  \item when $\zeta_{1}=(\theta_1+\theta_2)/(C_1+2)> \tilde{e}_2$,  the equilibrium gateway-independent delivered prices $\{p_i^\star\}_{i=1,2}$ satisfy
\begin{align}
p_i^{\star}&=\zeta_{2}=\frac{\theta_1+\theta_2}{C_1+C_2+2}-c, &\forall i\in\{1,2\},\label{DANE-Eq1}
\end{align}
according to Theorem \ref{T5}.
\end{itemize}
Note that Proposition \ref{T3} ensures the existence of a single-operator PCE, but Theorem \ref{T5} does not guarantee the existence of a multi-operator PCE. This motivates us to further characterize the sufficient and necessary conditions of  \eqref{DANE-Eq1} being a multi-operator PCE.

%

%



 \subsubsection{Sufficient and Necessary Conditions of Existence}
To check whether \eqref{DANE-Eq1} is a PCE (defined in Definition \ref{D-PCE}), we first consider the following notations.
Let $\bs{\Delta p}$ be the arbitrarily small price changes for MNO $i$ at \eqref{DANE-Eq1}, $\Delta d_j (\bs{\Delta p})$ be the corresponding change in user $j$'s demand, and $\Delta V_i(\bs{\Delta p})$ be the corresponding change in MNO $i$'s profit.
According to the definition of the PCE,
\eqref{DANE-Eq1} is an exact PCE if neither of the MNO can improve its profit by selecting any small price deviation, i.e., 
\begin{align}
\Delta V_i(\bs{\Delta p})=\sum_{j\in\{1,2\}}(\Delta d_j(\bs{\Delta p}) &\zeta_2+\Delta p_{j}d_j)\leq 0,\label{DeltaVV}\\~&\forall \bs{\Delta p}, \forall i\in\{1,2\}\nonumber.
\end{align}

 In order to obtain $\Delta d_j (\bs{\Delta p})$, we adopt the KKT conditions to (UPM) in \eqref{KKTf}-\eqref{KKTl} and have that for both $j\in\{1,2\}$ and for all $i,\bar{i}\in\{1,2\}, i\neq\bar{i}$
 \begin{subequations}
\begin{align}
U_j'(d_j(\zeta_2)+\Delta d_j(\bs{\Delta p}))-\zeta_2-c-\Delta p_j-\lambda_{i}+\mu_{j\leftarrow i}=0,\label{UPND1}\\
U_j'(d_j(\zeta_2)+\Delta d_j(\bs{\Delta p}))-\zeta_2-c-\lambda_{\bar{i}}+\mu_{j\leftarrow \bar{i}}=0,\label{UPND2}\\
U_{\bar{j}}'(d_j(\zeta_2)+\Delta d_j(\bs{\Delta p}))-\zeta_2-c-\Delta p_{\bar{j}}-\lambda_{i}+\mu_{\bar{j}\leftarrow i}=0,\label{UPND3}\\
U_{\bar{j}}'(d_j(\zeta_2)+\Delta d_j(\bs{\Delta p}))-\zeta_2-c-\lambda_{\bar{i}}+\mu_{\bar{j}\leftarrow \bar{i}}=0.\label{UPND4}
\end{align}
\end{subequations}
Similar to the arguments in the proof of Theorem \ref{T1}, under the MNO $i$'s best decision $\bs{\Delta{p}}^*$, the dual variable to downlink $i$'s capacity constraint $\lambda_i^*=0$. Otherwise, MNO $i$ can also set $\Delta p_j'\leftarrow\Delta p_j+\lambda_i^*$ to increase its profit.

Without loss of generality, we choose $\Delta p_j< \Delta p_{\bar{j}}$. By \eqref{UPND1}-\eqref{UPND4}, we have that $\mu_{j\leftarrow \bar{i}}^*=0$ and $\mu_{\bar{j}\leftarrow i}^*=0$.

It follows that
\begin{itemize}
  \item If $d_j(\zeta_2)<C_{\bar{i}}$, then user $j$ cannot fully utilize the downlink $\bar{i}$ alone. Hence we have that $\mu_{\bar{j}\leftarrow \bar{i}}^*=0$. We have
\begin{align}
\lambda_{\bar{i}}^*= \Delta p_{\bar{j}}.
\end{align}
That is, user $\bar{j}$ experiences a shadow price equal to $\Delta p_{\bar{j}}$ and her change
in demand $\Delta d_{\bar{j}}=\Delta p_{\bar{j}} \frac{\partial d_{\bar{j}}(\zeta_2)}{\partial p_{\bar{j}}}$. Hence, it follows that
  \begin{align}
\Delta V_i(\bs{\Delta p})= &\left(\frac{\partial d_j(\zeta_2)}{\partial p_j} \Delta p_j\right)(\zeta_{2}-e_i)+\Delta p_j d_j\nonumber\\
&+\left(\frac{\partial d_{\bar{j}}}{\partial p_{\bar{j}}} \rho\Delta p_j\right)(\zeta_{2}-e_i)+ \rho \Delta p_jd_{\bar{j}}, \label{IE2}
\end{align}
where $\rho\triangleq\Delta p_{\bar{j}}/\Delta p_j\in(-\infty,1]$.
  \item Otherwise, user $j$ can fully utilize the downlink $\bar{i}$ and we thus have  $\mu_{\bar{j}\leftarrow \bar{i}}^*>0$ and $\mu_{j\leftarrow i}^*=0$. In this case, we have
\begin{align}
\lambda_{\bar{i}}^*= \Delta p_{j}.
\end{align}
Thus,
\begin{align}
\Delta V_i(\bs{\Delta p})= &\left(\frac{\partial d_j(\zeta_2)}{\partial p_j}+\frac{\partial d_{\bar j}(\zeta_2)}{\partial p_{\bar j}} \right)\Delta p_j(\zeta_{2}-e_i)\nonumber\\
&+\Delta p_jC_{i}.\label{IE3}
\end{align}

\end{itemize}
 Combining \eqref{DeltaVV}, \eqref{IE2}, \eqref{IE3},  $d_i(\zeta_2)=\theta_i/(\zeta_2+c)-1$, and $\partial d_j(\zeta_2)/\partial p_j=-\theta_i/(\zeta_2+c)^2$ for the utility in \eqref{loguu},
  it follows that the pricing profile in \eqref{DANE-Eq1} is a multi-operator PCE if and only if
\begin{align}
\begin{cases} \frac{e_i\theta_1}{\zeta_2^2}\leq C_{\bar{i}}+1\\
\frac{e_i(\theta_1+\theta_2)}{\zeta_2^2}\leq C_{\bar{i}}+2\end{cases} \forall i, \bar{i}\in\{1,2\}, i\neq \bar{i}.
\end{align}
Therefore, we can summarize the above conditions in Fig. \ref{Chara}.

\subsection{Free-Tethering Competitive Pricing}\label{FTCOMP}
Similar to Section \ref{FTPCoop}, we also derive conditions under which the gateway-independent PCE $\bs{h}^\star$ leads to zero tethering prices, i.e., $t_{i\leftarrow j}^\star=0$ for all $i,j\in\mathcal{I}$. We summarize the conditions in the following corollary:
\begin{corollary}
	A PCE leads to zero tethering prices ($t_{i\leftarrow j}^\star=0$ for all $ i,j\in\mathcal{I}$) when  the following conditions hold
	\begin{itemize}
		\item for a single-operator PCE $(\zeta_{{g^{\rm thr}}-1}\leq \tilde{e}_{g^{\rm thr}})$: the users' cellular downlink energy costs are the same, i.e., $c_{i}^\text{Down}=c_{j}^\text{Down}$ for all $i,j\in\mathcal{I}$, and users have isoelastic utility functions.
		\item for a multi-operator PCE $(\zeta_{{g^{\rm thr}}-1}> \tilde{e}_{g^{\rm thr}})$: a multi-operator PCE exists and the users' cellular downlink energy costs are the same, i.e., $c_{i}^\text{Down}=c_{j}^\text{Down}$ for all $i,j\in\mathcal{I}$.
	\end{itemize}	
\end{corollary}

For a single-operator PCE, the conditions are the same as the cooperative pricing scheme, since we obtain these two pricing schemes in a similar fashion.
In addition, different from Corollary \ref{coro1}, zero-tethering prices at a multi-operator PCE do not rely on users' isoelastic utility functions. This is because the competition among MNOs intrinsically leads to the client-independent delivered prices, as suggested by Proposition \ref{T5}.

\subsection{Proof of Theorem \ref{T11}}

We will prove that, when any multi-operator PCE $\bs{p}^\star$ (satisfying $p_{i}^\star=\zeta_{\hat{s}}$ for all $i\in\mathcal{I}$), the quantity profile ($\bs{q}^\star$ satisfying
$q_n^\star=\sum_{\ell=1, \ell\notin\mathcal{I}_n}^{\hat{s}}{C_\ell}$ for all $n\in\mathcal{N}$ suggested by Theorem \ref{T5}) is a QCE.
We prove this that by showing that, if it is impossible to improve an MNO $n$'s profit by increasing its pricing strategy in a PCG, and it is also impossible to improve its profit by changing its quantity strategy in the corresponding QCG.

\subsubsection{Characterization of Changes in  Users' Demand}
We first prove the following lemma, which characterizes the small change of the total users' demand when an MNO $n$ slightly changes its hybrid prices:
 \begin{lemma}\label{C111}
At a multi-operator PCE of a PCG described in Theorem \ref{T5}, if MNO $n$ increases all its hybrid prices $\{h_{i\leftarrow j}^*\}_{i\in\mathcal{I}, j\in\mathcal{I}_n}$ by an arbitrarily small amount $\Delta p$, there is a decease in total users' demand of $\Delta p\cdot \sum_{i\in\mathcal{I}}\frac{\partial d_i(\zeta_{\hat{s}})}{\partial p}$.
\end{lemma}
\begin{IEEEproof}
 Suppose that MNO $n$ changes its PCE hybrid prices $\{h_{i\leftarrow j}^*\}_{j\in\mathcal{I}}$ to $\{h_{i\leftarrow j}'\}_{j\in\mathcal{I}}$, where $h_{i\leftarrow j}'=h_{i\leftarrow j}^*+\Delta p$ for all $i\in\mathcal{I},~j\in\mathcal{I}_n.$ Through Theorem \ref{T5}, we have $h_{i\leftarrow j}^*+c_{j}=\zeta_{\hat{s}}$ for some $\hat{s}$.
 According to Theorem \ref{T1} and the KKT conditions for (UPM) in \eqref{KKTf}-\eqref{KKTl}, the users' optimal primal-dual solution $(\bs{x}',\bs{\lambda}')$ satisfies
 \begin{subequations}
\begin{align}
d_i'&=\sum_{j\in\mathcal{I}}x_{i\leftarrow j}',~\forall i\in\mathcal{I},\\
U_i'(d_i)&=\begin{cases}\zeta_{\hat{s}}+\lambda_j'+\Delta p,~\forall j\in\mathcal{I}_n,\\
\zeta_{\hat{s}}+\lambda_j',~~~~~~~~\forall j\notin\mathcal{I}_n,
\end{cases}\label{Dp}\\
\sum_{j\in\mathcal{I}}x_{i\leftarrow j}'&\leq C_j,~\forall j\in\mathcal{I},\\
\lambda_j'&=\begin{cases}0,~~~~~\forall j\in\mathcal{I}_n,\\
\Delta p,~~~\forall j\notin\mathcal{I}_n.\label{Dp1}\end{cases}
\end{align}
 \end{subequations}
Hence,  \eqref{Dp} and \eqref{Dp1} show that every user experiences a delivered price increase of $\Delta p$, which implies that
\begin{equation}
\lim_{\Delta p\rightarrow 0}\frac{ \sum_{i\in\mathcal{I}}d_i(\zeta_{\hat{s}}+\Delta p)-\sum_{i\in\mathcal{I}}d_i(\zeta_{\hat{s}})}{\Delta p}=\sum_{i\in\mathcal{I}}\frac{\partial d_i(\zeta_{\hat{s}})}{\partial p}.
\end{equation}
This completes the proof of  Lemma \ref{C111}.
\end{IEEEproof}

\subsubsection{Characterization of the QCE}
We now prove the remaining part of Theorem \ref{T11}.
To show $\{q_n^\star\}_{n\in\mathcal{N}}$ is the QCE of the QCG, it is sufficient to show
\begin{align}
&\frac{\partial (q_n^\star\pi(\boldsymbol{q}^\star))}{\partial q_n}\nonumber\\
=&\pi\left(\sum_{\ell=1}^{\hat{s}}C_\ell\right)+q_n^\star\frac{\partial \pi(\boldsymbol{q}^\star)}{\partial q_n}\begin{cases}\geq \tilde{e}_i,~~\forall i\leq\hat{s},\\
\leq \tilde{e}_i,~~\forall i>\hat{s}.\end{cases}\label{T111}
\end{align}
Note that we have $\pi\left(\sum_{\ell=1}^{\hat{s}}C_\ell\right)=\zeta_{\hat{s}}$, $\tilde{e}_{\hat{s}}\leq \zeta_{\hat{s}}\leq \tilde{e}_{\hat{s}+1}$, and $z_i^\star=0$ for all $i>\hat{s}$ due to Proposition \ref{T3}.
Hence, \eqref{T111} holds for all $i>\hat{s}$.

\subsubsection{A Multi-operator PCE is a QCE}
Next, we will prove that \eqref{T111} holds for all cases of $i\leq \hat{s}$.
Since at a multi-operator PCE of the PCG, no MNO can increase its profit by changing the prices. Thus, since any MNO $n$ cannot improve its profit by increasing its equilibrium hybrid prices by an arbitrarily small amount of $\Delta p$, Lemma \ref{C111} yields: for all $i\leq \hat{s}$
\begin{align}
&\hspace{-0.5cm}~~\frac{\partial[(p-\tilde{e}_i)(\sum_{\ell\in\mathcal{I}} d_\ell(\cdot)-\sum_{\ell=1, \ell\notin\mathcal{I}_n}^{\hat{s}}{C_\ell})]}{\partial p}\Big|_{p=\zeta_{\hat{s}}}\nonumber\\
&\hspace{-0.5cm}=\sum_{\ell\leq \hat{s}, \ell\in \mathcal{I}_{n}}{C_\ell}+\sum_{\ell\in\mathcal{I}}\frac{\partial \left(  d_\ell(\zeta_{\hat{s}})\right)}{\partial p}(\zeta_{\hat{s}}-\tilde{e}_i)&\leq 0\label{MOPCE2}.
\end{align}
By Proposition \ref{T3}, it follows that $\sum_{\ell\leq \hat{s}, \ell\in \mathcal{I}_{n}}{C_\ell}=q_n^\star$. Multiplying both sides of \eqref{MOPCE2} by $\partial \pi(\cdot)/\partial q_n$, we have:
\begin{align}
q_n^\star\frac{\partial \pi(\cdot)}{\partial q_n}+\sum_{\ell\in\mathcal{I}}\frac{\partial  d_\ell(\cdot)}{\partial p}\frac{\partial \pi(\cdot)}{\partial q_n}(p-\tilde{e}_i)\geq 0,~~~\forall i\leq \hat{s}\label{AAAAA2}.
\end{align}
Note that
\begin{align}
&\sum_{\ell\in\mathcal{I}}\frac{\partial  d_\ell(\cdot)}{\partial p}\left(\frac{\partial \pi(\cdot)}{\partial (\sum_{\ell\in\mathcal{I}} d_\ell(\cdot))}\frac{\partial (\sum_{\ell\in\mathcal{I}} d_\ell(\cdot))}{\partial q_n}\right)\nonumber\\
=&\sum_{\ell\in\mathcal{I}}\frac{\partial  d_\ell(\cdot)}{\partial p}\frac{\partial \pi(\cdot)}{\partial q_n}\nonumber\\
=&1,\label{AAAAA3}
\end{align}
because $\sum_{\ell\in\mathcal{I}}d_\ell(\cdot)$ is the inverse function of $\pi(\cdot)$.
Combining \eqref{AAAAA2} and \eqref{AAAAA3},  we have, for all  users $i\leq \hat{s}$,
\begin{align}
\frac{\partial (q_n\pi(\boldsymbol{q}^\star))}{\partial q_n}=\pi\left(\sum_{\ell=1}^{\hat{s}}C_\ell\right)+q_n^\star\frac{\partial \pi(\boldsymbol{q}^\star)}{\partial q_n}\geq \tilde{e}_i~\nonumber.
\end{align}
Therefore,  no MNO can improve its profit by changing its quantity strategy when all MNO $n$'s quantity strategy is $q_n^\star=\sum_{\ell=1, \ell\notin\mathcal{I}_n}^{\hat{s}}{C_\ell}$.
This completes  the proof of  Theorem \ref{T11}.

\subsection{Proof of Proposition \ref{PP5}}\label{Limit}

\begin{figure}
	\begin{centering}
		\includegraphics[scale=.4]{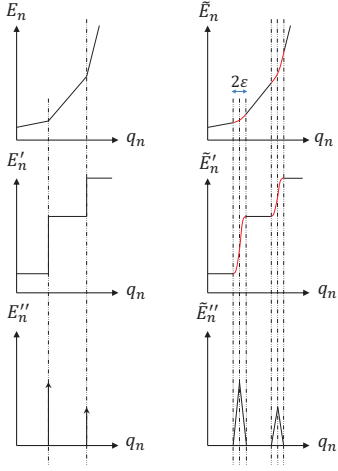}
		\caption{Comparison of the first order and second order derivatives of the piece-wise linear cost function $E_n(q_n)$ and its smoothed approximation $\tilde{E}_n(q_n)$.}
		\label{CostFunctions}
	\end{centering}
\end{figure}

 To prove Proposition \ref{PP5},
 let Game A denote the QCG where MNOs have smoothed operational cost functions $\tilde{E}_n(q_n)$ in \eqref{smooth}, where $\varepsilon$ is the term with an arbitrary small value. Let Game B denote the QCG with MNOs having the non-smoothed operational cost function ${E}_n(q_n)$ in \eqref{EEE}.\footnote{Fig. \ref{CostFunctions} demonstrates the difference between  ${E}_n(q_n)$ in \eqref{EEE} and $\tilde{E}_n(q_n)$ in \eqref{smooth}.}
 We then define $$\varphi_n(c)\triangleq\arg\max_{q_n\in\mathcal{Q}_n}[q_n\pi(b)+q_n^2/2\pi'(b)-{E}_n(q_n)],$$ and $$\tilde\varphi_n(c,\varepsilon)\triangleq\arg\max_{q_n\in\mathcal{Q}_n}[q_n\pi(b)+q_n^2/2\pi'(b)-\tilde{E}_n(q_n,\varepsilon)].$$ Next, we prove the limit of $\tilde{\varphi}_n(c,\varepsilon)$ as $\varepsilon$ approaches $0$ is $\varphi_n(c)$. Finally, we prove that the limit of a QCE of Game A is a QCE of Game B.




\subsubsection{The Limit of $\tilde{\varphi}_n(c,\varepsilon)$} 

By the maximum theorem, since $q_n\pi(q_n+q_{-n})-\tilde{E}_n(q_n,\varepsilon)$ is strictly quasi-concave in $q_n$ and $\mathcal{Q}_n$ is convex for all $\varepsilon$, $\tilde{\varphi}_n(c,\varepsilon)$ is continuous in $\varepsilon$.
Since both Game A and Game B satisfy the unimodal condition in \cite{IterCour}, functions $\tilde{\varphi}_n(c,\varepsilon)$ and $\varphi_n(c)$ are unique-valued, for any $c$ and $\varepsilon$.
The continuity of $\tilde{\varphi}_n(c,\varepsilon)$ leads to
\begin{align}
&~~~~\lim_{\varepsilon\rightarrow 0}\tilde{\varphi}_n(c,\varepsilon)\nonumber\\
&=
\lim_{\varepsilon\rightarrow 0}\arg\max_{q_n\in\mathcal{Q}_n}[q_n\pi(b)+q_n^2/2\pi'(b)-\tilde{E}_n(q_n,\varepsilon)]\nonumber\\
&=
\arg\max_{q_n\in\mathcal{Q}_n}[q_n\pi(b)+q_n^2/2\pi'(b)-\lim_{\varepsilon\rightarrow 0}\tilde{E}_n(q_n,\varepsilon)]\nonumber\\
&=\varphi_n(c). \label{83}
\end{align}

\subsubsection{The limit of a QCE of Game A}
In Game A, there always exists a QCE (by Proposition \ref{P6}).  Lemma \ref{L3} suggests that $\tilde{\boldsymbol{q}}^*(\varepsilon)$ is a QCE of Game A if and only if there exists $\tilde{c}^*(\varepsilon)$ such that
\begin{equation}
\sum_{n\in\mathcal{N}}\tilde{q}_n(\varepsilon)=\tilde{c}^*(\varepsilon)=\sum_{n\in\mathcal{N}}\tilde{\varphi}_{n}\left(\tilde{c}^*(\varepsilon),\varepsilon\right)
\end{equation}
and
\begin{equation}
\tilde{q}_{n}^*(\varepsilon)=\tilde{\varphi}_{n}\left(\tilde{c}^*(\varepsilon),\varepsilon\right),~\forall n\in\mathcal{N}.
\end{equation}
 That is, $\tilde{c}^*(\varepsilon)$ is a fixed point of $\sum_{n\in\mathcal{N}}\tilde{\varphi}_{n\in\mathcal{N}}\left(\tilde{c}^*(\varepsilon) ,\varepsilon\right)$.
It follows that
\begin{equation}
\lim_{\varepsilon\rightarrow 0}\tilde{c}^*(\varepsilon)=\lim_{\varepsilon\rightarrow 0}\sum_{n\in\mathcal{N}}\tilde{\varphi}_n\left(\lim_{\varepsilon\rightarrow 0}\tilde{c}^*(\varepsilon),\varepsilon\right)\overset{(a)}{=}\sum_{n\in\mathcal{N}}\varphi_n(\lim_{\varepsilon\rightarrow 0}\tilde{c}^*\left(\varepsilon)\right), \label{86}
\end{equation}
where equality $(a)$ is due to \eqref{83}.
Therefore, Lemma \ref{L3} and \eqref{86} imply that  $\lim_{\varepsilon\rightarrow 0}\tilde{c}^*(\varepsilon)$ is a fixed point of $\sum_{n\in\mathcal{N}}\varphi_n(\cdot)$ and  $\boldsymbol{q}^*$ is a QCE of Game B, where
\begin{equation}
q_n^*=\varphi_n\left(\lim_{\varepsilon\rightarrow 0}\tilde{c}^*(\varepsilon)\right),~~\forall n\in\mathcal{N}.
\end{equation}
This completes the proof.

\subsection{Proof for Lemma \ref{L3}} \label{P-L3}
By \eqref{profit4} and the strict quasi-concavity of each MNO's profit, $\bs{q}^\star$ is a QCE if and only if, for every MNO $n$, the following KKT conditions are satisfies: 
\begin{subequations}\label{P-L3-1}
	\begin{align}
	\hspace{-0.55cm}\pi\left(\sum_{\ell\in\mathcal{N}}q_\ell\right)+q_n\pi'\left(\sum_{\ell\in\mathcal{N}}q_\ell\right)-\tilde{E}_n'(q_n)-\hat\lambda_n+\hat\mu_n&= 0,\\
	\hat\lambda_n\left(q_n-\sum_{\ell\in\mathcal{I}_n}C_\ell\right)&=0,\\
	\hat\mu_nq_n&=0.
	\end{align}
\end{subequations}
The KKT conditions of $x=\varphi_n(b)$ lead to, for every MNO $n$, 
\begin{subequations}\label{P-L3-2}
	\begin{align}
	\pi\left(b\right)+x\pi'\left(b\right)-\tilde{E}_n'(x)-\tilde{\lambda}_n+\tilde{\mu}_n&= 0,\\
	\tilde{\lambda}_n\left(x-\sum_{\ell\in\mathcal{I}_n}C_\ell\right)&=0,\\
	\tilde{\mu}_nx&=0.
	\end{align}
\end{subequations}
Therefore, comparing \eqref{P-L3-1} to \eqref{P-L3-2}, we can see $\bs{q}^\star$ is a QCE if and only if $q_n^\star=\varphi_n(\sum_{l}q_l^\star)$. This happens if and only if  $\sum_{l}q_l^\star=\Phi(\sum_{l}q_l^\star)$.

\subsection{Proof of Proposition \ref{P6}}\label{P-P6}


We use the Brouwer's fixed-point theorem to prove the existence of variable $b$ such that $b=\Phi(b)$. To do this, we will prove that (i) $\Phi(b)$ is continuous on $\left[0, \sum_{i\in\mathcal{I}}C_i\right]$ and (ii) $\Phi(b)$ maps $\left[0, \sum_{i\in\mathcal{I}}C_i\right]$ into itself.

%
%

\subsubsection{Continuity of $\Phi(b)$}
 Define $f_n(b)\triangleq x\pi(b)+x^2\pi'(b)/2-\tilde{E}_n(x)$ for all $n\in\mathcal{N}$. Function $f_n(b)$ is continuous because (i) $\pi(\cdot)$ is continuous, which is because its inverse function $D(\cdot)=\sum_{i\in\mathcal{I}}d_i(\cdot)$ is continuous; (ii) $\pi'(\cdot)=1/(\sum_{i\in\mathcal{I}}d_i'(\cdot))$ is continuous; and (iii) $\tilde{E}'_n(x)$ is continuous due to the smoothed approximation.

By the maximum theorem, if $f_n(b)$ is continuous in $b$ and strictly concave in $x$, then $\varphi(b)$ is continuous in $b$. Therefore, $\varphi_n(b)$ is continuous.

\subsubsection{Function $\Phi(b)$ is Self-Mapping}
 Recall that $\varphi_n(b)$ is lower-bounded by $0$ and upper-bounded by $\sum_{\ell\in\mathcal{I}_n}C_\ell$, which ensures that $0\leq \Phi(b)\leq \sum_{\ell\in\mathcal{I}} C_\ell$. This indicates that $\Phi(b)$ is a mapping that maps $[0,\sum_{\ell\in\mathcal{I}} C_\ell]$ into itself.

Combining the above discussions, we conclude that there exists a $b\in[0,\sum_{\ell\in\mathcal{I}} C_\ell]$ such that $b=\Phi(b)$ by Brouwer's fixed-point theorem.

\subsection{Proof for Proposition \ref{P7}} \label{P-P7}
 According to \cite{moulin1984dominance}, for a game where each player's payoff is concave and continuous in its strategy and continuous in others' strategies, there exists a unique QCE if
 \begin{equation}
 -\frac{\partial^2 V_n}{\partial q_n^2}(\bs{q})>\sum_{\bar{n} \neq n}\Big|\frac{\partial^2 V_{\bar{n}}}{\partial q_n\partial q_{\bar{n}}}(\bs{q})\Big|,~\forall n\in\mathcal{N}. \label{moulin}
 \end{equation}
 If there are only two MNOs, we have
\begin{align}
-\frac{\partial^2 V_n}{\partial q_n^2}&=-q_n\frac{\partial^2 \pi}{\partial q_n^2}-2\frac{\partial \pi}{\partial q_n}+\frac{\partial^2 \tilde{E}_n}{\partial q_n^2},~\forall n\in\{1,2\},\\
\Big|\frac{\partial^2 V_{\bar{n}}}{\partial q_n\partial q_{\bar{n}}}\Big|&=\Big|-q_n\frac{\partial^2 \pi}{\partial q_n^2}-\frac{\partial \pi}{\partial q_n}\Big|,~\forall n\in\{1,2\}.
\end{align}
Since $\pi(x)x$ is quasi-concave, $\pi(x)$ is twice continuously differentiable and decreasing, and $\tilde{E}_n$ is twice continuously differentiable and convex, we see that \eqref{moulin} always holds, and thus the game with two MNOs admits a unique QCE.




\subsection{Proof for Proposition \ref{P8}} \label{P-P8}

According to \cite{meanvalue}, Algorithm \ref{Algo2} converges to the fixed point of $f(x)$ if (i) $f(x)$ is continuous and (ii) $f(x)$ maps $[0,1]$ into itself. Note that $\Phi(b)$ maps $[0,\sum_{i\in\mathcal{I}}C_i]$ into itself, and  we can normalize the input and output $\Phi(b)$ to satisfy condition (ii). We have already proved these two properties in Appendix \ref{P-P6}.

\subsection{Supplementary Numerical Results} \label{SNR}


 In this subsection, we will study the performances in terms of users' payoff  for the next two sets of experiments.

\begin{figure}[!t]
	\begin{centering}
		\subfigure [] {\includegraphics[scale=.29]{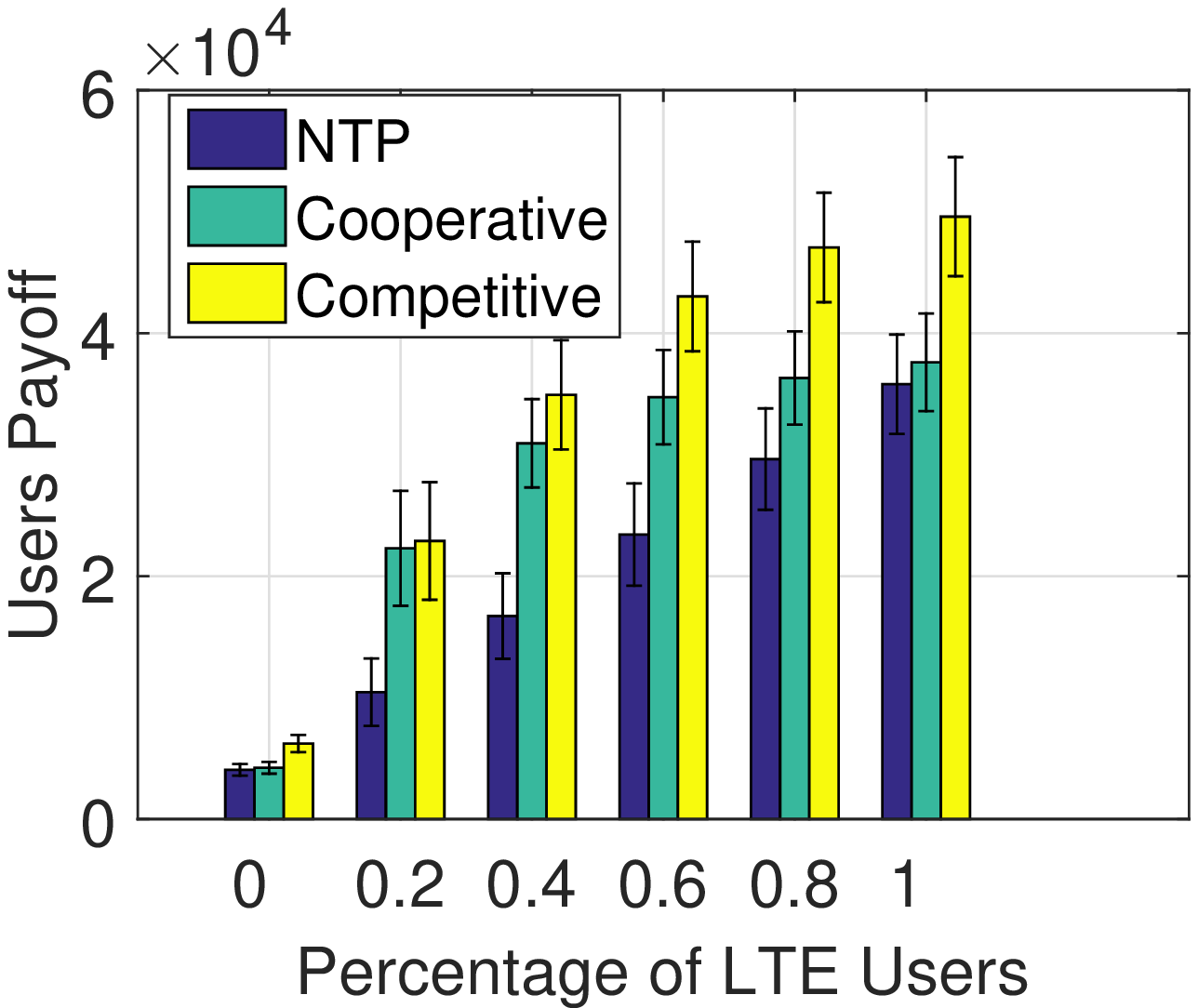}}
		\subfigure []{\includegraphics[scale=.29]{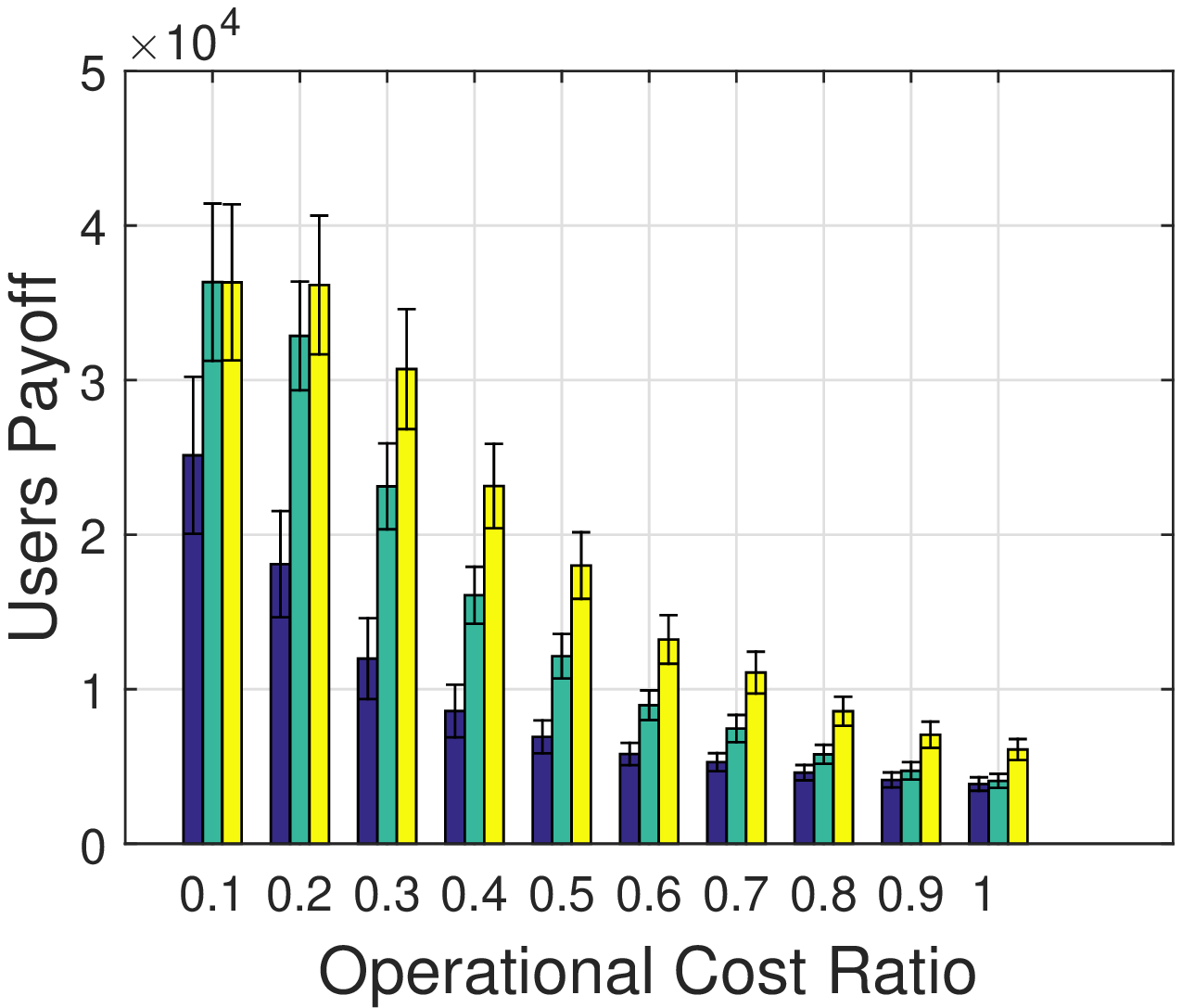}}
		\caption{Users' profit versus (a) the percentage of the LTE users and (b) the ratio of the means of LTE operational cost and 3G operational cost (the LTE operational cost distribution $e_{\rm LTE}\sim TN(\eta\times 350,\eta\times 40)$). }
		\label{f7}
	\end{centering}
\end{figure}

\textbf{Percentage of LTE users}: Fig. \ref{f7}(a) shows  a similar trend regarding the network heterogeneity in terms of the percentage of LTE users. As shown in Fig. \ref{f6}(a), when the percentage of LTE users is between $20\%$-$60\%$, the cooperative schemes' benefits in terms of users' payoff  increases are large, compared with the NTP scheme.  However, when the percentage is $0\%$ or $100\%$, the benefit in terms of the users' payoff is small, compared with the NTP scheme.
This implies that when users are homogeneous, each users' payoff gained from the MCA due to resource pooling is limited. The competitive MNOs further improves the 
the users' payoff.


\textbf{LTE operational cost}: Finally, we study the LTE operational cost's impact on users' payoff in Fig. \ref{f7}(b). Specifically, we consider an operational cost ratio $\eta\in[0,1]$ and set $e_{\rm LTE}\sim TN(\eta\times 350,\eta\times 40)$. Therefore,  a larger $\eta$ leads to a LTE operational cost closer to the 3G operational cost. We show that as $\eta$ increases, users' payoff improvements for both proposed schemes decrease, compared with the NTP scheme in Fig. \ref{f6}(b).

\end{document}